\documentclass[aps,prd,twocolumn,twoside,superscriptaddress,floatfix, nofootinbib]{revtex4-1}

\usepackage{graphicx}
\usepackage{ifthen,amsmath,amssymb, bm}
\usepackage{graphicx}
\usepackage[dvipsnames]{xcolor}
\usepackage{hyperref}
\usepackage{float}
\usepackage{aas_macros}
\usepackage[export]{adjustbox}
\usepackage{bookmark}

\newcommand{\beq}{\begin{equation}}
\newcommand{\eeq}{\end{equation}}

\begin{document}

\title{Precision Kinematic Sunyaev--Zel’dovich Measurements Across Halo Mass and Redshift with DESI DR2 and ACT DR6: Part II. Bright Galaxy Survey and Emission-Line Galaxies}

\author{B.~Hadzhiyska}
\email{boryanah@ast.cam.ac.uk}
\affiliation{Institute of Astronomy, Madingley Road, Cambridge, CB3 0HA, UK}
\affiliation{Kavli Institute for Cosmology Cambridge, Madingley Road, Cambridge, CB3 0HA, UK}
\author{S.~Ferraro}
\affiliation{Lawrence Berkeley National Laboratory, 1 Cyclotron Road, Berkeley, CA 94720, USA}
\affiliation{Berkeley Center for Cosmological Physics, Department of Physics, University of California, Berkeley, CA 94720, USA}
\author{F.~J.~Qu}
\affiliation{Kavli Institute for Particle Astrophysics and Cosmology, Stanford University, 452 Lomita Mall, Stanford, CA, 94305, USA}
\affiliation{Department of Physics, Stanford University, 382 Via Pueblo Mall, Stanford, CA, 94305, USA}
\affiliation{SLAC National Accelerator Laboratory, 2575 Sand Hill Road, Menlo Park, California 94025, USA}
\author{B.~Ried~Guachalla}
\affiliation{Kavli Institute for Particle Astrophysics and Cosmology, Stanford University, 452 Lomita Mall, Stanford, CA, 94305, USA}
\affiliation{Department of Physics, Stanford University, 382 Via Pueblo Mall, Stanford, CA, 94305, USA}
\affiliation{SLAC National Accelerator Laboratory, 2575 Sand Hill Road, Menlo Park, California 94025, USA}
\author{E.~Schaan}
\affiliation{Kavli Institute for Particle Astrophysics and Cosmology, Stanford University, 452 Lomita Mall, Stanford, CA, 94305, USA}
\affiliation{Department of Physics, Stanford University, 382 Via Pueblo Mall, Stanford, CA, 94305, USA}
\affiliation{SLAC National Accelerator Laboratory, 2575 Sand Hill Road, Menlo Park, California 94025, USA}

\author{J.~Aguilar}
\affiliation{Lawrence Berkeley National Laboratory, 1 Cyclotron Road, Berkeley, CA 94720, USA}
\author{S.~Ahlen}
\affiliation{Department of Physics, Boston University, 590 Commonwealth Avenue, Boston, MA 02215 USA}
\author{D.~Bianchi}
\affiliation{Dipartimento di Fisica ``Aldo Pontremoli'', Universit\`a degli Studi di Milano, Via Celoria 16, I-20133 Milano, Italy}
\affiliation{INAF-Osservatorio Astronomico di Brera, Via Brera 28, 20122 Milano, Italy}
\author{D.~Brooks}
\affiliation{Department of Physics \& Astronomy, University College London, Gower Street, London, WC1E 6BT, UK}
\author{F.~J.~Castander}
\affiliation{Institut d'Estudis Espacials de Catalunya (IEEC), c/ Esteve Terradas 1, Edifici RDIT, Campus PMT-UPC, 08860 Castelldefels, Spain}
\affiliation{Institute of Space Sciences, ICE-CSIC, Campus UAB, Carrer de Can Magrans s/n, 08913 Bellaterra, Barcelona, Spain}
\author{E.~Chaussidon}
\affiliation{Lawrence Berkeley National Laboratory, 1 Cyclotron Road, Berkeley, CA 94720, USA}
\author{T.~Claybaugh}
\affiliation{Lawrence Berkeley National Laboratory, 1 Cyclotron Road, Berkeley, CA 94720, USA}
\author{A.~de la Macorra}
\affiliation{Instituto de F\'{\i}sica, Universidad Nacional Aut\'{o}noma de M\'{e}xico,  Circuito de la Investigaci\'{o}n Cient\'{\i}fica, Ciudad Universitaria, Cd. de M\'{e}xico  C.~P.~04510,  M\'{e}xico}
\author{Arjun~Dey}
\affiliation{NSF NOIRLab, 950 N. Cherry Ave., Tucson, AZ 85719, USA}
\author{Biprateep~Dey}
\affiliation{Department of Astronomy \& Astrophysics, University of Toronto, Toronto, ON M5S 3H4, Canada}
\affiliation{Department of Physics \& Astronomy and Pittsburgh Particle Physics, Astrophysics, and Cosmology Center (PITT PACC), University of Pittsburgh, 3941 O'Hara Street, Pittsburgh, PA 15260, USA}
\author{P.~Doel}
\affiliation{Department of Physics \& Astronomy, University College London, Gower Street, London, WC1E 6BT, UK}
\author{J.~E.~Forero-Romero}
\affiliation{Departamento de F\'isica, Universidad de los Andes, Cra. 1 No. 18A-10, Edificio Ip, CP 111711, Bogot\'a, Colombia}
\affiliation{Observatorio Astron\'omico, Universidad de los Andes, Cra. 1 No. 18A-10, Edificio H, CP 111711 Bogot\'a, Colombia}
\author{E.~Gaztañaga}
\affiliation{Institut d'Estudis Espacials de Catalunya (IEEC), c/ Esteve Terradas 1, Edifici RDIT, Campus PMT-UPC, 08860 Castelldefels, Spain}
\affiliation{Institute of Cosmology and Gravitation, University of Portsmouth, Dennis Sciama Building, Portsmouth, PO1 3FX, UK}
\affiliation{Institute of Space Sciences, ICE-CSIC, Campus UAB, Carrer de Can Magrans s/n, 08913 Bellaterra, Barcelona, Spain}
\author{Satya~{Gontcho A Gontcho}}
\affiliation{University of Virginia, Department of Astronomy, Charlottesville, VA 22904, USA}
\author{G.~Gutierrez}
\affiliation{Fermi National Accelerator Laboratory, PO Box 500, Batavia, IL 60510, USA}
\author{J.~Guy}
\affiliation{Lawrence Berkeley National Laboratory, 1 Cyclotron Road, Berkeley, CA 94720, USA}
\author{K.~Honscheid}
\affiliation{Center for Cosmology and AstroParticle Physics, The Ohio State University, 191 West Woodruff Avenue, Columbus, OH 43210, USA}
\affiliation{Department of Physics, The Ohio State University, 191 West Woodruff Avenue, Columbus, OH 43210, USA}
\affiliation{The Ohio State University, Columbus, 43210 OH, USA}
\author{C.~Howlett}
\affiliation{School of Mathematics and Physics, University of Queensland, Brisbane, QLD 4072, Australia}
\author{D.~Huterer}
\affiliation{Department of Physics, University of Michigan, 450 Church Street, Ann Arbor, MI 48109, USA}
\affiliation{University of Michigan, 500 S. State Street, Ann Arbor, MI 48109, USA}
\author{M.~Ishak}
\affiliation{Department of Physics, The University of Texas at Dallas, 800 W. Campbell Rd., Richardson, TX 75080, USA}
\author{R.~Joyce}
\affiliation{NSF NOIRLab, 950 N. Cherry Ave., Tucson, AZ 85719, USA}
\author{R.~Kehoe}
\affiliation{Department of Physics, Southern Methodist University, 3215 Daniel Avenue, Dallas, TX 75275, USA}
\author{T.~Kisner}
\affiliation{Lawrence Berkeley National Laboratory, 1 Cyclotron Road, Berkeley, CA 94720, USA}
\author{A.~Kremin}
\affiliation{Lawrence Berkeley National Laboratory, 1 Cyclotron Road, Berkeley, CA 94720, USA}
\author{O.~Lahav}
\affiliation{Department of Physics \& Astronomy, University College London, Gower Street, London, WC1E 6BT, UK}
\author{M.~Landriau}
\affiliation{Lawrence Berkeley National Laboratory, 1 Cyclotron Road, Berkeley, CA 94720, USA}
\author{L.~Le~Guillou}
\affiliation{Sorbonne Universit\'{e}, CNRS/IN2P3, Laboratoire de Physique Nucl\'{e}aire et de Hautes Energies (LPNHE), FR-75005 Paris, France}
\author{A.~Leauthaud}
\affiliation{Department of Astronomy and Astrophysics, UCO/Lick Observatory, University of California, 1156 High Street, Santa Cruz, CA 95064, USA}
\affiliation{Department of Astronomy and Astrophysics, University of California, Santa Cruz, 1156 High Street, Santa Cruz, CA 95065, USA}
\author{M.~Manera}
\affiliation{Departament de F\'{i}sica, Serra H\'{u}nter, Universitat Aut\`{o}noma de Barcelona, 08193 Bellaterra (Barcelona), Spain}
\affiliation{Institut de F\'{i}sica d’Altes Energies (IFAE), The Barcelona Institute of Science and Technology, Edifici Cn, Campus UAB, 08193, Bellaterra (Barcelona), Spain}
\author{P.~Martini}
\affiliation{Center for Cosmology and AstroParticle Physics, The Ohio State University, 191 West Woodruff Avenue, Columbus, OH 43210, USA}
\affiliation{Department of Astronomy, The Ohio State University, 4055 McPherson Laboratory, 140 W 18th Avenue, Columbus, OH 43210, USA}
\affiliation{The Ohio State University, Columbus, 43210 OH, USA}
\author{A.~Meisner}
\affiliation{NSF NOIRLab, 950 N. Cherry Ave., Tucson, AZ 85719, USA}
\author{R.~Miquel}
\affiliation{Instituci\'{o} Catalana de Recerca i Estudis Avan\c{c}ats, Passeig de Llu\'{\i}s Companys, 23, 08010 Barcelona, Spain}
\affiliation{Institut de F\'{i}sica d’Altes Energies (IFAE), The Barcelona Institute of Science and Technology, Edifici Cn, Campus UAB, 08193, Bellaterra (Barcelona), Spain}
\author{S.~Nadathur}
\affiliation{Institute of Cosmology and Gravitation, University of Portsmouth, Dennis Sciama Building, Portsmouth, PO1 3FX, UK}
\author{N.~Palanque-Delabrouille}
\affiliation{IRFU, CEA, Universit\'{e} Paris-Saclay, F-91191 Gif-sur-Yvette, France}
\affiliation{Lawrence Berkeley National Laboratory, 1 Cyclotron Road, Berkeley, CA 94720, USA}
\author{W.~J.~Percival}
\affiliation{Department of Physics and Astronomy, University of Waterloo, 200 University Ave W, Waterloo, ON N2L 3G1, Canada}
\affiliation{Perimeter Institute for Theoretical Physics, 31 Caroline St. North, Waterloo, ON N2L 2Y5, Canada}
\affiliation{Waterloo Centre for Astrophysics, University of Waterloo, 200 University Ave W, Waterloo, ON N2L 3G1, Canada}
\author{F.~Prada}
\affiliation{Instituto de Astrof\'{i}sica de Andaluc\'{i}a (CSIC), Glorieta de la Astronom\'{i}a, s/n, E-18008 Granada, Spain}
\author{I.~P\'erez-R\`afols}
\affiliation{Departament de F\'isica, EEBE, Universitat Polit\`ecnica de Catalunya, c/Eduard Maristany 10, 08930 Barcelona, Spain}
\author{G.~Rossi}
\affiliation{Department of Physics and Astronomy, Sejong University, 209 Neungdong-ro, Gwangjin-gu, Seoul 05006, Republic of Korea}
\author{L.~Samushia}
\affiliation{Abastumani Astrophysical Observatory, Tbilisi, GE-0179, Georgia}
\affiliation{Department of Physics, Kansas State University, 116 Cardwell Hall, Manhattan, KS 66506, USA}
\affiliation{Faculty of Natural Sciences and Medicine, Ilia State University, 0194 Tbilisi, Georgia}
\author{E.~Sanchez}
\affiliation{CIEMAT, Avenida Complutense 40, E-28040 Madrid, Spain}
\author{E.~F.~Schlafly}
\affiliation{Space Telescope Science Institute, 3700 San Martin Drive, Baltimore, MD 21218, USA}
\author{D.~Schlegel}
\affiliation{Lawrence Berkeley National Laboratory, 1 Cyclotron Road, Berkeley, CA 94720, USA}
\author{J.~Silber}
\affiliation{Lawrence Berkeley National Laboratory, 1 Cyclotron Road, Berkeley, CA 94720, USA}
\author{D.~Sprayberry}
\affiliation{NSF NOIRLab, 950 N. Cherry Ave., Tucson, AZ 85719, USA}
\author{G.~Tarl\'{e}}
\affiliation{University of Michigan, 500 S. State Street, Ann Arbor, MI 48109, USA}
\author{B.~A.~Weaver}
\affiliation{NSF NOIRLab, 950 N. Cherry Ave., Tucson, AZ 85719, USA}
\author{R.~Zhou}
\affiliation{Lawrence Berkeley National Laboratory, 1 Cyclotron Road, Berkeley, CA 94720, USA}
\author{H.~Zou}
\affiliation{National Astronomical Observatories, Chinese Academy of Sciences, A20 Datun Road, Chaoyang District, Beijing, 100101, P.~R.~China}

\begin{abstract}
We present the first high-significance spectroscopic stacked kinetic Sunyaev--Zel’dovich (kSZ) measurements of circumgalactic gas profiles for both Bright Galaxy Survey (BGS) and Emission Line Galaxy (ELG) tracers, combining DESI Data Release 2 with ACT Data Release 6. Using reconstructed line-of-sight velocities from the DESI galaxies and high-resolution ACT temperature maps, we detect the kSZ signal at high significance, reaching signal-to-noise ratios of up to $\sim$9 for BGS and $\sim$7.5 for ELGs in optimal stellar-mass selections. Together with the LRG measurements presented in Paper I, these constitute the most significant kSZ detections from any spectroscopic survey to date. We perform the analysis in both real and harmonic space, obtaining consistent results. By splitting both tracers into stellar-mass bins, we study the scaling of the kSZ amplitude with galaxy properties. Combining the kSZ measurements with ACT Data Release 6 (DR6) CMB lensing maps enables a joint calibration of the galaxy-halo connection and the gas fractions of host halos. For the BGS galaxies, we observe low gas fractions around the virial radius relative to standard expectations, likely attributable to active galactic nuclei (AGN) activity. We find some evidence for higher-mass halos retaining a larger fraction of their baryons, consistent with more efficient feedback in lower-mass systems. For the ELG sample, dominated by blue, star-forming galaxies, we provide the first detection of the gas distribution in ELG host halos. The ELGs appear to exhibit relatively high gas fractions, which points to the possibility of weaker feedback (due to e.g. low AGN and supernova feedback activity) at their mass scale. Finally, we present generalized Navarro-Frenk-White (GNFW) fits to the harmonic-space measurements, providing a compact parametrization of gas profiles for forward modeling in large-scale structure analyses. 
\end{abstract}
\maketitle

\section{Introduction} 
\label{sec:intro}

Baryons represent a fundamental yet elusive component of the cosmic matter budget. While they constitute roughly 15\% of the total matter density, their spatial distribution remains far less well characterized than that of dark matter \citep{Fukugita:2004ee}. The majority of baryons are thought to reside outside the stellar and cold gas phases, in the circumgalactic medium (CGM) and intracluster medium (ICM), where they are subject to complex astrophysical processes such as stellar feedback and active galactic nuclei (AGN) outflows \citep{Cen_2006}. These processes redistribute gas on scales of order one to several virial radii, shaping the thermodynamic state of halos while simultaneously imprinting baryonic effects on cosmological observables. Inadequate modeling of baryons leads to systematic uncertainties in weak lensing, clustering, and galaxy-CMB lensing cross-correlations \citep{2019OJAp....2E...4C, Semboloni:2011fe}, with potential biases in key parameters such as $S_8$ and the sum of neutrino masses \citep{2019JCAP...03..020S,2022MNRAS.516.5355A}.

A powerful observational tool for probing diffuse baryons is the Sunyaev--Zel’dovich (SZ) effect, which arises from the scattering of cosmic microwave background (CMB) photons off free electrons. The thermal SZ (tSZ) effect, proportional to the integrated electron pressure, has been widely used to study hot gas in massive clusters \citep{1999PhR...310...97B}. By contrast, the kinematic SZ (kSZ) effect results from the Doppler shift induced by electrons with bulk line-of-sight velocities relative to the CMB rest frame. Unlike the tSZ, the kSZ signal depends only on the electron momentum field, making it sensitive to the total ionized baryon distribution, regardless of temperature or metallicity \citep{2019SSRv..215...17M}. This enables kSZ measurements to directly constrain the abundance and spatial distribution of baryons, including in the diffuse outskirts of halos, a regime which has traditionally been challenging to access. 

Over the past decade, a variety of statistical methods have been developed to detect the kSZ effect, including the pairwise momentum estimator \citep{Hand2012, Soergel_2016, 2017JCAP...03..008D, Calafut_2021}, projected-field approaches \citep{Dore:2003ex,2016PhRvL.117e1301H,2016PhRvD..94l3526F}, velocity reconstruction templates \citep{Ho:2009iw, Harscouet:2025pwl}, and stacking analyses around galaxy and cluster samples \citep{ACTPol:2015teu,Schaan21,2024arXiv240707152H,2025arXiv250319870R, Mallaby-Kay_2023}. Together, these techniques have established the kSZ effect as a sensitive probe of baryons on halo mass scales down to $\sim 10^{13} M_\odot$. 

Importantly, the signal amplitude encodes the line-of-sight optical depth, linking the observed kSZ temperature decrement to the free electron density of the host halos. Stacking analyses in particular offer a powerful way to extract the average kSZ signal around large galaxy samples, allowing us to map baryon profiles as a function of scale, mass, and redshift. The wide mass and redshift lever arm that this series of papers provides is essential both for the study of feedback and for calibration of baryon effects in weak lensing: although the sensitivity of the matter or lensing power spectrum on mildly non-linear scales is dominated by cluster-mass halos \cite{Lucie-Smith:2025hgj} (with an obvious dependence on scale cuts), the group-sized halos likely retain a significant contribution of suppression to the matter power, given the large feedback in this mass range \cite{Miller:2025gwh}.

The Bright Galaxy Sample (BGS, \cite{Hahn:2022dnf}) of the Dark Energy Spectroscopic Instrument (DESI) provides a unique opportunity for kSZ studies. As the lowest-redshift DESI tracer, BGS occupies the regime most sensitive to cosmological growth through both galaxy clustering as well as weak lensing studies, rendering the study of baryonic feedback extremely useful. The dense spectroscopic sampling of DESI's BGS over thousands of square degrees allows precise localization of galaxies and robust stacking measurements. Combined with the latest high-resolution CMB maps from the Atacama Cosmology Telescope (ACT DR6), the stacked kSZ study on BGS enables new low-redshift tests of the baryon distribution and its connection to large-scale structure. 

In addition to the BGS, we also consider the DESI emission-line galaxy (ELG, \cite{Raichoor:2022jab}) sample. ELGs are actively star-forming galaxies identified primarily through their strong [O II] and [O III] emission lines. They typically reside in lower-mass halos than BGS or LRGs and have a higher mean redshift ($z \sim 1$), making them an important but relatively understudied population. While no detection of the kSZ signal has yet been achieved for ELGs, their role is crucial: by probing the gas content of lower-mass halos at earlier cosmic times, ELGs provide a complementary window into how baryonic feedback operates across different mass and redshift regimes, filling in a missing piece of the overall picture.

In this work, we present a measurement of the stacked kSZ signal around DESI BGS galaxies and ELGs using ACT DR6 temperature maps, while in a companion paper \cite{FrankLRG}, we measure the same signal around DESI Luminous Red Galaxies (LRGs, \cite{LRG.TS.Zhou.2023}). This analysis builds upon recent kSZ detections around photometric BGS as well as photometric and spectroscopic DESI LRGs \citep{Hadzhiyska:2024ecq, 2025arXiv250319870R}, but extends to a spectroscopic, lower-redshift sample with different halo mass and environmental properties. Our study has three primary goals: (1) to map the average baryon distribution around BGS and ELG galaxies out to several virial radii at different masses; (2) to present simple parametrized forms of the gas density profiles of their halo hosts; and (3) to assess the implications of our results for baryonic feedback modeling in cosmological analyses. By targeting the BGS, we probe the gas physics of group-sized halos at $z \lesssim 0.4$, providing a new low-redshift anchor for kSZ studies and supplying invaluable information about baryonic feedback at low redshifts to weak lensing studies. The ELGs on the other hand, inform a completely new regime of low-mass star-forming galaxy groups, the gas distribution around which has hitherto remained out of reach.

To measure the gas-to-mass ratio in our samples, and hence assess the strength of feedback, we also measure the stacked CMB lensing signal, similar to \cite{Hadzhiyska:2025mvt}. We do not attempt a full joint fit of CMB lensing and kSZ data, focusing instead on measuring the ratio of the kSZ and CMB lensing as a proxy to the gas-to-mass ratio. An additional reason for including a CMB lensing cross-correlation measurement is that the halo occupation distributions (HODs) of the galaxy (sub)samples used in this work have not been previously constrained. An early comparison of the kSZ signal to small-scale galaxy-galaxy lensing (GGL) was performed in \cite{2021PhRvD.103f3514A} and further studied in \cite{Sunseri:2025hhj} with the halo model, and in \cite{2025MNRAS.540..143M} with the use of state-of-the-art hydrodynamical simulations. A combined analysis of cosmic shear and the kSZ effect was conducted in \cite{2024MNRAS.534..655B, Siegel:2025frt, Bigwood:2025kur}.

This paper is organized as follows. In Section~\ref{sec:data}, we describe the DESI BGS and ELG datasets as well as ACT CMB maps. Section~\ref{sec:methods} outlines the methodology used to extract the stacked kSZ profiles. Section~\ref{sec:results} presents our measurement and comparison to theoretical models. In Section~\ref{sec:discussion}, we discuss the astrophysical and cosmological implications of our findings and summarize our conclusions.

\section{Data}
\label{sec:data}

\subsection{Dark Energy Spectroscopic Instrument}
\label{sec:desi}


The Dark Energy Spectroscopic Instrument (DESI) is a robotic, fiber-fed, highly multiplexed spectroscopic surveyor that operates on the Mayall 4-meter telescope at Kitt Peak National Observatory \citep{DESI2022.KP1.Instr}. DESI, which can obtain simultaneous spectra of almost 5000 objects over a $\sim3^\circ$ field \citep{ Corrector.Miller.2023, FiberSystem.Poppett.2024}, is conducting an eight-year survey of about $17{,}000,\mathrm{deg}^2$ of the sky. The full survey will lead to 63 million spectroscopically-confirmed galaxies and quasars, compared to the initial forecasts of 39 million \citep{DESI2016b.Instr}. The sheer scale of the DESI experiment necessitates multiple supporting software pipelines and products \citep{Spectro.Pipeline.Guy.2023,SurveyOps.Schlafly.2023}.

Using the First Data Release \citep[DR1][]{DESI2024.I.DR1}, cosmological results were obtained from the full-shape analysis \citep{DESI2024.VII.KP7B}. Cosmological analysis has commenced with the upcoming DR2 \citep{DESI.DR2.DR2}.

Target selection for DESI is based on the DESI Legacy Imaging Surveys \citep{2017PASP..129f4101Z,2019AJ....157..168D}.
Machine-learning classification, colour selections, and imaging quality cuts are combined to identify bright galaxies (BGS), luminous red galaxies (LRGs), emission-line galaxies (ELGs), and quasars \citep{2023AJ....165...50M}.
The present work focuses on the BGS BRIGHT sample with an absolute magnitude limit cut of ${M_r < -20.2}$ (BGS\_BRIGHT-20.2) and the ELG standard sample (ELG\_LOPnotQSO) as delivered in DESI DR2. The ELGs are selected via several color-space cuts and target star-forming galaxies in the range between $z = 0.8$ and 1.5 \citep{2023AJ....165..126R}.
Both samples provide high-density tracers with exquisitely well-measured spectroscopic redshifts, well-suited for cross-correlations with CMB-derived velocity fields.

For both tracers, we additionally match to the DESI DR2 value-added catalog of physical properties derived using the \textsc{CIGALE} code \citep{2024A&A...691A.308S,2025arXiv250609143S}.
\textsc{CIGALE} fits galaxy spectral energy distributions from the optical to the infrared using stellar population synthesis, delayed star-formation histories with optional bursts, nebular emission, dust attenuation and re-emission, and AGN templates.
The fits incorporate the $g,r,z$ bands as well as $W1$-$W4$ photometry from the Wide-field Infrared Survey Explorer (WISE, \cite{2010AJ....140.1868W}) and yield posterior distributions for stellar mass, star-formation rate, and AGN fraction. 
The DR2 Value Added Catalog (VAC) contains $\sim 26$ million galaxies across all primary DESI target types and assumes a WMAP7 cosmology.

Our final dataset consists of both ELG and BGS-BRIGHT galaxies with reliable spectroscopic redshifts, matched to CIGALE-derived physical properties from DESI DR2. 
All physical properties used in our analysis: stellar masses, star-formation rates, and AGN fractions are taken directly from the DR2 CIGALE VAC \citep{2024A&A...691A.308S}.

\subsection{Atacama Cosmology Telescope}

\label{sec:act}

For the CMB component of this analysis, we use the harmonic-space Internal Linear Combination (hILC) maps from the Atacama Cosmology Telescope (ACT) Data Release 6 \citep{2024PhRvD.109f3530C}\footnote{The ACT DR6 products can be found under \url{https://lambda.gsfc.nasa.gov/product/act/act_dr6.02/act_dr6.02_maps_coadd_get.html}.}.
ACT was a 6-meter telescope located in northern Chile that mapped the millimeter sky between 2007 and 2022.
DR6 includes multifrequency observations from 2017-2022 at \texttt{f090}, \texttt{f150}, and \texttt{f220}, covering approximately one-third of the sky with arcminute resolution.
This work uses only the night-time portion of the data from the first five seasons, as recommended for kSZ analyses.

The fidelity of the hILC maps derives from combining frequency channels in harmonic space to suppress foregrounds (most notably, dust and radio emission) while retaining the kSZ signal.
All maps are delivered in the CAR projection.
We use the dr6.01 release, the same version used in the ACT SZ and kSZ companion analyses.

We apply a conservative mask that removes extended radio sources, unobserved regions, the Galactic plane, and all detected ACT DR6 clusters.
Massive clusters can bias pairwise and cross-correlation kSZ measurements by contributing strong tSZ residuals; we thus mask all ACT DR6 clusters with ${\rm SNR}>6$ \citep{2025arXiv250721459A}. Throughout this work, we adopt the hILC map as our fiducial CMB temperature field.
In Appendix~\ref{app:hilc_vs_single}, 
we show that the kSZ cross-correlation signal is consistent between the hILC map and the single-frequency \texttt{f090} and \texttt{f150} maps.
The agreement of these independent measurements confirms that our kSZ estimator is robust to foreground contamination and insensitive to the details of the ILC construction.

We additionally make use of the publicly available CMB lensing convergence ($\kappa$) maps from the Atacama Cosmology Telescope Data Release 6 (DR6), described in \cite{ACT:2023kun,ACT:2023dou,ACT:2023ubw,ACT:2023oei}.
The DR6 lensing reconstruction uses multifrequency temperature and polarization data from the 2017--2021 ACT observing seasons at \texttt{f090} and \texttt{f150}, covering roughly one third of the sky at arcminute resolution.
Lensing is reconstructed using a standard quadratic estimator applied to the combined temperature+polarization maps, with profile-hardened estimators \citep{Sailer:2020lal,ACT:2023ubw} to suppress biases from thermal SZ, CIB, and point-source foregrounds.
The public release provides the reconstructed $\kappa$ field in terms of spherical harmonic coefficients ($a_{\ell m}$) together with the appropriate analysis mask.
Following the DR6 pipeline, the reconstruction includes a conservative multipole filtering that removes poorly measured large-scale modes and suppresses small-scale modes dominated by foregrounds or noise, yielding a high-signal-to-noise map that is well suited for cross-correlation analyses with large-scale structure. The final $\kappa$ map is low-pass filtered at $L = 3000$.

\subsection{Properties of the galaxy tracers}

\begin{figure*}[t]
\centering
\includegraphics[width=0.95\linewidth]{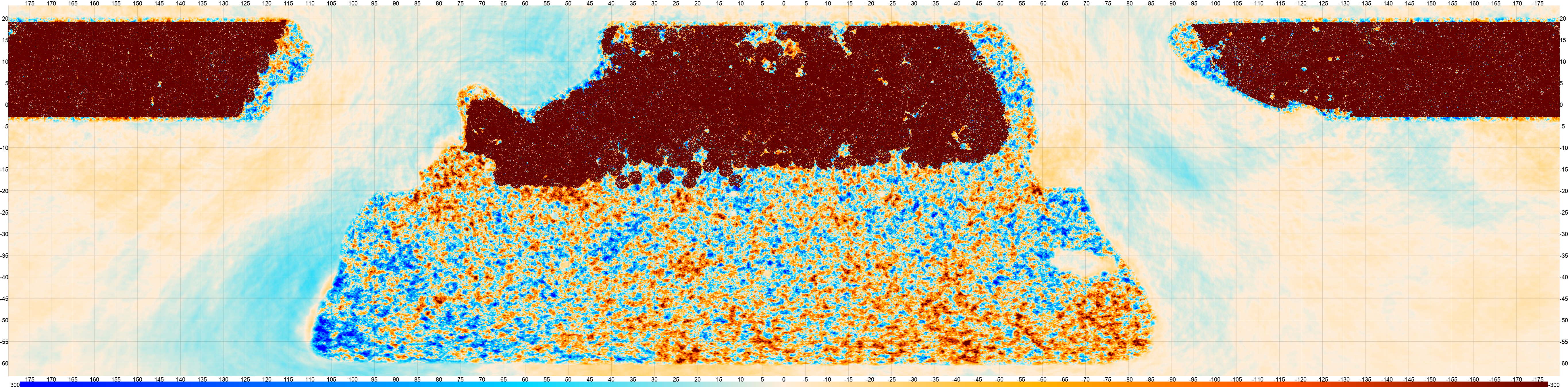}
\includegraphics[width=0.95\linewidth]{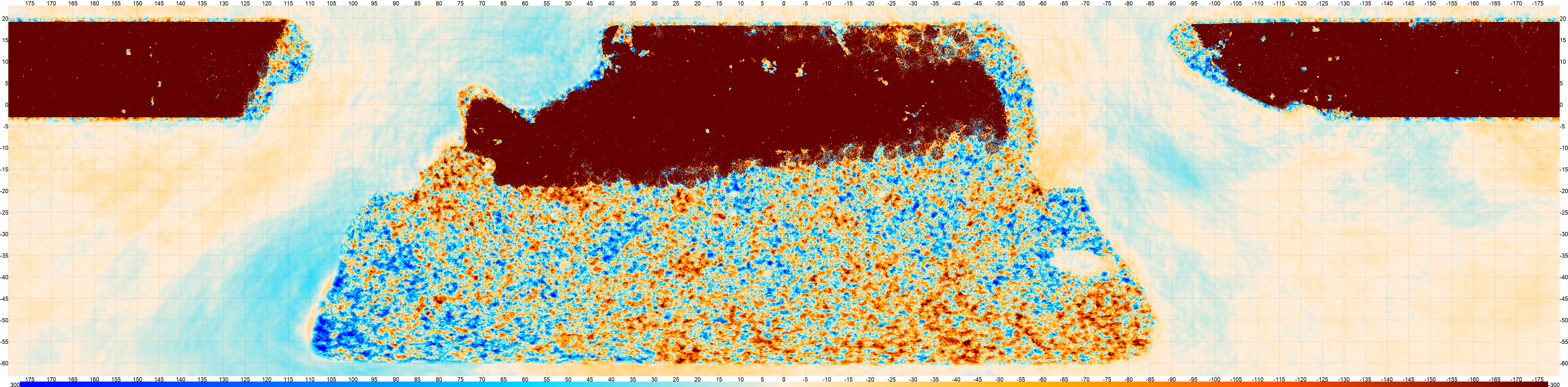}
\caption{
Overlap between the \textsc{DESI} DR2 footprint and the ACT DR6 temperature map as RA (deg) and DEC (deg). 
The top panel shows the distribution of Bright Galaxy Sample (BGS) targets, and the 
bottom panel shows Emission Line Galaxies (ELGs), plotted as individual points over 
the ACT DR6 CMB temperature map. Both DESI samples correspond to Data Release~2, 
and the CMB map is taken from ACT DR6. The DESI footprint includes 
the characteristic petal-shaped geometries imposed by the 5000-fiber focal plane, 
which determine where repeated passes were taken on the sky. DR2 represents a 
substantially more complete footprint than DR1, particularly for BGS, although some 
incompleteness remains near the survey edges. The total sky area overlapping with 
ACT is $\sim 5{,}000$--$6{,}000\,\mathrm{deg}^2$, and is not expected to change 
significantly with the final DR3 footprint.
}
\label{fig:overlap}
\end{figure*}

Fig.~\ref{fig:overlap} shows the sky coverage for the two DESI tracers used in this work: BGS (top) and ELG (bottom), plotted as individual galaxies over the ACT temperature map. The footprint features the familiar ``petal'' structures produced by DESI’s 5000-fiber robotized focal plane, whose geometry dictates the pattern of overlapping tiles across the sky. Although DR2 is markedly more complete than the original DR1, especially for the bright-time high-priority BGS targets, some incompleteness persists along the survey boundaries where additional passes will continue to accumulate in later releases. The region in common with ACT spans roughly 5,000-6,000 deg$^2$, and even with the completion of DR3 this overlap will remain largely unchanged. This shared area forms the basis for our stacked kSZ measurements around DESI galaxies.

\begin{figure}[t]
\centering
\includegraphics[width=0.95\linewidth]{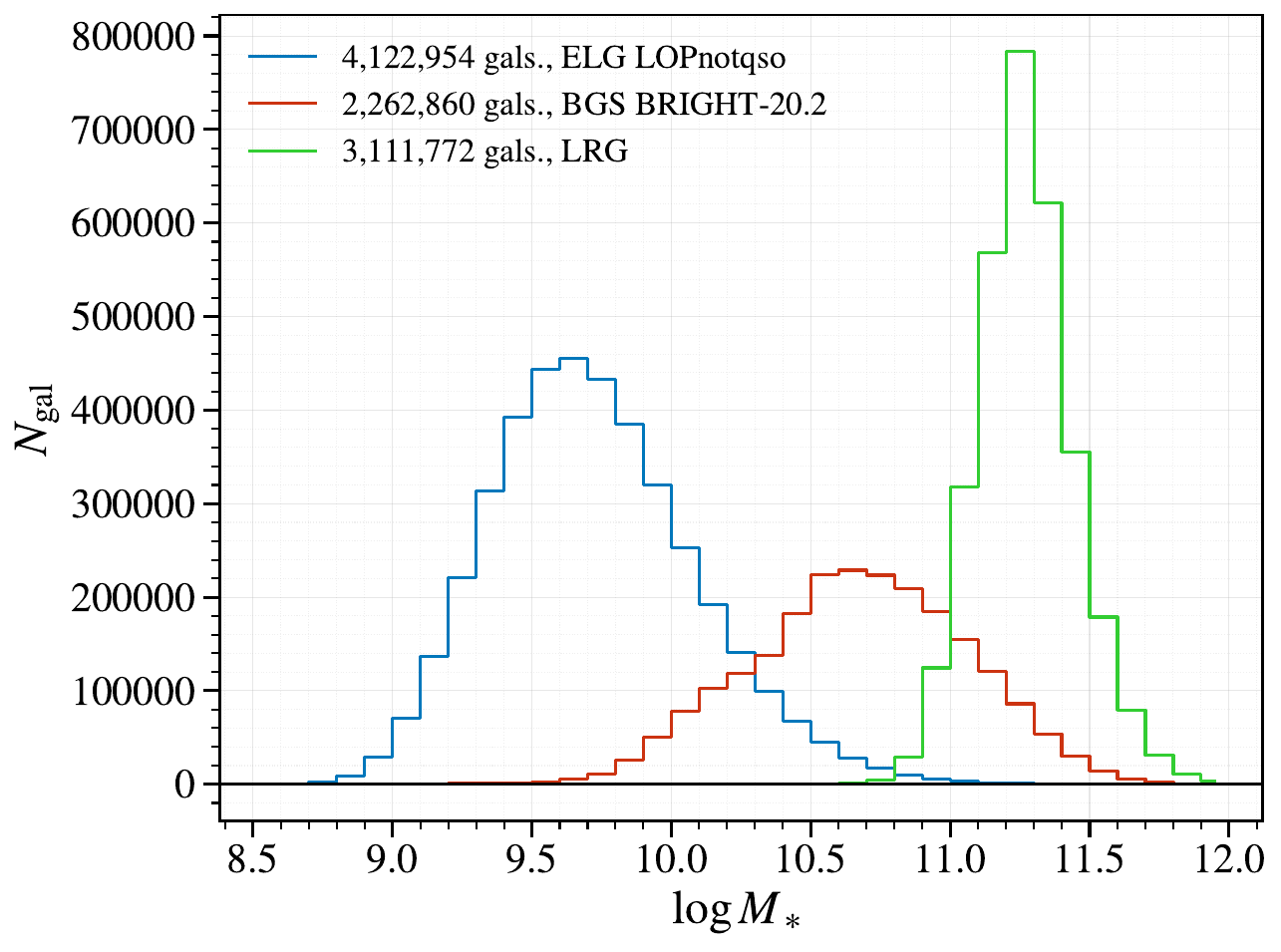}
\caption{
Stellar mass histograms for the DESI DR2 Bright Galaxy Sample (BGS), Emission Line Galaxies (ELGs), and Luminous Red Galaxies (LRGs). Stellar masses are taken from the CIGALE value--added catalog, which provides SED--based estimates using optical photometry together with \textit{WISE} mid--infrared data. The combined samples span the range $9 \lesssim \log_{10}(M_\star/M_\odot) \lesssim 12$. 
The ELG, BGS, and LRG selections populate different regions of this range: the ELG distribution peaks near $\log M_\star \approx 9.7$, the BGS near $\log M_\star \approx 10.75$, and the LRGs near $\log M_\star \approx 11.3$. 
The BGS sample shows the widest mass distribution, while ELGs and LRGs occupy narrower intervals with little overlap. No additional selection beyond the standard DESI DR2 targeting is imposed. The kSZ signal around LRGs is studied in detail in the companion paper \citep{Qu2026}.
}
\label{fig:logm}
\end{figure}

Fig.~\ref{fig:logm} presents the stellar mass distributions for the DESI DR2 BGS, ELG, and LRG targets used in this analysis and the companion paper \citep{Qu2026}. The stellar masses originate from the CIGALE value-added catalog, which provides SED-based mass estimates  using DESI photometry in combination with WISE measurements. All three tracers cover a combined span of nearly three orders of magnitude in stellar mass, from approximately $10^9$ to $10^{12} \ M_\odot$. Throughout the paper, we use $M_\odot$ units when reporting stellar masses and $M_\odot/h$ when reporting halo masses.

Each tracer occupies a distinct portion of this range as a consequence of the DESI targeting definitions. The ELG distribution peaks at $\log M_\star \simeq 9.7$ and is concentrated in a relatively narrow interval. The BGS sample peaks at $\log M_\star \simeq 10.75$ and extends over a broader interval, reflecting the wide distribution of apparent magnitudes included in the selection. The LRG distribution peaks at $\log M_\star \simeq 11.3$ with a small tail toward lower masses. Across the three samples, the ELG and LRG mass intervals have minimal overlap, while the BGS distribution spans the intermediate region.

\begin{figure}[t]
\centering
\includegraphics[width=0.95\linewidth]{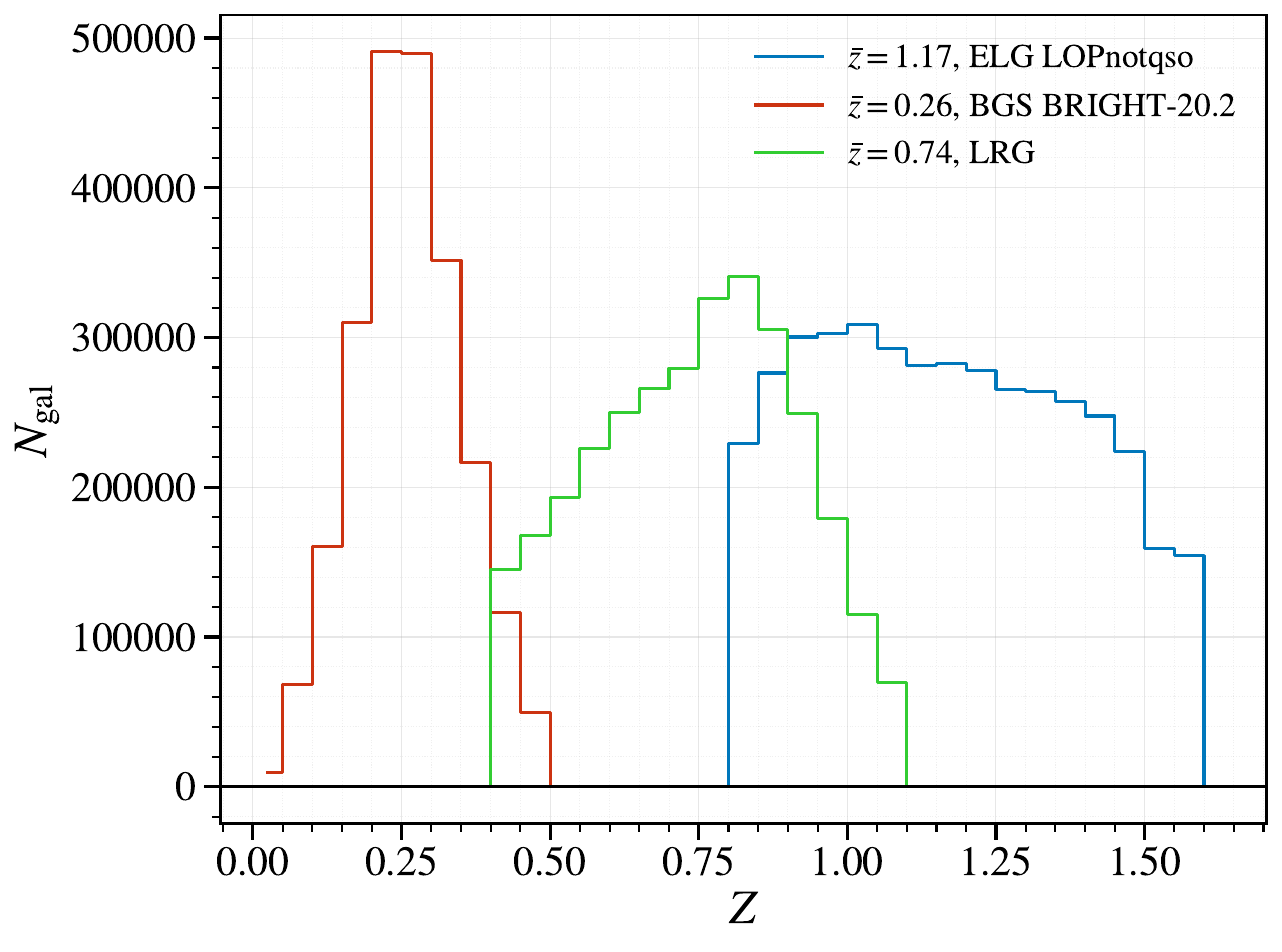}
\caption{
Redshift distributions for the DESI DR2 Bright Galaxy Sample (BGS), Luminous Red Galaxies (LRGs), and Emission Line Galaxies (ELGs), shown as the number of galaxies per redshift bin. 
The samples span the interval $0 \leq z \lesssim 1.7$ and together provide continuous coverage across this range. 
The BGS distribution peaks sharply at $z \simeq 0.3$ and contains $2.26$ million galaxies. 
The LRG distribution extends from $z \simeq 0.5$ to $z \simeq 1.1$ and includes $3.1$ million galaxies. 
The ELG sample reaches the highest redshifts, up to $z \simeq 1.7$, with $4.1$ million galaxies and a mean redshift of $z \simeq 1.17$. 
For reference, the mean redshifts of the BGS and LRG samples are $z \simeq 0.26$ and $z \simeq 0.74$, respectively. 
No volume normalization is applied; the figure displays the raw number of galaxies in each redshift bin as selected in DESI DR2.
}
\label{fig:dndz}
\end{figure}

Fig.~\ref{fig:dndz} shows the redshift distributions of the DESI DR2 BGS, LRG, and ELG samples expressed as the number of galaxies per redshift bin. The three selections together span a wide redshift interval, from the local Universe at $z \simeq 0$ to beyond $z \simeq 1.5$, with each tracer occupying a distinct range that reflects its targeting strategy.

The BGS sample is concentrated at low redshift and peaks around $z \simeq 0.3$, where it contains the largest number of galaxies per bin. Its distribution is relatively narrow and approximately Gaussian in shape, a consequence of selecting the brightest nearby galaxies down to the BGS magnitude threshold. These magnitude limits produce a steep fall-off toward higher redshifts, since only the intrinsically brightest systems remain above the flux limit at $z \gtrsim 0.4$.

The LRGs populate an intermediate redshift range, with most objects lying between $z \simeq 0.5$ and $z \simeq 1.1$. Their distribution is broader and flatter than that of the BGS sample when shown simply as counts per redshift bin. Although this figure does not account for the varying comoving volume per bin, the LRG sample maintains a relatively uniform comoving number density of around $5.4 \times 10^{-4} \ ({\rm Mpc}/h)^{-3}$ across its targeted interval, yielding a mean redshift of $z \simeq 0.74$ (see Part I for the analysis of LRG data \citep{Qu2026}).

The ELG sample extends to the highest redshifts, reaching up to $z \simeq 1.7$ with a mean redshift of $z \simeq 1.17$. ELGs have the highest intrinsic comoving number density among the DESI tracers of around $8 \times 10^{-4} \ ({\rm Mpc}/h)^{-3}$ (similarly to the LRGs, this value is constant over a large range of redshifts), but here the distribution reflects only the raw number of observed galaxies per redshift bin. The rise and decline of the histogram follow from the DESI ELG selection and the instrument sensitivity as a function of redshift.

Across the three samples, the combination of low-redshift BGS, intermediate-redshift LRGs, and high-redshift ELGs provides near-continuous coverage over the redshift and mass range relevant for understanding the effect on baryons on weak lensing observables.

\section{Methods}
\label{sec:methods}

This section describes the methods used to measure and interpret the kinematic Sunyaev--Zel'dovich (kSZ) signal around DESI DR2 galaxies using ACT DR6 temperature maps.  We begin with a brief theoretical introduction to the kSZ observable, then describe the two complementary estimators used in this work (real-space, CAP filtered stacks and a harmonic-space template cross-correlation).  We explain how the velocity field is reconstructed, how covariances are estimated for each estimator, and how the measurements are modelled using a halo model with a generalized NFW (GNFW) electron profile.  Finally, we describe the parameter inference procedure used to constrain the GNFW parameters.

\subsection{The kSZ signal}
\label{sec:ksz_theory}

The kinematic Sunyaev Zel'dovich (kSZ) effect arises from the Doppler shift that CMB photons acquire when Thomson scattered by free electrons that have a bulk line-of-sight velocity relative to the CMB rest frame.  The fractional CMB temperature fluctuation produced along a line of sight \(\hat{\mathbf{n}}\) is:
\begin{eqnarray}
\frac{\Delta T_{\rm kSZ}(\hat{\mathbf{n}})}{T_{\rm CMB}}
= - \int d\chi a^{-2}(\chi)\, \sigma_T\, 
n_e(\chi\hat{\mathbf{n}},z)\; \frac{v_r(\chi\hat{\mathbf{n}},z)}{c}\; e^{-\tau(\chi)}
\label{eq:ksz_full}
\end{eqnarray}
where:
\begin{itemize}
  \item \(T_{\rm CMB}=2.7255~\mathrm{K}\) is the mean CMB temperature,
  \item \(\chi\) is comoving radial distance and \(z=z(\chi)\) the corresponding redshift,
  \item \(a(\chi)\equiv 1/(1+z)\) is the scale factor,
  \item \(\sigma_T\) is the Thomson scattering cross-section,
  \item \(n_e(\mathbf{x},z)\) is the comoving free electron number density,
  \item \(v_r \equiv \mathbf{v}\cdot\hat{\mathbf{n}}\) is the line-of-sight component of the electron bulk velocity field (positive for receding flows),
  \item \(c\) is the speed of light, and
  \item \(\tau(\chi)=\int_0^\chi d\chi'\, a^{-2}(\chi')\,\sigma_T\, n_e(\chi'\hat{\mathbf{n}},z')\) is the optical depth from the observer to \(\chi\).
\end{itemize}

In the regime relevant for groups and clusters, the medium is optically thin (\(\tau\ll1\)), so the attenuation factor \(e^{-\tau}\) in Eq.~\eqref{eq:ksz_full} can be dropped to very high accuracy.  For a single halo \(i\) with line-of-sight optical depth \(\tau_i\) and LOS velocity \(v_i^{\rm halo}\) the kSZ temperature shift is therefore well approximated by
\begin{equation}
\Delta T_{\rm kSZ}^{(i)} \simeq -\,T_{\rm CMB}\, \tau_i\,\frac{v_i^{\rm halo}}{c}.
\label{eq:ksz_single_halo}
\end{equation}
Equation~\eqref{eq:ksz_single_halo} makes explicit that the kSZ signal has the same blackbody spectrum as the primary CMB and therefore cannot be separated from primary CMB fluctuations using multi-frequency spectral information (in contrast to the thermal SZ effect or foregrounds such as dust).  The typical RMS one-dimensional halo peculiar velocity is \(\sigma_v \sim 300\ \mathrm{km\,s^{-1}}\) (weakly dependent on mass and redshift, see e.g. Ref.~\cite{Guachalla:2023lbx}); this scale sets the order of magnitude for stacked kSZ signals after averaging over many halos. We assume this value of $\sigma_v$ throughout when converting from ``CMB units'' into physical quantities (i.e., optical depth). A more accurate estimate of \(\sigma_v\) can be obtained by using numerical simulations and defining a realistic mock catalog of DESI galaxies.

\subsection{Real-space estimator}
\label{sec:real_space}

We build a real-space estimator based on compensated aperture photometry (CAP) to extract the kSZ temperature around individual galaxies and then stack the velocity-weighted results.  CAP is robust to large-scale primary CMB fluctuations (as well as random uncorrelated signal along the line-of-sight) and provides a simple, profile-agnostic measure of the mean enclosed kSZ signal as a function of aperture radius.

Intuitively, one can think of these curves as showing roughly a cumulative gas density profile. At large radii, we expect the profiles to become shallower as more and more of the gas is enclosed. In the absence of a two-halo term, they should asymptote to a constant, set by the mean halo mass and baryon fraction. It is useful to compare the value of the CAP filter for the gas at the virial radius with that of the CAP filter for the matter at the same radius as a way of estimating how much of the gas has been expelled.

\subsubsection{Thumbnail extraction and the CAP filter}
\label{sec:cap_def}

For each galaxy \(i\) we extract a small thumbnail of the CMB temperature map centred at the galaxy position and apply a compensated top-hat (disk-minus-ring) filter of characteristic angular radius \(\theta_d\).  The CAP-filtered temperature for galaxy \(i\) at filter scale \(\theta_d\) is
\begin{equation}
\mathcal{T}_i(\theta_d) \;=\; \int d^2\hat{\mathbf{n}}\; \delta T(\hat{\mathbf{n}})\, W_{\theta_d}(\theta),
\end{equation}
where \(\delta T(\hat{\mathbf{n}})\) is the CMB temperature fluctuation field and the filter \(W_{\theta_d}(\theta)\) is
\begin{equation}
W_{\theta_d}(\theta) =
\begin{cases}
+1, & 0 \le \theta < \theta_d,\\[4pt]
-1, & \theta_d \le \theta \le \sqrt{2}\,\theta_d,\\[4pt]
0, & \theta > \sqrt{2}\,\theta_d,
\end{cases}
\end{equation}
so that the annulus \([\theta_d,\sqrt{2}\theta_d]\) has the same area as the inner disk.  This compensated choice suppresses large-scale modes (e.g. primary CMB) while preserving sensitivity to structure on scales \(\lesssim \theta_d\).

In this work, we evaluate the CAP filter at 17 logarithmically-spaced values between \(\theta_d=1'\) and \(14'\) (the set of filter radii used for the stacks; note that at large radii, there is almost no information due to the primary CMB, so we limit the largest radii shown to $\sim10'$).  We convert between angular radius and proper physical radius at the mean redshift \(z_i\) of each sample when producing physical plots; however, the raw CAP measurements are naturally expressed in angular units. We discard galaxies within $8.5'$ of the CMB mask to avoid contamination from point and extended sources in the CMB map.

\subsubsection{Velocity weighting and stacking}
\label{sec:cap_stack}

Each galaxy has an associated reconstructed line-of-sight velocity \(v_{{\rm rec},i}\).  We adopt the velocity-weighted, uniform-mean estimator (a variant of the inverse-variance motivated weighting used in previous work, e.g. \cite{ACTPol:2015teu}).  Our stacked kSZ estimate at aperture radius \(\theta_d\) is
\begin{equation}
\widehat{T}_{\rm kSZ}(\theta_d)
\;=\;
-\,\frac{1}{r}\,\frac{\sigma_v^{\rm rec}}{c}\,
\frac{\displaystyle\sum_{i=1}^{N_{\rm gal}} \mathcal{T}_i(\theta_d)\,\bigl(v_{{\rm rec},i}/c\bigr)}
{\displaystyle\sum_{i=1}^{N_{\rm gal}} \bigl(v_{{\rm rec},i}/c\bigr)^2 } \,,
\label{eq:cap_estimator}
\end{equation}
where:
\begin{itemize}
  \item \(v_{{\rm rec},i}\) is the reconstructed LOS velocity for galaxy \(i\),
  \item \(\sigma_v^{\rm rec}\) is the RMS of the reconstructed velocities in the sample,
  \item \(r\) is the cross-correlation coefficient between the reconstructed velocities and the true halo LOS velocities, \(r\equiv \langle v_{\rm true}v_{\rm rec}\rangle/(\sigma_v^{\rm true}\sigma_v^{\rm rec})\), with $v_{\rm true}$ being the host halo velocity,
  \item \(N_{\rm gal}\) is the number of galaxies in the stack.
\end{itemize}
The prefactor \(1/r\) corrects for the imperfect fidelity of the reconstruction (if \(v_{\rm rec}=v_{\rm true}\), then \(r=1\) and no correction is needed).  The estimator in Eq.~\eqref{eq:cap_estimator} returns a measured temperature in units of \(\mu{\rm K}\) (or \(\mu{\rm K}\,\mathrm{arcmin}^2\) when multiplied by the appropriate solid angle factor), and may be converted to an optical-depth estimate by dividing by $T_{\rm CMB} \ (\sigma_v^{\rm true}/c)$.

\subsubsection{Velocity reconstruction}
\label{sec:reconstruction}

The reconstructed velocities used in Eq.~\eqref{eq:cap_estimator} are derived from the galaxy density field under the linearized continuity equation.  In configuration space, the continuity relation in the presence of redshift-space distortions can be written as
\begin{equation}
\nabla\cdot\mathbf{v} + \frac{f}{b}\nabla\cdot\bigl[(\mathbf{v}\cdot\hat{\mathbf{n}})\,\hat{\mathbf{n}}\bigr] \;=\; -\,a H f\,\frac{\delta_g}{b},
\label{eq:continuity_rsd}
\end{equation}
where \(a\) is the scale factor, \(H\) the Hubble parameter, \(f\equiv d\ln D/d\ln a\) the linear growth rate, \(b\) the large-scale galaxy bias, and \(\delta_g\) the galaxy overdensity.  To obtain a practical estimator for the peculiar velocity field, we solve Eq.~\eqref{eq:continuity_rsd} in Fourier space for the scalar potential (or equivalently the Lagrangian displacement \(\boldsymbol{\psi}\)) and then compute the associated line-of-sight component.

For this analysis, we use modified reconstruction settings for the BGS and the fiducial DESI DR2 Baryon Acoustic Oscillations (BAO) settings for the ELGs. For the BGS, we apply a Multigrid Poisson solver to the catalog on a Cartesian mesh, and a Gaussian smoothing of scale \(R_{\rm smooth}=12.5\ h^{-1}\mathrm{Mpc}\) to the density field prior to reconstruction using the public code \textsc{pyrecon}\footnote{\url{https://github.com/cosmodesi/pyrecon}}. This smoothing scale reduces small-scale shot noise and optimizes the fidelity of the large-scale velocity modes that dominate the stacked kSZ signal and assumes \textit{Planck} 2018 cosmology. Our reconstruction is similar in spirit to BAO reconstruction methods, but our end product is an estimate of the mean flow of matter. Compared with the fiducial DESI DR2 analysis \citep{2025PhRvD.112h3515A}, our sample selection includes a larger fraction of faint BGS galaxies and ELGs because we aim to reduce noise by increasing the number density. A two-dimensional histogram of the reconstructed velocities of both samples are shown in App.~\ref{app:velz}.

To estimate the reconstruction fidelity \(r\) and the RMS velocity \(\sigma_v^{\rm rec}\), we adopt the results from \cite{Guachalla:2023lbx, Hadzhiyska:2023nig}, which use mock catalogs drawn from the \textsc{AbacusSummit} simulations processed through the same selection and reconstruction pipeline as the DESI target survey galaxies.  In practice \(r\) depends mainly on tracer density, masks, redshift, redshift quality, and the smoothing scale; we adopt fiducial values \(r_{\rm fid, BGS}=0.64\) and \(r_{\rm fid, ELG}=0.55\) in our baseline analysis and propagate a nominal \(\sim10\%\) uncertainty on \(r\) in the interpretation of amplitudes for the full sample. We note, however, that the correlation coefficient $r$ is likely redshift and subsample-dependent. Therefore, our results should be interpreted as a measurement of $(r/r_{\rm{fid}}){T}^{\rm true}_{\rm kSZ}$, which is the quantity that is directly measured from the data. Any interpretation of the amplitude of the signal should consider that the uncertainty on $r$ likely dominates our uncertainty on the amplitude (especially for the subsamples that have not been studied in \cite{Guachalla:2023lbx, Hadzhiyska:2023nig}). Characterizing the subsamples in realistic mocks (for example, various stellar mass and redshift bins, and including the effects of the complex mask and fiber collisions in DESI) would be necessary to reduce the uncertainty on $r$ and achieve more definitive conclusions about the amplitude of the signal. Given that HODs and masses are not readily available for our subsamples, we leave a full characterization of $r$ to future work.

In principle, the velocity reconstruction coefficient $r$ could exhibit a mild
scale dependence. However, end-to-end tests using DESI-like mock catalogs (see App. D of Ref.~\citep{2024arXiv240707152H}, and also Refs.~\citep{Guachalla:2023lbx,Hadzhiyska:2023nig}) and
light-cone kSZ simulations show that the recovered kSZ optical depth profile
matches the true profile up to an overall multiplicative constant identified
with $r$. When true (non-reconstructed) velocities are used, a very small
scale dependence appears at large radii dominated by primary CMB fluctuations,
but this effect is below the $0.2\sigma$ level and vanishes once reconstruction
is applied (see App. D of Ref.~\citep{2024arXiv240707152H}). This behavior is consistent with analytic expectations that
reconstruction suppresses non-linear velocity contributions, thereby restoring
effective scale-independence over the angular scales relevant for our analysis.
We therefore treat $r$ as an overall amplitude parameter, with any residual
scale dependence well below our current statistical uncertainties.

\subsubsection{Real-space covariance: block bootstrap}
\label{sec:real_cov}

To estimate uncertainties for the configuration-space stacks, we use a block bootstrap computed by partitioning the survey footprint into non-overlapping $2^\circ\times 2^\circ$ patches using \textsc{HEALPix} at resolution $N_{\rm side}=32$ (pixel scale $\approx1.83^\circ$), yielding $N_{\rm patch}\approx1837$ occupied spatial blocks. Our largest aperture radius is $\sim10$ arcmin, corresponding to an angular scale roughly sixty times smaller than the linear extent of a typical patch. The patches are therefore much larger than the scales probed by the stacking estimator, justifying the assumption that the bootstrap blocks are approximately independent realizations for covariance estimation. By resampling on patches rather than individual galaxies, the estimator accounts for spatial correlations induced by galaxy clustering, which would otherwise cause galaxy-level bootstraps to underestimate the covariance.

\begin{figure}[t]
\centering
\includegraphics[width=0.85\linewidth]{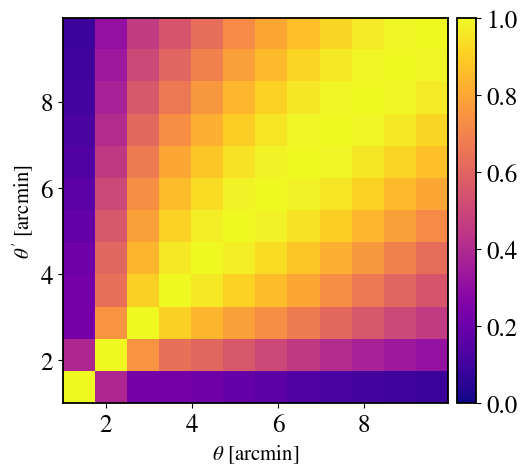}
\caption{Correlation matrix obtained from the block bootstrap covariance for the CAP-filtered, real-space stacked kSZ profiles of the BGS sample.  The covariance was estimated with $N_{\rm boot}=10{,}000$ resamplings of $N_{\rm patch}\approx1837$ \textsc{HEALPix} $2^\circ\times2^\circ$ spatial blocks and computed using Eq.~\eqref{eq:bootstrap_cov}.  The matrix is shown in correlation form, \(\mathrm{Corr}_{ab}=\mathrm{Cov}_{ab}/\sqrt{\mathrm{Cov}_{aa}\mathrm{Cov}_{bb}}\), and reveals strong correlations between filter radii at large radii arising from the compensating ring in the CAP filter.}
\label{fig:cap_cov}
\end{figure}

For each bootstrap realization $b\in\{1,\dots,N_{\rm boot}\}$ with $N_{\rm boot}=10{,}000$, we resample $N_{\rm patch}$ patches with replacement to construct a resampled galaxy catalog, from which we compute the CAP stack $\widehat{T}^{(b)}_{\rm kSZ}(\theta_d)$. The bootstrap estimate of the covariance is then
\begin{eqnarray}
\widehat{\mathrm{Cov}}_{ab}^{\rm BB}
\;=\;& \frac{1}{N_{\rm boot}-1} \sum_{b=1}^{N_{\rm boot}}
\Bigl[\,\widehat{T}^{(b)}_{\rm kSZ}(\theta_a)-\overline{T}_{\rm kSZ}(\theta_a)\Bigr]\,
\nonumber \\
&\times \Bigl[\,\widehat{T}^{(b)}_{\rm kSZ}(\theta_b)-\overline{T}_{\rm kSZ}(\theta_b)\Bigr],
\label{eq:bootstrap_cov}
\end{eqnarray}
where \(\overline{T}_{\rm kSZ}(\theta)\equiv (1/N_{\rm boot})\sum_b \widehat{T}^{(b)}_{\rm kSZ}(\theta)\) is the mean over bootstrap realizations and indices \(a,b\) label filter radii. The velocity dispersion $\sigma_{v_r}$ is held fixed to its full-sample value across all realizations to match the real measurement.
The correlation matrix estimated from the block bootstrap for the BGS is shown in Fig.~\ref{fig:cap_cov}.  The matrix displays substantial off-diagonal correlations at large aperture radii due to the extended response of the CAP filter (the disk-ring combination correlates neighboring apertures and introduces long-range covariance).  We therefore use the full bootstrap covariance when fitting models to the CAP stacks. In App.~\ref{app:cl_ksz_cov}, we show the 2D real-space stacks for both BGS and ELG tracers, which show the kSZ effect (and thus mean optical depth) around these objects.

\subsection{Harmonic-space estimator}
\label{sec:harmonic_space}

The harmonic-space approach computes the cross-spectrum between a reconstructed momentum map derived from the galaxy catalog and the ACT CMB temperature map.  This estimator naturally complements the real-space stacks and is particularly convenient when connecting to lensing or other harmonic-space probes. Here we summarize the formalism, similar to \cite{Harscouet:2025pwl, FrankLRG}.

\subsubsection{Momentum template construction}
\label{sec:momentum_template}

We construct a pixelized momentum (momentum-per-pixel) map \(\pi(\hat{\mathbf{n}})\) on the sphere (notation \(\pi\) is used for the momentum-projection field) as follows.  Galaxies are binned into a CAR-projected map of 0.5 arcmin resolution
and each pixel receives the sum of reconstructed LOS velocities of galaxies that fall in that pixel, normalized to form an intensive quantity:
\begin{equation}
\hat{\pi}(\hat{\mathbf{n}}) \;=\; \frac{1}{\Omega_{\rm pix}\,\overline{n}_g^{2D}}
\sum_{i\in \mathrm{pix}(\hat{\mathbf{n}})} \frac{v_{{\rm rec},i}}{c},
\label{eq:pi_map}
\end{equation}
where \(\Omega_{\rm pix}\) is the pixel solid angle and \(\overline{n}_g^{2D}\) is the mean angular galaxy density (galaxies per steradian).  Equation~\eqref{eq:pi_map} can be written as a line-of-sight projection of the 3D velocity-weighted density,
\begin{align}
\hat{\pi}(\hat{\mathbf{n}})
&= \int d\chi \;\frac{dp}{d\chi}\; \frac{v_r(\chi,\hat{\mathbf{n}})}{c} (1+\delta_g(\chi, \hat{\mathbf{n}})),
\end{align}
where \(dp/d\chi\) encodes the normalized radial selection and the angular normalization and $\delta_g(\chi, \hat{\mathbf{n}})$ is the galaxy overdensity field.

We then expand \(\hat{\pi}(\hat{\mathbf{n}})\) and the CMB temperature map \(\Theta(\hat{\mathbf{n}})=\Delta T(\hat{\mathbf{n}})/T_{\rm CMB}\) in spherical harmonics,
\begin{equation}
\hat{\pi}(\hat{\mathbf{n}})=\sum_{\ell m}\pi_{\ell m}Y_{\ell m}(\hat{\mathbf{n}}),\qquad
\Theta(\hat{\mathbf{n}})=\sum_{\ell m}\Theta_{\ell m}Y_{\ell m}(\hat{\mathbf{n}}),
\end{equation}
and compute the pseudo cross-spectrum
\begin{equation}
\widetilde{C}_\ell^{\hat{\pi} \Theta}
\;=\; \frac{1}{2\ell+1}\sum_{m=-\ell}^{\ell}\pi_{\ell m}\,\Theta_{\ell m}^\ast.
\end{equation}

As in the real-space stacking, we discard galaxies within $8.5'$ of the CMB mask to avoid contamination from point and extended sources in the CMB map. Additionally, we construct a binary mask using the randoms corresponding to the ELG and BGS samples. This constitutes the mask of the momentum field.

\subsubsection{Pseudo-\(C_\ell\) correction with \texttt{NaMaster}}
\label{sec:namaster}

Survey masks, beam convolution, and pixelization couple harmonic modes and bias the direct pseudo-\(C_\ell\) estimators.  We use the public \texttt{NaMaster} package \cite{Alonso_2019} to compute the mode-coupling (workspace) for the particular pair of masks (momentum map mask and ACT temperature mask), decouple the pseudo-spectra, and produce unbiased estimates of the bandpowers \(C_{\ell_b}^{\pi\Theta}\).  In brief, the pseudo-\(C_\ell\) formalism relates the expectation value of the observed (masked) pseudo-spectrum \(\widetilde{C}_\ell\) to the true full-sky spectrum via a mode-coupling matrix \(M_{\ell\ell'}\):
\begin{equation}
\langle \widetilde{C}_\ell\rangle = \sum_{\ell'} M_{\ell\ell'}\,C_{\ell'}.
\end{equation}
\texttt{NaMaster} constructs \(M_{\ell\ell'}\) from the two masks and returns a binned, decoupled estimator and an analytic estimate of the Gaussian covariance that includes the effect of the mask.

\subsubsection{Harmonic-space covariance}
\label{sec:covariance_harmonic}

For the harmonic-space analysis we use the analytic Gaussian covariance computed by \texttt{NaMaster}.  In bandpower space (bins labelled by \(b\) with width \(\Delta\ell\)) the (approximate) Gaussian diagonal variance for the cross-spectrum is
\begin{equation}
{\rm Var}\bigl(C_{\ell_b}^{\hat{\pi}\Theta}\bigr)
\;=\;
\frac{1}{(2\ell_b+1)\Delta\ell\, f_{\rm sky}}
\Bigl[\,C_{\ell_b}^{\hat{\pi}\hat{\pi}}\,C_{\ell_b}^{\Theta\Theta} + \bigl(C_{\ell_b}^{\pi\Theta}\bigr)^2\Bigr],
\label{eq:knox}
\end{equation}
with \(C_{\ell_b}^{\pi\pi}\) the measured momentum auto-spectrum, \(C_{\ell_b}^{\Theta\Theta}\) the (beam+noise convolved) measured CMB temperature auto-spectrum, and \(f_{\rm sky}\) the observed sky fraction. $C_{\ell_b}^{\pi\Theta}$ is measured using \textsc{AbacusSummit}-based kSZ maps and galaxy mock catalogs (see App. C and D of Ref.~\citep{2024arXiv240707152H}). The full \texttt{NaMaster} covariance generalises Eq.~\eqref{eq:knox} to include mode-coupling from masks and the exact binning window function; we use the latter for all parameter inference in harmonic space and provide the simple Knox form as an intuitive check.  The analytic \texttt{NaMaster} covariance is nearly diagonal for our masks and binning and follows closely the Knox approximation (see App.~\ref{app:cl_ksz_cov}). 
This fact allows visual inspection of fits to be more informative compared with the real-space equivalent (see also \cite{Harscouet:2025pwl, FrankLRG}).

\subsection{Theoretical modelling}
\label{sec:theory}

This section describes the modelling pipeline used to predict the cross-correlation between galaxies and electrons and the relation between the measured \(\pi\Theta\) cross-spectrum and the electron-galaxy cross-spectrum \(C_\ell^{\tau g}\).

\subsubsection{Harmonic estimators of the optical depth}
\label{sec:pi_to_taug}

On small scales, and assuming statistical isotropy of the velocity field, the momentum-temperature cross-spectrum can be related to the optical-depth-galaxy cross-spectrum by a velocity factor \cite{Ma:2001xr, 1986ApJ...306L..51O}.  Denoting by \(C_\ell^{\tau g}\) the angular cross-power spectrum between the projected optical-depth field \(\tau(\hat{\mathbf{n}})\) and the galaxy overdensity, one finds
\begin{equation}
C_\ell^{\pi\Theta} \;\simeq\; -\; r\, \sigma_{\rm true}\,\sigma_{\rm rec}\; C_\ell^{\tau g},
\label{eq:pi_theta_relation}
\end{equation}
where:
\begin{itemize}
  \item \(r\) is the cross-correlation coefficient between reconstructed and true LOS velocities defined above,
  \item \(\sigma_{\rm true}\) is the RMS true halo LOS velocity, which we set to 300 km/s for ease here,
  \item \(\sigma_{\rm rec}\) is the RMS of the reconstructed LOS velocities,

\end{itemize}
Equation~\eqref{eq:pi_theta_relation} is used to connect the measured \(\pi\Theta\) bandpowers to \(C_\ell^{\tau g}\), which we model with a halo model as described below.

\subsubsection{Limber approximation}
\label{sec:cl_from_p}

In the Limber approximation, the angular cross-spectrum between the projected optical-depth (or electron column) and galaxies is
\begin{equation}
C_\ell^{\tau g} \;=\; \int_0^\infty \frac{d\chi}{\chi^2}\; W_\tau(\chi)\, W_g(\chi)\; P_{eg}\!\Bigl(k=\frac{\ell+1/2}{\chi},\,z(\chi)\Bigr),
\label{eq:cl_taug_limber}
\end{equation}
where \(P_{eg}(k,z)\) is the electron-galaxy cross-power spectrum, with the radial kernels
\begin{align}
W_\tau(\chi) &= \sigma_T\, a^{-2}(\chi)\, \bar{n}_e(\chi), \\
W_g(\chi)   &= \frac{dp}{d\chi},
\end{align}
where \(\bar{n}_e\) is the mean comoving electron density and \(dp/d\chi\) is the normalized radial selection function of the galaxy sample.

\subsubsection{Halo model for the electron-galaxy cross-power spectrum}
\label{sec:halo_model}

We model the 3D electron-galaxy cross-spectrum \(P_{eg}(k,z)\) with a halo-model decomposition into one-halo and two-halo components:
\begin{align}
P_{eg}^{1h}(k,z) &= \int dM\; \frac{dn}{dM}(M,z)\; \tilde{u}_e(k|M,z)\; \tilde{u}_g(k|M,z), \label{eq:peg_1h}\\
P_{eg}^{2h}(k,z) &= \Biggl[\int dM\; \frac{dn}{dM}(M,z)\; b(M,z)\; \tilde{u}_e(k|M,z)\Biggr]\times \nonumber\\
&
\Biggl[\int dM\; \frac{dn}{dM}(M,z)\; b(M,z)\; \tilde{u}_g(k|M,z)\Biggr]\; P_L(k,z).
\label{eq:peg_2h}
\end{align}
Here \(\tilde{u}_e(k|M,z)\) is the Fourier transform of the ionized gas density profile, $\rho_{\rm gas,free}$, 
and \(\tilde{u}_g(k|M,z)\) is the Fourier transform of the galaxy profile within halos (centrals and satellites), $\rho_g$, which is well approximated by an NFW profile.  \(dn/dM\) is the halo mass function, \(b(M,z)\) the halo bias, and \(P_L(k,z)\) the linear matter power spectrum. For more details, see Ref.~\cite{2023JCAP...03..039B}. Throughout this work, we use the halo model and gas implementation of \texttt{class-sz}\footnote{\url{https://github.com/CLASS-SZ/class_sz}} \cite{Bolliet_2023, Bolliet:2025oqo}.

For the galaxy profile we use standard halo occupation distribution (HOD) prescriptions matched to each tracer:
\begin{itemize}
  \item For the BGS we adopt the standard Zheng-like HOD \cite{Zheng:2004id} (central smooth step function + satellite power-law) calibrated to each of the BGS sub-samples:
  \begin{align}
    \bar{n}_{\mathrm{cent}}^{\mathrm{BGS}}(M) & = \frac{1}{2}\mathrm{erfc} \left[\frac{\log_{10}(M_{\mathrm{cut}}/M)}{\sqrt{2}\sigma}\right], \label{equ:zheng_hod_cent}\\
    \bar{n}_{\mathrm{sat}}^{\mathrm{BGS}}(M) & = \left[\frac{M-\kappa M_{\mathrm{cut}}}{M_1}\right]^{\alpha}\bar{n}_{\mathrm{cent}}^{\mathrm{BGS}}(M),
    \label{equ:zheng_hod_sat}
\end{align}
where the five parameters characterizing the model are $M_{\mathrm{cut}}, M_1, \sigma, \alpha, \kappa$: $M_{\mathrm{cut}}$ characterizes the minimum halo mass to host a central galaxy; $M_1$ characterizes the typical halo mass that hosts one satellite galaxy; $\sigma$ describes the steepness of the transition from 0 to 1 in the number of central galaxies; $\alpha$ is the power law index on the number of satellite galaxies; $\kappa M_\mathrm{cut}$ gives the minimum halo mass to host a satellite galaxy.
Satellites are distributed following an NFW profile with concentration \(c(M)\) from the Tinker 2008 \cite{Tinker:2008ff} concentration-mass relation.
  \item Because ELGs are predominantly younger, bluer galaxies, their HOD is different from that of LRGs and BGS. For this sample, we adopt the \texttt{AbacusHOD} \cite{Yuan:2021izi} implementation of the high-mass quenched (HMQ) model based on Refs.~\citep{2021MNRAS.504.4667A,2020MNRAS.499.5486A,2021MNRAS.502.3599H}, for which the centrals HOD is given by:
  \begin{eqnarray}
    &\bar{n}_{\mathrm{cent}}^{\mathrm{ELG}}(M) =  2 A \phi(M) \Phi(\gamma M)  + \nonumber \\  
    & \frac{1}{2Q} \left[1+\mathrm{erf}\left(\frac{\log_{10}{M_h}-\log_{10}{M_{\mathrm{cut}}}}{0.01}\right) \right],  \label{eq:NHMQ}
\end{eqnarray}
where
\begin{eqnarray}
&\phi(x) =\mathcal{N}(\log_{10}{ M_{\mathrm{cut}}},\sigma_M), \label{eq:NHMQ-phi}\\
&\Phi(x) = \int_{-\infty}^x \phi(t) \, dt = \frac{1}{2} \left[ 1+\mathrm{erf} \left(\frac{x}{\sqrt{2}} \right) \right], \label{eq:NHMQ-Phi}\\
&A =p_{\rm max}  -1/Q.
\label{eq:alam_hod_elg}
\end{eqnarray}
For the ELG satellites, we adopt the same form as the BGS samples.
\end{itemize}
We hold the HOD parameters fixed when fitting the GNFW gas parameters (see Section~\ref{sec:gnfw_fit}); the best-fit HODs are obtained from independent fits to the \(\kappa(\theta)\) measurements using the emulator described in Sec.~\ref{sec:kappa_modeling}.

\subsubsection{GNFW model for the gas density}
\label{sec:gnfw}

We model the 3D ionized gas density with a generalized Navarro--Frenk-White (GNFW) profile:
\begin{align}
\rho_{\rm GNFW}(r) &= \rho_0\,
\Biggl(\frac{r}{x_c\,r_{200c}}\Biggr)^{\gamma}
\Biggl[1+\Biggl(\frac{r}{x_c\,r_{200c}}\Biggr)^\alpha\Biggr]^{-(\beta+\gamma)/\alpha},
\label{eq:gnfw_rho}\\
\rho_{\rm gas,free}(r) &= f_b\,\rho_{\rm cr}(z)\; \rho_{\rm GNFW}(r).
\end{align}
We adopt fixed \(\gamma=-0.5\) and \(x_c=0.7\) and free the parameters \(\{\rho_0,\alpha,\beta\}\) in the fits.  This is done to reduce degeneracies between the parameters, some of which are poorly constrained. The fixed values are chosen as being close to the best-fit values of all samples used in this study. As this is a parametric fit, the purpose of this model is to provide a good fit to the data. We provide an accompanying script with the settings chosen for the class-sz run. 
We also allow the amplitude of the two-halo term, i.e. Eq.~\ref{eq:peg_2h}, to vary, introducing a multiplicative parameter, $A_{k2h}$, due to the difficulty of capturing the one-to-two-halo transition regime.

\subsection{GNFW parameter inference}
\label{sec:gnfw_fit}

\subsubsection{Likelihood and priors}
\label{sec:likelihood_priors}

For each stellar-mass selection, we compare the measured bandpowers in harmonic space, \(C_\ell^{\pi\Theta}\), after converting to an equivalent \(C_\ell^{\tau g}\) via Eq.~\eqref{eq:pi_theta_relation}, to the model predictions from the halo model described above.  We adopt a Gaussian likelihood for the vector of observed bandpowers \(\mathbf{d}\):
\begin{equation}
\ln \mathcal{L}(\mathbf{d} \mid \boldsymbol{\theta})
= -\frac{1}{2}\,\bigl(\mathbf{d}-\mathbf{m}(\boldsymbol{\theta})\bigr)^{\!\mathrm{T}}
\mathbb{C}^{-1}\,
\bigl(\mathbf{d}-\mathbf{m}(\boldsymbol{\theta})\bigr),
\label{eq:gauss_likelihood}
\end{equation}
where \(\boldsymbol{\theta}\) denotes the GNFW parameters (and the two-halo amplitude \(A_{k2h}\) when allowed to vary), \(\mathbf{m}(\boldsymbol{\theta})\) the model bandpowers computed with \texttt{class\_sz}, and \(\mathbb{C}\) is the covariance matrix obtained from \texttt{NaMaster}.

We adopt broad, conservative priors for the GNFW parameters.  In the baseline analysis, we use:
\begin{align*}
\log_{10}\rho_0 &\in [5,\;15],\\
\alpha &\in [0.15,\;0.25],\\
\beta &\in [3.0,\;9.0],\\
\log_{10}A_{k2h} &\in [-1,\;1] \quad (\text{log-uniform prior}). 
\end{align*}
These choices are intentionally broad so as to ensure that the choice of prior does not affect the posterior.

\subsubsection{Sampling}
\label{sec:sampling}

Posterior sampling is performed with the nested-sampling package \texttt{dynesty} \cite{Speagle_2020} to estimate the posterior distributions and Bayesian evidence.  For each fit we run \texttt{dynesty} with \texttt{nlive}=500 using dynamic nested sampling until convergence.  From the posterior samples, we compute marginalized parameter constraints (shown in App. \ref{app:corner}), and the model bandpowers plotted in the main text.

\subsection{HOD parameter inference}
\label{sec:kappa_modeling}


In this work, we adopt the same framework as Ref.~\cite{Hadzhiyska:2025mvt,2025arXiv251014135H}. Below, we summarize the main parts of this model.

We infer the halo occupation distribution (HOD) parameters by fitting the stacked CMB lensing convergence profiles, $\kappa(\theta)$, measured around galaxies, using the ACT DR6 lensing convergence maps. The data vector, $\mathbf{d}_\kappa$, is obtained by azimuthally averaging the stacked convergence maps into annular bins ranging from $0'$ to $5.5'$ in linear bins of width $1.5'$.

Model predictions are constructed from the galaxy-matter cross-power spectrum, $P_{gm}(k,z)$, measured in the \textsc{AbacusSummit} simulations. For a given set of HOD parameters entering $\tilde{u}_g$ in Eqs.~\eqref{eq:peg_1h}--\eqref{eq:peg_2h}, the simulated $P_{gm}$ is projected into the angular galaxy-convergence power spectrum $C_\ell^{g\kappa}$ using the Limber approximation, and subsequently transformed into a real-space convergence profile $\kappa(\theta)$ via an inverse harmonic transform. The model profiles are filtered and binned in an identical manner to the data, ensuring a consistent forward model of the measurement pipeline.

Direct evaluation of the model during inference is accelerated using a Gaussian Process emulator trained on \textsc{AbacusSummit} mocks. The emulator is constructed using a Latin Hypercube sampling of five HOD parameters and predicts the binned $\kappa(\theta)$ profiles at the mean redshift of each lens sample. 


For the BGS\_BRIGHT-20.2 sample, we adopt the standard vanilla HOD parameterization for central and satellite galaxies (Eqs.~\ref{equ:zheng_hod_cent} and \ref{equ:zheng_hod_sat}). This choice is motivated by the fact that no dedicated HOD fits currently exist for that specific BGS sample and the stellar-mass splits used in this work.

For the ELG sample, we instead adopt the HMQ HOD model (Eq.~\ref{eq:NHMQ}). A direct HOD fit to our ELG measurements is poorly constrained, as the ELG one-halo term is pushed to very small angular scales that lie beyond the multipole range currently accessible to the ACT CMB lensing maps ($\ell \gtrsim 3000$). As a result, the ELG HOD parameters cannot be robustly inferred from our data alone. 

We therefore fix the ELG HOD parameters to the best-fit values obtained for the same ELG sample in Ref.~\citep{2025JCAP...01..132G}, adopting the HMQ parameters $\log M_{\rm cut}=11.27$, $\alpha=0.42$, $\kappa=3.41$, $\log M_1=15.50$, $\sigma=0.22$, $\gamma=6.79$, $p_{\rm max}=0.02$, and $Q=100$ (Table~2 of that work). This corresponds to a mean halo mass of $\log (M_{\rm halo}/h^{-1} M_\odot) \approx 12.2$.

We constrain the HOD parameters by comparing the measured and predicted convergence profiles using a Gaussian likelihood, with the covariance matrix estimated from the stacked $\kappa(\theta)$ measurements. Posterior sampling is performed using dynamic nested sampling with \texttt{dynesty}, adopting uniform priors consistent with the emulator training range. The resulting HOD constraints are subsequently fixed when modeling the stacked kSZ signal, thereby reducing degeneracies between galaxy occupation and gas profile parameters. 

\subsection{Remarks on covariance approximations}
\label{sec:cov_validity}

We use block bootstrap covariances for the CAP stacks and the analytic \texttt{NaMaster} covariance for harmonic fits. The bootstrap procedure captures large-scale sample variance and mask effects but can underestimate variance on the largest scales if a small number of large survey modes dominate; our use of \(N_{\rm patch}\sim1000\) mitigates this risk by producing many independent subsamples while keeping each subregion large enough to contain modes comparable to the largest apertures.  The \texttt{NaMaster} Gaussian covariance accounts for mask-induced mode coupling and is accurate when non-Gaussian (connected) contributions are subdominant. Given current and near-future precision of the kSZ measurements, \citep{Harscouet:2025pwl} finds that non-Gaussian corrections are subdominant.

\subsection{Interpretation and systematic checks}
\label{sec:interpretation}

The measured amplitudes are proportional to \(r\,\sigma_{\rm true}\,\sigma_{\rm rec}\) through Eq.~\eqref{eq:pi_theta_relation}.  Uncertainty in \(r\) therefore translates directly into an uncertainty on the inferred optical depth (or cumulative gas mass), and we account for this in our quoted final errors by propagating a \(\sim10\%\) systematics uncertainty on \(r\) (in addition to the statistical covariance).  We perform a suite of null tests (random catalogs, velocity shuffles) to verify that contaminating signals such as tSZ leakage, foreground residuals, or map-making artefacts do not bias our results; these tests are shown in Section~\ref{sec:cl_ksz_null}.

\subsection{Signal-to-noise ratio definition}
\label{sec:snr}

We quantify the significance of the kSZ detection using a likelihood-based signal-to-noise ratio (SNR), defined as
\begin{equation}
    {\rm SNR} \equiv \sqrt{\chi^2_{\rm null} - \chi^2_{\rm bf}} \, ,
\end{equation}
where $\chi^2_{\rm null}$ is the $\chi^2$ evaluated under the null hypothesis of zero kSZ signal, and $\chi^2_{\rm bf}$ is the minimum $\chi^2$ obtained by fitting a fixed template with a single free amplitude parameter. A fit using the four GNFW parameters is performed for each galaxy sample (ELG or BGS) as well as the mass subsamples.

\section{Results}
\label{sec:results}

\subsection{Stacked kSZ}

\begin{figure}
\centering
\includegraphics[width=0.95\linewidth]{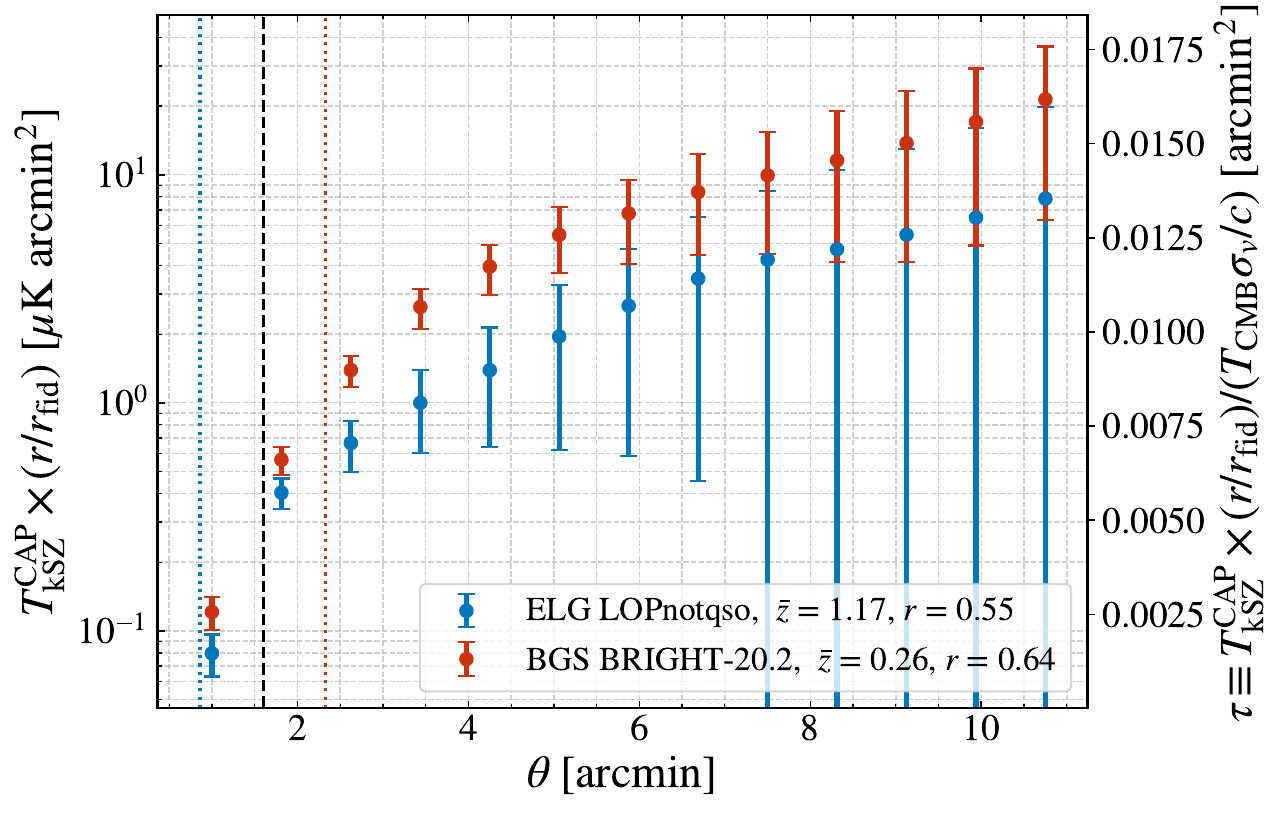}
\includegraphics[width=0.95\linewidth]{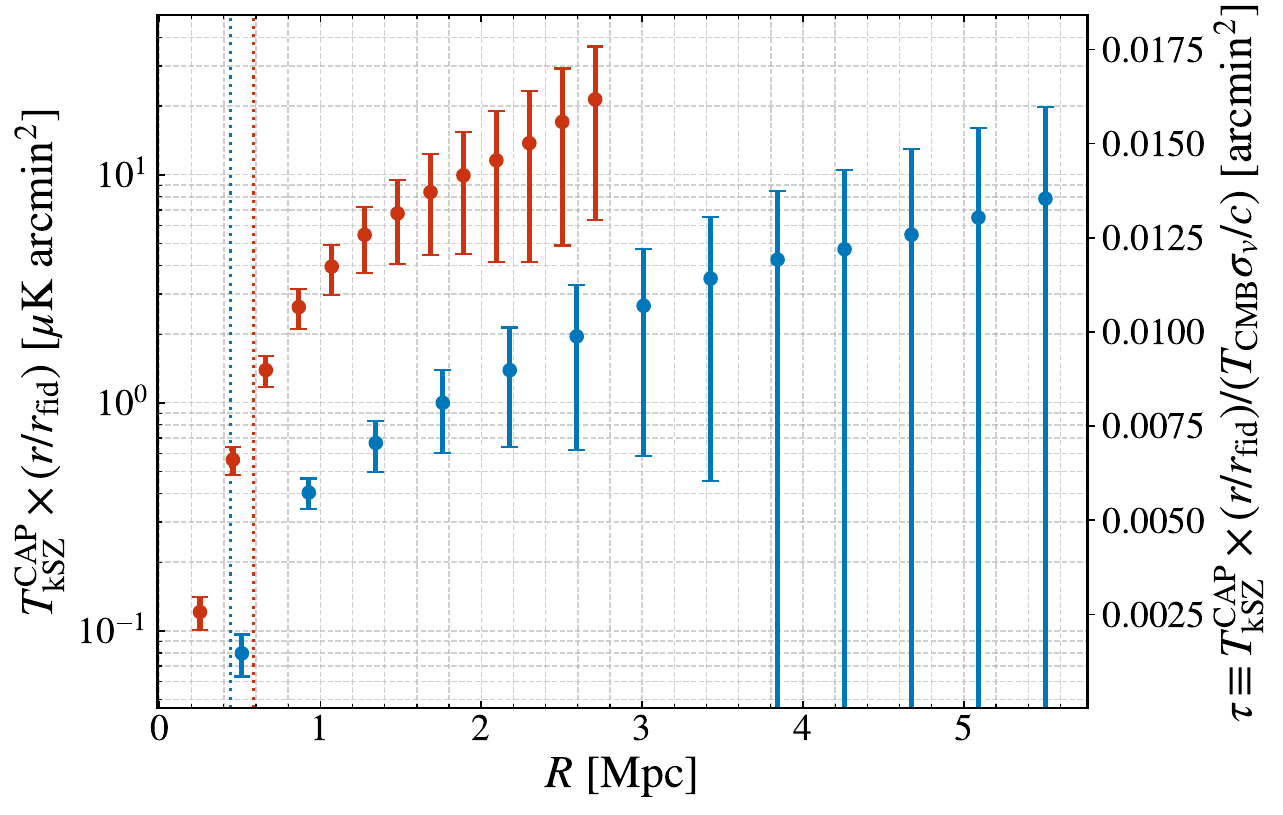}
\caption{
Stacked kSZ profiles for the DESI DR2 Bright Galaxy Sample (BGS) and Emission Line Galaxies (ELGs). 
The top panel shows the measured temperature signal $\Delta T_{\mathrm{kSZ}}$ as a function of angular separation $\theta$ (in arcminutes) and the bottom panel shows them as a function of the physical distance from the center, $R$. 
A vertical line marks the ACT DR6 beam full width at half maximum of $1.6'$ for reference. 
The bottom panel shows the same measurements, but with the horizontal axis converted from angular scale to proper physical distance (in Mpc) assuming a \textit{Planck}~2018 cosmology. 
We adopt a fiducial reconstruction coefficient of $r=0.55$ and $r=0.64$ for the ELG and BGS samples, respectively. 
Both values carry an estimated $\sim 10\%$ uncertainty; the profiles shown here are rescaled by $r/r_{\rm fid}$. 
Dotted vertical lines indicate the virial radii of the typical host halos of BGS and ELG galaxies (accounting for the ACT beam). 
A significant fraction of the inferred gas signal lies outside the virial radius (i.e. the cumulative value of the profile keeps on increasing), consistent with previous measurements of the circumgalactic gas distribution (e.g., \cite{Schaan2021,2024arXiv240707152H,2025arXiv250319870R}). 
}
\label{fig:ksz_stack}
\end{figure}

Fig.~\ref{fig:ksz_stack} presents the stacked kSZ measurements for the DESI DR2 BGS and ELG samples.
The units of $\mu$K ${\rm arcmin}^2$ reflect the direct output of the stacking procedure, which combines the ACT DR6 temperature map with reconstructed line-of-sight velocities and measures the signal in concentric disks and rings around each galaxy.
A vertical line marks the ACT beam size of $1.6'$ to indicate the angular scales on which beam smoothing becomes important, and dotted lines show the effective virial radius of each sample. As the CMB map is convolved with a 1.6 arcmin (FWHM) Gaussian beam, we can estimate the beam-convolved mean virial radius to be $\sqrt{((1.6^2/(8 \ln 2)+\theta_{\rm vir}^2}$, where $\theta_{\rm vir}$ is the mean virial radius (in arcmin) corresponding to each sample (and estimated based on its mean halo mass).

While angular units are natural for the map-based measurement, they do not provide an intuitive comparison between different tracers that lie at very different redshifts.
The BGS and ELG samples differ in mean redshift by nearly a factor of five, so the same angular separation probes very different physical scales.
For this reason, the bottom panel converts the horizontal axis into proper megaparsecs (Mpc) using \textit{Planck}~2018 cosmology.
Proper rather than comoving distances are used because the underlying quantities of interest: gas pressure, density, and the gravitational potential governing the halo environment, are most naturally expressed in proper units.
This conversion allows the two galaxy samples to be compared on a physically meaningful scale, independent of their redshift separation.

The stacked kSZ signal depends on the accuracy of the reconstructed velocity field.
We adopt fiducial reconstruction coefficients of $r=0.64$ for the BGS sample and $r=0.55$ for the ELGs, reflecting the different redshift ranges and galaxy number densities associated with each tracer.
Both coefficients carry an uncertainty of order 10\%.
The curves shown in the figure are therefore displayed in terms of $T_{\mathrm{kSZ}}(r/r_{\rm fid})$, so that improved estimates of $r$ may be applied without redoing the stacking measurement.
A corresponding optical-depth profile is shown on the right-hand axis of each panel, using an assumed velocity rms of $300\ \mathrm{km\ s^{-1}}$. 

When expressed in proper physical units (bottom panel), the differences between the two samples become clear.
The BGS signal has a higher overall amplitude, which follows naturally from the stellar-mass distributions presented in Fig.~\ref{fig:logm}: BGS galaxies occupy more massive halos on average than the ELGs, leading to a larger optical depth and therefore a stronger kSZ signal.
The ELG profile appears to be less steep around the halo boundary (1-2 Mpc) compared to BGS. Combining the kSZ profiles with predictions of the (dark) matter distribution around these galaxies can help us disentangle the effects of baryonic feedback from those coming from the clustering properties of the sample (e.g., two-halo term and satellite effects). We attempt to incorporate clustering (and lensing) information in subsequent sections. Such joint analysis is crucial in understanding the impact of feedback on halos with lower virial masses and deciphering whether their lower potentials make it easier for the halo to expel gas into the surrounding circumgalactic medium, or whether their lower mass makes it unlikely that AGN activity drives the redistribution of gas.

Dashed vertical lines indicate the approximate virial radii of the typical host halos of BGS and ELG galaxies.
In both cases, we see that a fraction of the observed kSZ signal arises at scales extending beyond the virial boundary. This agrees with findings from previous observational and theoretical studies showing that a fraction of the ionized gas resides outside the virial radius in the form of extended circumgalactic and intra-group media (e.g., \cite{Schaan2021,Amodeo2021,2024arXiv240707152H,2025arXiv250319870R}).
The profiles shown here provide an empirical demonstration of this distribution using DESI-ACT cross-correlations. Thus, we define stellar-mass-selected subsamples using lower mass thresholds rather
than disjoint mass ranges. While splitting into narrow mass bins is possible,
the cumulative selection preserves the standard functional form of the HOD (used for the BGS),
which is required for consistent modeling of the galaxy–halo connection.
Using cumulative thresholds, therefore, allows for robust HOD inference while maintaining sufficient signal-to-noise in each subsample.

\subsection{Stellar mass evolution}
Next, we split both the ELG and BGS samples according to stellar masses derived from the \textsc{CIGALE} value-added catalogs. For the ELGs we use two bins, $\log_{10}(M_\star/M_\odot) > 9$ and $> 9.5$. For the BGS we adopt a finer binning to reflect the larger sample size and mass range: $\log_{10}(M_\star/M_\odot) > 10$, $10.5$, $11.0$, and $11.25$. The mean redshift of each subsample (reported in the legends) increases with stellar mass. This trend arises because the BGS and ELG selections impose magnitude cuts: at higher redshift, only intrinsically more massive (and therefore more luminous) galaxies satisfy the flux limits. As a result, mass and redshift become positively correlated in the observed sample.

\begin{figure}[t]
    \centering
    \includegraphics[width=0.95\linewidth]{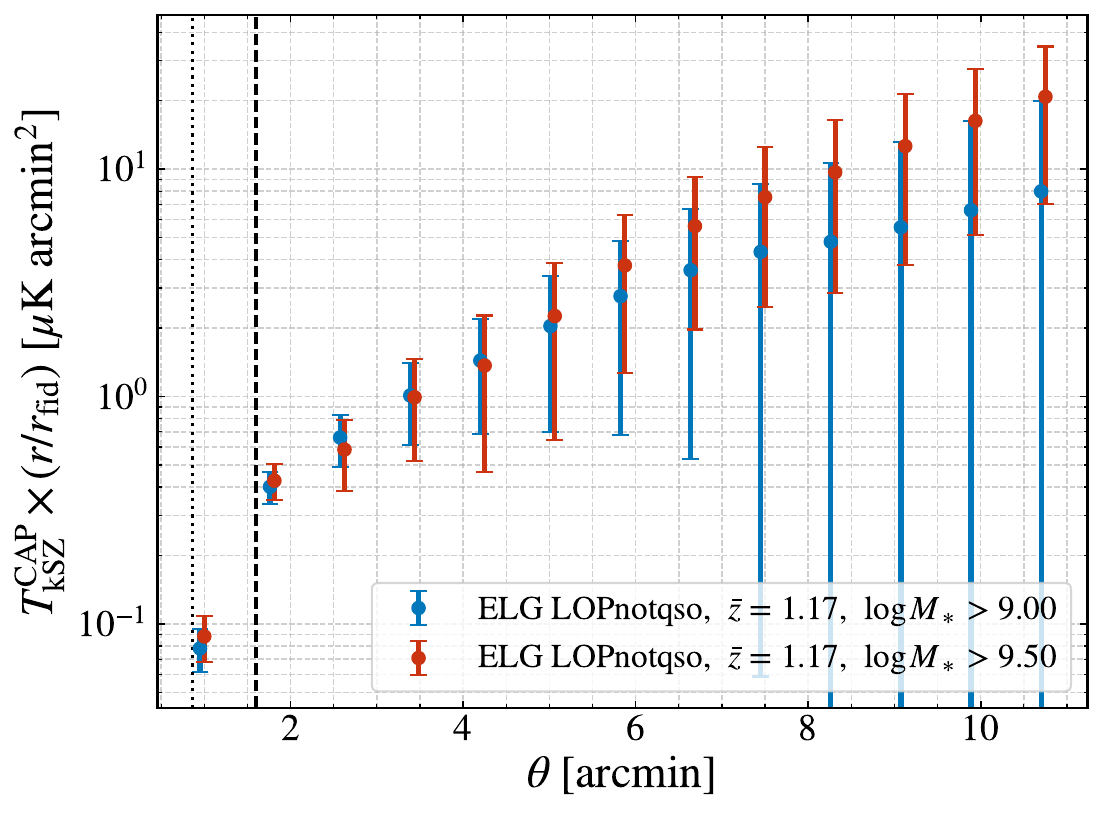}
    \includegraphics[width=0.95\linewidth]{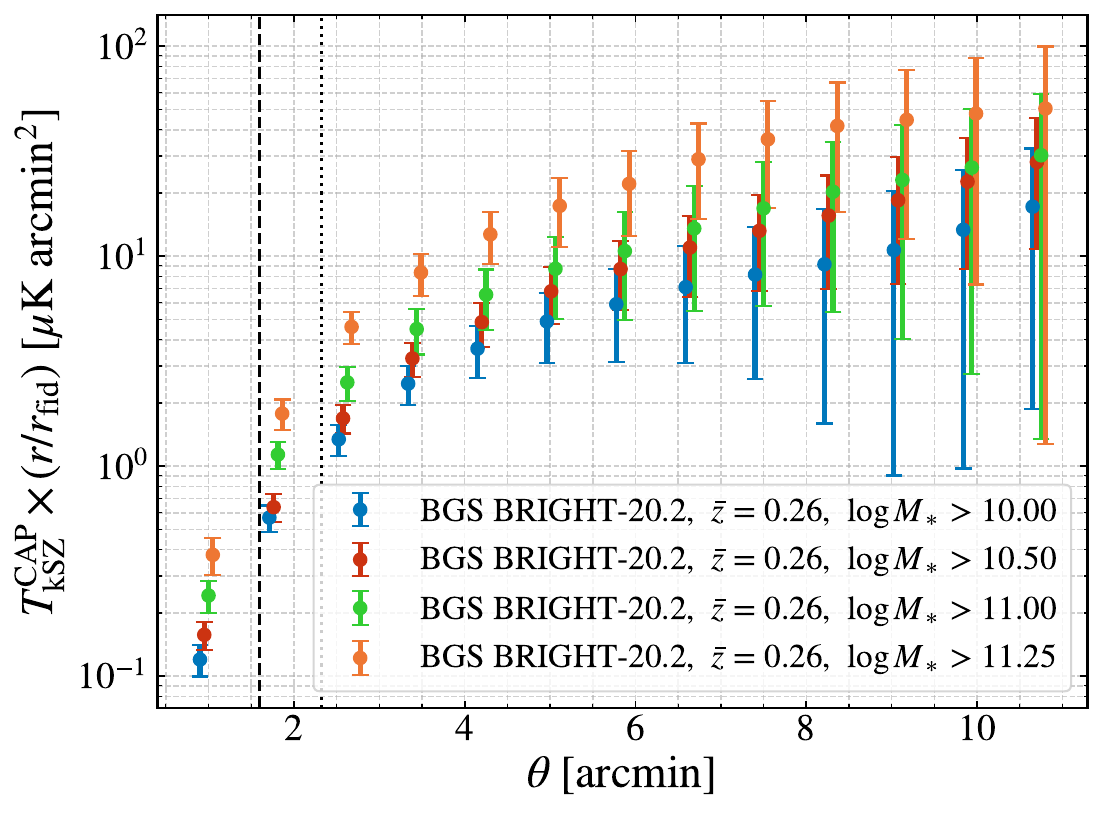}
    \caption{
    Cumulative stellar-mass splits for ELG and BGS kSZ measurements.
    We divide each galaxy sample into subsamples based on stellar mass, using \textsc{CIGALE}-derived estimates. For ELGs we adopt two bins, $\log_{10}(M_\star/M_\odot) > 9$ and $> 9.5$, while for BGS we use five thresholds spanning $\log_{10}(M_\star/M_\odot) > 10$ to $> 11.25$. Higher stellar-mass subsamples exhibit slightly larger kSZ amplitudes, consistent with the expected scaling with total gas (and hence halo) mass. The corresponding SNRs exhibit only a mild dependence on stellar mass, reflecting the balance between increasing halo mass and decreasing sample size. The dashed black line indicates the FWHM of the ACT DR6 hILC map, 1.6$'$, whereas the dotted line corresponds to the virial radius (accounting for the ACT beam effect) of the given galaxy tracer. Notably, the first two points of the BGS sample are within the virial radius; none of the ELG points are within the virial radius.
    }
    \label{fig:ksz_stack_mass}
\end{figure}

In Fig.~\ref{fig:ksz_stack_mass}, we observe an increase in the signal amplitude with stellar mass in both tracers. This behavior is physically expected: the kSZ signal is proportional to the electron momentum along the line of sight, and in stacked measurements, the amplitude scales roughly with the total gas mass enclosed within the aperture. Since gas mass broadly increases with halo mass, and halo mass correlates with stellar mass, higher-mass galaxy samples naturally yield higher kSZ amplitudes. This trend is modest because the relationship between stellar mass and halo mass is not one-to-one and has substantial scatter; moreover, the changes in mean halo mass are moderate, as we apply only a lower mass threshold.

For the BGS, the signal-to-noise ratio does not necessarily peak at a single stellar-mass threshold, but instead reflects a trade-off between sample size and halo mass. Selecting higher-mass galaxies increases the characteristic halo mass, which boosts the kSZ signal, but simultaneously reduces the number of objects in the sample. Since the stacked kSZ signal-to-noise scales approximately as $S/N \propto \sqrt{N}\, M_{\rm halo}$, these competing effects lead to only modest variations in the total SNR across stellar-mass selections. In practice, restricting to higher-mass subsamples yields at most a mild improvement relative to the full BGS sample. For the ELGs, a similar trend is observed, though the variation in SNR is weaker due to the narrower stellar-mass range and the fact that ELGs typically reside in lower-mass halos.

The SNR values corresponding to each mass selection are summarized in Table~\ref{tab:gnfw_results}.

\subsection{Stacked CMB lensing and HOD inference}

\begin{figure}[t]
\centering
\includegraphics[width=0.95\linewidth]{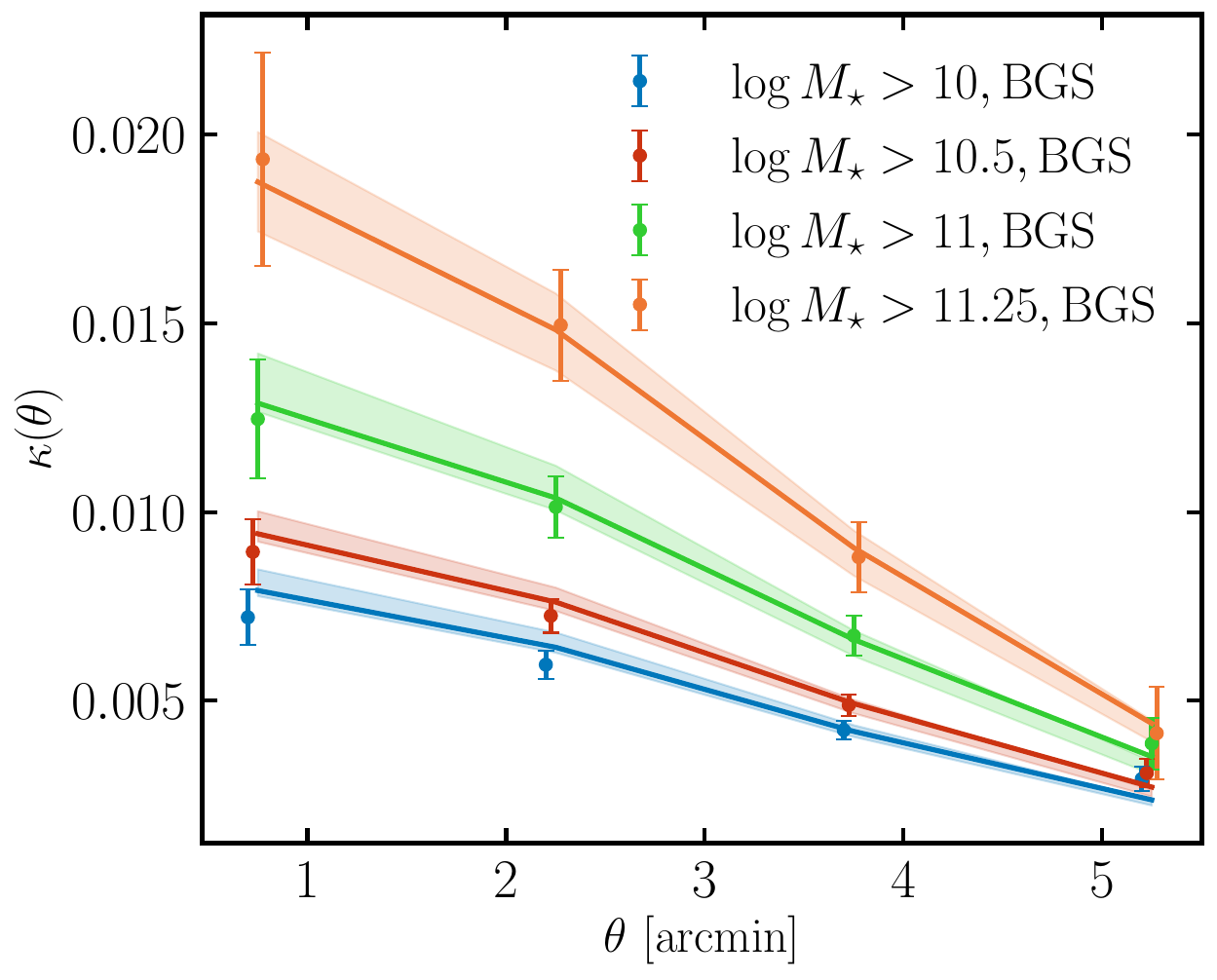}
\caption{
Measurement of the CMB lensing convergence $\kappa$ around DESI BGS 
galaxies, split into the same stellar-mass bins used in Fig.~\ref{fig:ksz_stack_mass}. The stellar-mass estimates are taken from the \textsc{CIGALE} VACs. The $\kappa$ field is reconstructed from the ACT DR6 lensing map, which includes a harmonic-space cut at $L=3000$ to limit small-scale foreground contamination. 
The points show the stacked measurements of $\kappa(\theta)$, and the solid curves indicate the best-fit predictions from our emulator. The emulator maps a five-parameter HOD model to $\kappa(\theta)$ and is trained on simulated lensing data (see Section~\ref{sec:kappa_modeling}). The best-fit HOD parameters and derived quantities are listed in Table~\ref{tab:hod_results}.
The amplitude of $\kappa(\theta)$ increases monotonically with stellar mass, consistent with the expectation that $\kappa \propto$ total mass along the line of sight. 
}
\label{fig:kappa}
\end{figure}
Here we present small-scale CMB lensing measurements for the purpose of calibrating the gas-to-mass ratio of our samples, following the procedure outlined in \cite{Hadzhiyska:2025mvt}. In Fig.~\ref{fig:kappa}, we study the measurement of the CMB lensing convergence, $\kappa$, around DESI BGS galaxies,
split into the same stellar-mass bins described in Fig.~\ref{fig:ksz_stack_mass}. 
As before, the stellar-mass bins for BGS are $\log_{10}(M_\star/M_\odot) > 10$, $10.5$, $11.0$, and $11.25$.

The observable $\kappa$ is particularly informative because it can be interpreted as a weighted projection of the matter overdensity along the line of sight. In other words, higher $\kappa$ corresponds to more matter -- both dark and baryonic, along the line of sight to the galaxies being stacked.

A key feature of the ACT DR6 lensing reconstructed maps used here is the harmonic-space cut at $L = 3000$. This cutoff is imposed to mitigate contamination from small-scale foregrounds. As a result, the reconstructed $\kappa$ field is smoothed on small angular scales, which suppresses information in the deeply non-linear one-halo regime. Despite this, the measurement retains high fidelity in the $1$-$2$ halo transition and in the two-halo regime. These scales contain crucial information about the galaxy-matter cross-correlation, including the galaxy bias and mean halo masses.

Alongside the measurements, we also plot the best-fit predictions from our $\kappa$ emulator. The emulator was trained on simulated $\kappa$ measurements (see Section~\ref{sec:kappa_modeling}), mapping a five-parameter HOD model to the predicted $\kappa(\theta)$. 
We use \texttt{dynesty} to infer the posterior distribution of the HOD parameters and their uncertainties. The resulting best-fit HOD parameters, together with derived quantities such as the mean halo mass, are listed in Table~\ref{tab:hod_results}.

The amplitude of $\kappa(\theta)$ increases systematically across the higher stellar-mass bins, as expected: since $\kappa \propto$ total mass along the line of sight, more massive galaxies, on average, reside in more massive halos and therefore produce a stronger convergence signal. This trend mirrors what we observed in the stacked kSZ profiles.

Finally, because $\kappa(\theta)$ reaches peak values of only order $\sim 0.01$, the weak-lensing approximation underlying quadratic estimators is well satisfied. The small magnitude of the signal also highlights the usefulness of stacking in revealing the average matter distribution around large samples of galaxies.

\begin{table*}
\begin{tabular}{l|ccccc}
\hline
Param. & ${\rm BGS, \ All}$ & ${{\rm log}M_\star > 10}$ & ${{\rm log}M_\star > 10.5}$ & ${{\rm log}M_\star > 11}$ & ${{\rm log}M_\star > 11.25}$ \\
\hline
$\log M_{\rm cut}$ & $11.89^{+0.489}_{-0.594}$ & $11.875^{+0.532}_{-0.616}$ & $11.818^{+0.712}_{-0.529}$ & $12.309^{+0.432}_{-0.343}$ & $12.589^{+0.27}_{-0.325}$ \\ \vspace{0.2cm}
$\log M_1$ & $13.202^{+0.364}_{-0.717}$ & $13.149^{+0.404}_{-0.796}$ & $12.867^{+0.624}_{-0.583}$ & $12.831^{+0.591}_{-0.576}$ & $12.675^{+0.587}_{-0.465}$ \\ \vspace{0.2cm}
$\sigma_{\log M}$ & $0.545^{+0.322}_{-0.374}$ & $0.534^{+0.325}_{-0.362}$ & $0.579^{+0.301}_{-0.381}$ & $0.532^{+0.318}_{-0.352}$ & $0.405^{+0.361}_{-0.274}$ \\ \vspace{0.2cm}
$\alpha$ & $1.379^{+0.368}_{-0.518}$ & $1.389^{+0.362}_{-0.51}$ & $1.402^{+0.327}_{-0.465}$ & $1.357^{+0.356}_{-0.514}$ & $1.322^{+0.362}_{-0.449}$ \\ \vspace{0.2cm}
$\kappa$ & $1.186^{+0.868}_{-0.816}$ & $1.209^{+0.862}_{-0.819}$ & $1.257^{+0.833}_{-0.851}$ & $1.251^{+0.856}_{-0.85}$ & $1.209^{+0.873}_{-0.818}$ \\
\hline
$\bar{n} \times 1000$ & $8.476^{+6.873}_{-4.237}$ & $8.879^{+8.674}_{-4.814}$ & $11.394^{+7.955}_{-7.912}$ & $6.831^{+5.995}_{-4.506}$ & $5.958^{+2.656}_{-4.098}$ \\ \vspace{0.2cm}
$f_{\rm sat}$ & $0.15^{+0.224}_{-0.063}$ & $0.165^{+0.27}_{-0.073}$ & $0.279^{+0.212}_{-0.157}$ & $0.364^{+0.213}_{-0.177}$ & $0.535^{+0.2}_{-0.227}$ \\ \vspace{0.2cm}
$\log \bar{M}_{\rm h}$ & $13.239^{+0.043}_{-0.074}$ & $13.258^{+0.043}_{-0.07}$ & $13.336^{+0.056}_{-0.061}$ & $13.511^{+0.059}_{-0.081}$ & $13.682^{+0.049}_{-0.095}$ \\
\hline
$\chi^2_{\rm null}$ & 747.189 & 751.449 & 711.989 & 390.593 & 227.484 \\
\hline
\end{tabular}
\caption{
Best-fit halo occupation distribution (HOD) parameters inferred from the stacked CMB lensing $\kappa(\theta)$ measurements using the Gaussian-process emulator and \texttt{dynesty} nested sampling (see Sec.~\ref{sec:kappa_modeling}). 
We report the median and central 68\% credible intervals for the five HOD parameters, along with derived quantities: the comoving galaxy number density $\bar{n}$, satellite fraction $f_{\rm sat}$, and mean host halo mass $\log \bar{M}_{\rm h}$ in units of $M_\odot/h$. 
Results are shown for all BGS stellar-mass-selected samples. 
The final row lists the minimum $\chi^2$ values of the best-fitting models; each measurement has four effective degrees of freedom.
}
\label{tab:hod_results}
\end{table*}

Table~\ref{tab:hod_results} summarizes the HOD parameter constraints obtained by fitting the stacked CMB lensing convergence profiles, $\kappa(\theta)$, using the emulator described in Ref.~\citep{Hadzhiyska:2025mvt} and a Gaussian likelihood explored with \texttt{dynesty}. 
Uncertainties correspond to the central 68\% credible intervals marginalized over the full five-dimensional HOD parameter space.

For the BGS samples, increasing the stellar-mass threshold leads to a systematic increase in the inferred mean host halo mass, as expected. 
While the individual HOD parameters exhibit substantial degeneracies -- particularly among $\log M_{\rm cut}$, $\log M_1$, and $\sigma_{\log M}$, the mean halo mass $\log \bar{M}_{\rm h}$ is comparatively well constrained across all samples. 
This reflects the fact that the stacked lensing signal is primarily sensitive to the overall mass scale of the galaxy population rather than the detailed partitioning between central and satellite galaxies. 

We see that the mean halo mass of the samples (the best-constrained property of the sample) is comparable to that inferred for the LRG sample (see Ref.~\citep{Hadzhiyska:2025mvt}) despite the lower stellar masses of the BGS objects (see Fig.~\ref{fig:logm}). We attribute this to the higher satellite fraction of the BGS samples. In other words, while the galaxies are less stellar massive, they live in larger halos as a result of structure evolution: halos grow and merge with each other leading to an increase in satellite occupation. We leave a more detailed study of the BGS and ELG HODs for future work.

The goodness-of-fit values, quantified by the minimum $\chi^2$ listed in the final row, indicate statistically acceptable fits given the four effective degrees of freedom per measurement. 
In the remainder of the paper, we fix the HOD parameters to their best-fit values when modeling the kSZ measurements with GNFW gas profiles, and we assess the impact of residual HOD uncertainties in targeted robustness tests. In addition to the statistical error, we add a $\sim 10\%$ systematic uncertainty due to residual foreground contamination and modeling uncertainties, as discussed in Ref.~\cite{Hadzhiyska:2025mvt}.

\subsection{Gas fractions}

\begin{figure}[t]
\centering
\includegraphics[width=0.95\linewidth]{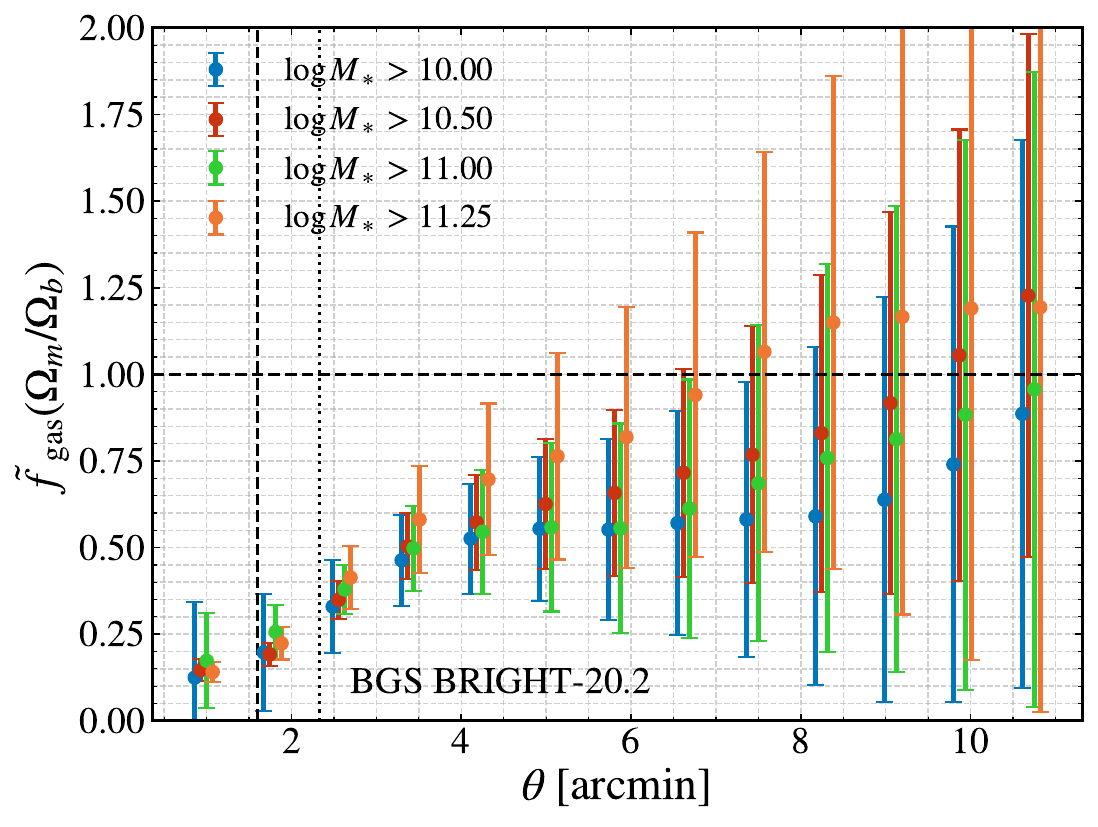}
\includegraphics[width=0.95\linewidth]{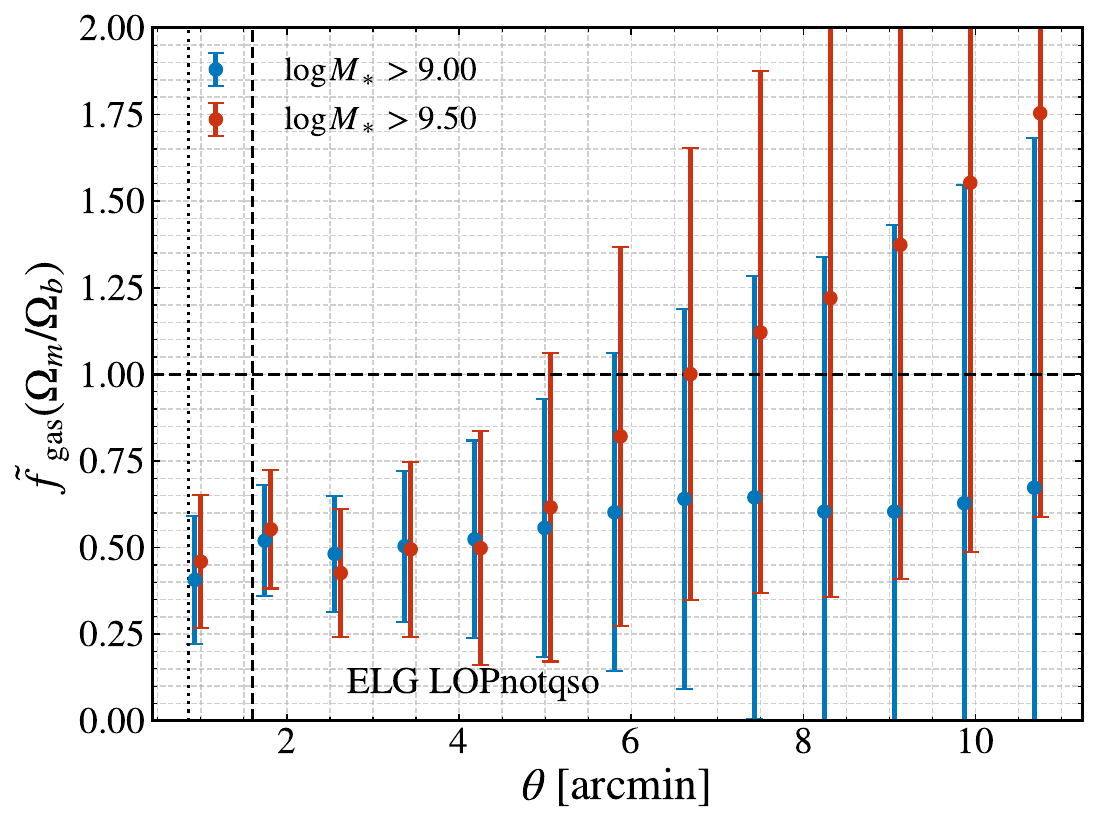}
\caption{
Ratio of the CAP-filtered kSZ and CMB-lensing signals for DESI BGS (top) and ELG (bottom) galaxies split by stellar mass. 
BGS samples use the stellar-mass thresholds $\log_{10}(M_\star/M_\odot)>10,\ 10.5,\ 11,$ and $11.25$; ELGs use $\log_{10}(M_\star/M_\odot)>9, 9.5$, with stellar masses taken from the {\sc CIGALE} VACs.
We adopt the fiducial velocity-density correlation coefficients $r_{\rm fid, ELG}=0.55$ and $r_{\rm fid, BGS}=0.64$. 
The vertical dotted lines mark the characteristic $R_{\rm vir}$ for each sample, while the vertical dashed line denotes the approximate scale of the ACT beam. For the BGS, the ratios imply low enclosed gas fraction, and a mild increase in $f_{\rm gas}$ with stellar mass, consistent with the decrease of the effects of feedback in the more massive host halos. The ELGs show systematically higher gas fractions than the BGS, which can be attributed to the relatively lower effect of AGN activity.  
Ratios approach unity at large radii, where the total gas content is recovered (except for the stellar fraction contribution). We add in quadrature the errors of the numerator and denominator to reflect the uncertainty on both the kSZ and the CMB lensing measurements.
}
\label{fig:fgas}
\end{figure}

To investigate how the gas content of halos varies across our stellar mass selections, we examine the ratio of the CAP-filtered kSZ and CMB-lensing signals, shown in Fig.~\ref{fig:fgas}. For each of the BGS and ELG samples, we adopt the same stellar-mass bins as in the previous sections (see e.g. caption of Fig.~\ref{fig:ksz_stack_mass}). As before, stellar masses are obtained from the {\sc CIGALE} SED-fitting VACs. The numerator, i.e. the CAP-filtered $T_{\rm kSZ}$ profiles, is measured directly by stacking the ACT DR6 temperature map at the galaxy positions. In contrast, the denominator, i.e. the CAP-filtered $\kappa$ profiles are predictions derived from the best-fit HOD parameters of the $\kappa(\theta)$ emulator fits described earlier. For each stellar-mass bin, we forward-model the expected CAP-filtered $\kappa$ curve and propagate the posterior 68\% uncertainty on the HOD parameters, which marginally inflates the errors on the ratio.


Following \cite{Hadzhiyska:2025mvt}, the resulting ratio, $\tilde{f}_{\rm gas}(\theta) \equiv T_{\rm kSZ}^{\rm CAP}/\kappa^{\rm CAP}$ (in appropriate units) represents a gas fraction, so that $\tilde{f}_{\rm gas} = \Omega_b/\Omega_m$ corresponds to gas and matter tracing each other exactly\footnote{Following a different convention, Ref. \cite{Hadzhiyska:2025mvt} normalized $\tilde{f}_{\rm gas} = 1$ corresponding to the cosmological abundance, while in this work, the cosmological abundance of baryons has $\tilde{f}_{\rm gas} = \Omega_b/\Omega_m$.}. This quantity is closely related to the gas-to-mass ratio within an aperture of radius $\theta$. The tilde denotes the fact that this is not exactly the 3D physical gas fraction within a sphere of radius $R$, $f_{\rm gas}(R)$, because of 2D projection, CAP filtering, and the smoothing effect of the CMB beam. Nonetheless, Ref.~\citep{Liu2026} shows 
that for a wide range of feedback scenarios, $\tilde{f}_{\rm gas}(\theta)$ is approximately equal to $\tilde{f}_{\rm gas}(R)$ (to within $\sim 15\%$), making it a very useful probe of the strength of feedback. In addition to being directly measurable from the data (unlike the 3D $f_{\rm gas}(R)$), it can also be extracted directly in simulations and compared to our observations. We have further normalized $\tilde f_{\rm gas}(R)$ by the baryon fraction such that if baryons followed dark matter (and all baryons were in the form of ionized gas), $\tilde f_{\rm gas} (\Omega_m/\Omega_b)$ would be 1. In reality, we expect that about 5\% to 15\% of the baryons are locked in stars or are in the form of neutral gas, which affects the overall normalization at the $\sim 10\%$ level.

For consistency with the rest of the analysis we adopt the fiducial velocity-reconstruction correlation coefficients $r_{\rm fid, ELG}=0.55$ and $r_{\rm fid, BGS}=0.64$ when computing $T_{\rm kSZ}^{\rm CAP}$, while noting that, as discussed in Sec \ref{sec:reconstruction}, this should be more correctly be interpreted as $(r/r_{\rm fid}) \tilde{f}_{\rm gas}$, and uncertainties on $r$ propagated in any interpretation of the signal. 

Several trends emerge clearly. Around the virial radius of each sample (indicated by the dashed vertical lines) the ratios correspond to relatively low enclosed gas fractions, typically $\tilde f_{\rm gas} (\Omega_m/\Omega_b) \sim 0.2 - 0.3$. Beyond ${\sim}2\text{-}3\,R_{\rm vir}$ the ratios rise toward unity for all samples, indicating that the total baryonic content is only fully recovered on these larger scales. We also see that, within the BGS, the higher stellar-mass selections yield slightly higher enclosed gas fractions near the halo boundary ($\sim2'$) than the lower-mass ones. This trend, albeit very mild, is consistent with a reduced impact of feedback in higher-mass halos whose deeper gravitational potentials retain gas more efficiently.

The gas fractions inferred for the BGS samples are relatively low,
$\tilde f_{\rm gas} (\Omega_m/\Omega_b) \sim 0.3$ near the virial radius, consistent with efficient feedback in
these systems. This is consistent with other kSZ measurements of halos with a similar mass, such as the DESI LRGs as measured with kSZ \cite{Hadzhiyska:2025mvt, Siegel:2025frt} and the X-ray gas fractions of low-redshift groups \cite{Popesso:2024ref}. Fast Radio Bursts (FRBs) are quickly emerging as a complementary probe of feedback, and early measurements also indicate efficient feedback \cite{Sharma:2026qxh, 2026arXiv260216781M}, although we caution that a detailed comparison between different works might be complicated by the different selection functions and is left to future work.

In contrast, we find that ELGs exhibit higher
gas fractions despite residing in lower-mass halos. If confirmed with improved
signal-to-noise and robust HOD modeling, this may allow us to probe the conjecture of reduced feedback efficiency near the
peak of the stellar-to-halo mass relation
($M_{\rm halo} \sim 10^{12}\,h^{-1}M_\odot$), where halos are too massive for
supernova feedback to dominate, yet not massive enough for strong AGN-driven
gas expulsion. This test would provide an important connection between
galaxy evolution, feedback processes, and the distribution of circumgalactic gas.

At large angular separations, the ratio of the kSZ- to $\kappa$-derived
profiles appear to increase slightly above unity for some high-mass samples.
However, these bins are highly correlated and dominated by primary CMB
uncertainty, and the apparent deviation corresponds to less than $0.1\sigma$
in significance. We therefore do not interpret this feature as physical.
In the absence of significant stellar or unbound gas contributions, the ratio
is expected to asymptote to unity on large scales, and the observed behavior
is fully consistent with statistical fluctuations in the correlated noise.

\subsection{Angular power spectrum}

\begin{figure}[t]
\centering
\includegraphics[width=0.95\linewidth]{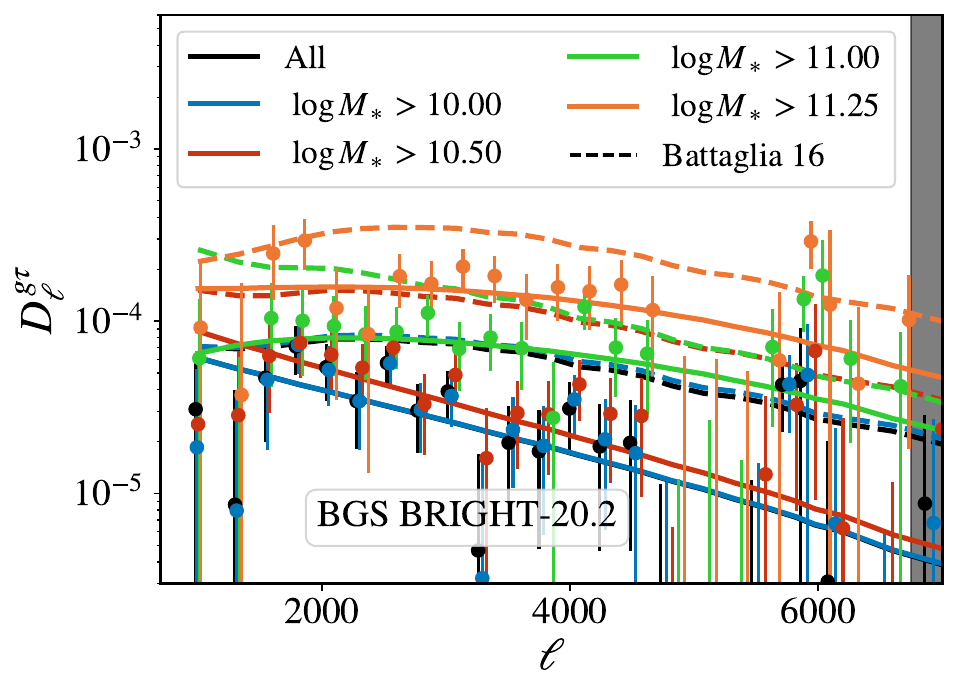}
\includegraphics[width=0.95\linewidth]{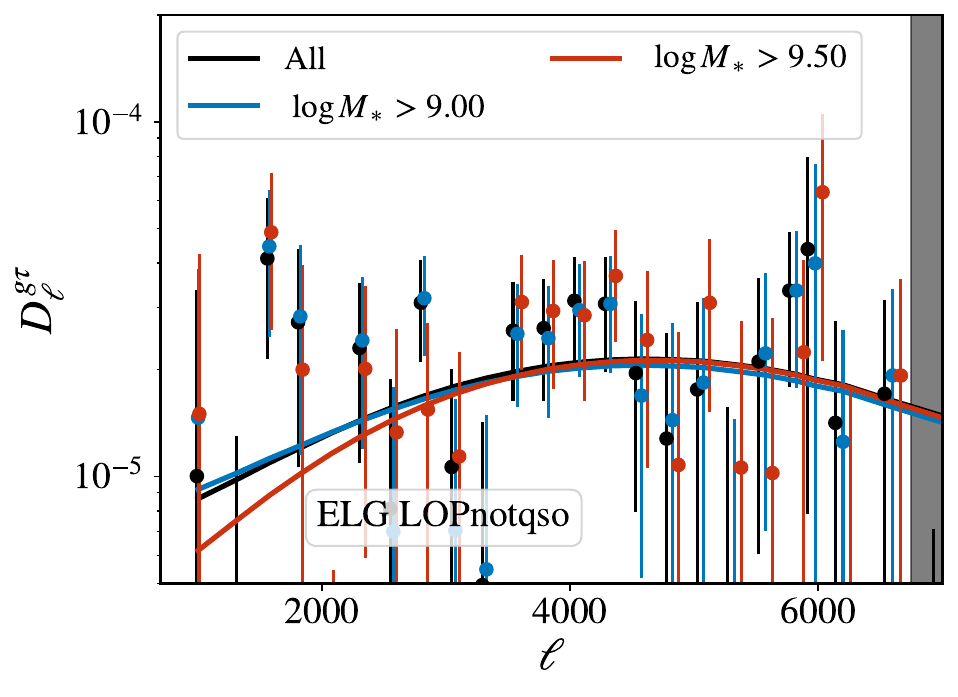}
\caption{
Harmonic-space kSZ measurements, $D_\ell^{\,g\tau}$, for DESI BGS (top) and ELG (bottom) stellar-mass selections. For BGS we use the same stellar-mass splits as in Fig.~\ref{fig:ksz_stack_mass}: 
$\log_{10}(M_\star/M_\odot)>10,\ 10.5,\ 11,$ and $11.25$; 
for ELGs we show $\log_{10}(M_\star/M_\odot)>9, \ 9.5$. We also show the full samples. 
The signal is scaled by $1/r$ and the covariance by $r^2$ for $r_{\rm fid, BGS}=0.64$ and $r_{\rm fid, ELG}=0.55$, consistent with the real-space analysis. 
Solid lines show the best-fit $C_\ell^{\,g\tau}$ curves obtained using the GNFW electron-pressure profile model. The dashed lines show the prediction of the Battaglia 2016 profile for the BGS galaxies \citep{2016JCAP...08..058B}\footnote{We note that the initial Battaglia 2016 fits were done on clusters using hydrodynamical simulations and that the predictions shown in this paper are extrapolated to group-sized halos. We refrain from doing this for the ELGs, where the extrapolation is even more extreme and the HOD more uncertain.}, which is lower than the data points, suggesting that the measurements point to stronger feedback than predicted by this standard fit (see also LRG companion paper \citep{Qu2026}). We fix the galaxy sample to the best-fit HOD parameters and vary four GNFW parameters. The best-fit parameters and uncertainties for each stellar-mass selection are summarized in Table~\ref{tab:gnfw_results}. The covariance is nearly diagonal and shown in Fig.~\ref{fig:cl_ksz_cov}. }
\label{fig:cl_ksz}
\end{figure}

In Fig.~\ref{fig:cl_ksz}, we present our final main kSZ result: the harmonic-space cross-correlation between galaxy density and electron momentum, $C_\ell^{\,g\tau}$. This statistic is the harmonic-space analogue of the real-space stacked kSZ profiles shown in the preceding sections. While the real-space stacks are particularly intuitive for visualizing the radial behavior of the signal, the harmonic-space quantity is directly comparable to many analyses performed entirely in multipole space, including most CMB lensing and galaxy-galaxy lensing measurements. Consequently, $C_\ell^{\,g\tau}$ provides a practical way to quantify the baryonic correction that would enter a purely harmonic-space weak-lensing analysis.

The computation of $D_\ell^{\,g\tau} \equiv C_\ell^{\,g\tau} \ell(\ell+1)/(2\pi)$ follows the ``template'' method described in Section~\ref{sec:harmonic_space}. Briefly, we construct a CMB temperature template from the ACT DR6 maps and cross-correlate it with the velocity-weighted DESI galaxy overdensity field. Mode-coupling, mask effects, and the associated covariance are handled with the \texttt{NaMaster} pseudo-$C_\ell$ formalism. The covariance used for the likelihood analysis is the full analytic \texttt{NaMaster} covariance matrix, which we find to be nearly diagonal for all samples; Fig.~\ref{fig:cl_ksz_cov} illustrates the minimal levels of inter-$\ell$ correlation. The signal-to-noise quoted for these measurements is shown in Table~\ref{tab:gnfw_results}.
While the real-space analysis relies on a completely different strategy for estimating the covariance (from independent regions of the map) and measuring a CAP filter, we verify that the resulting real-space SNRs agree well with those obtained in harmonic space. The agreement between these two conceptually different estimators (in that the real-space one is a windowed integral of the harmonic-space estimator) provides a useful cross-check of our entire pipeline.

Returning to Fig.~\ref{fig:cl_ksz}, the structure of the plot mirrors the earlier mass-split figures: the top panel shows the BGS in the same stellar-mass bins used throughout the analysis, while the bottom panel presents the ELG sample with $\log_{10}(M_\star/M_\odot)>9.5$. In each case, the points show the measured $C_\ell^{\,g\tau}$, with errors from the \texttt{NaMaster} covariance. As in real space, the amplitude displayed here depends on the assumed velocity-reconstruction correlation coefficient $r$: the signal is rescaled by $1/r$ and the covariance by $1/r^2$, where we adopt $r_{\rm fid, BGS}=0.64$ and $r_{\rm fid, ELG}=0.55$.

Overlaid on the data are the best-fit theoretical curves computed using a generalized NFW (GNFW) pressure profile. The galaxy distribution is fixed using the best-fit HOD parameters from the $\kappa(\theta)$ emulator analysis, ensuring a consistent treatment between the kSZ and lensing measurements. For each stellar-mass selection, we allow the primary GNFW shape parameters and a two-halo term amplitude for the electron field to vary, and compute the full prediction with \texttt{class\_sz}. We also show a comparison with the standard Battaglia 16 gas profile, exemplifying a large overall suppression (by a factor of $\sim$2-3) in the BGS signal relative to the Battaglia 16 prediction. This further corroborates the evidence for strong baryonic feedback. 

\begin{table*}
\centering
\begin{tabular}{lccccccc}
\hline\hline
Sample &
$\log_{10}\rho_0$ &
$\alpha$ &
$\beta$ &
$\log_{10}A_{\rm k2h}$ &
$\chi^2_{\rm null}$ &
$\chi^2_{\rm bf}$ &
SNR \\
\hline

ELG (all) &
$6.59 \pm 0.90$ &
$0.215 \pm 0.026$ &
$4.55 \pm 1.30$ &
$-0.72 \pm 0.23$ &
82.25 &
26.01 &
7.50 \\[3pt]

ELG ($\log M_\star>9.0$) &
$6.60 \pm 0.94$ &
$0.212 \pm 0.026$ &
$4.54 \pm 1.28$ &
$-0.71 \pm 0.24$ &
80.37 &
28.40 &
7.21 \\[3pt]

ELG ($\log M_\star>9.5$) &
$6.48 \pm 0.88$ &
$0.212 \pm 0.027$ &
$4.39 \pm 1.22$ &
$-0.72 \pm 0.25$ &
63.75 &
26.46 &
6.11 \\[3pt]

BGS (all) &
$9.33 \pm 1.25$ &
$0.172 \pm 0.020$ &
$7.27 \pm 1.24$ &
$0.74 \pm 0.22$ &
84.97 &
30.75 &
7.36 \\[3pt]

BGS ($\log M_\star>10.0$) &
$9.06 \pm 1.33$ &
$0.175 \pm 0.021$ &
$7.08 \pm 1.32$ &
$0.72 \pm 0.24$ &
79.67 &
28.48 &
7.15 \\[3pt]

BGS ($\log M_\star>10.5$) &
$9.18 \pm 1.41$ &
$0.176 \pm 0.021$ &
$7.04 \pm 1.39$ &
$0.63 \pm 0.28$ &
84.38 &
23.14 &
7.83 \\[3pt]

BGS ($\log M_\star>11.0$) &
$7.06 \pm 1.62$ &
$0.204 \pm 0.029$ &
$4.80 \pm 1.57$ &
$0.06 \pm 0.50$ &
98.24 &
28.07 &
8.38 \\[3pt]

BGS ($\log M_\star>11.25$) &
$7.57 \pm 1.64$ &
$0.200 \pm 0.028$ &
$5.17 \pm 1.70$ &
$0.33 \pm 0.47$ &
107.37 &
26.58 &
8.99 \\

\hline\hline
\end{tabular}
\caption{
Best-fit GNFW parameters obtained from the harmonic-space $C_\ell^{g e}$ analysis for the ELG and BGS samples.
For each tracer and stellar-mass selection, we report the posterior mean and symmetric $68\%$ confidence interval
for the four free GNFW parameters: $\log_{10}\rho_0$, $\alpha$, $\beta$, and $\log_{10}A_{\rm k2h}$.
We also list the $\chi^2$ values for the null hypothesis ($\chi^2_{\rm null}$),
the best-fit model ($\chi^2_{\rm bf}$), and the corresponding signal-to-noise ratio (SNR).
All fits have $\nu=30$ degrees of freedom, corresponding to the number of multipole bins minus
four free parameters. The remaining GNFW parameters are held fixed to
$x_c=0.7$ and $\gamma=-0.5$.
}
\label{tab:gnfw_results}
\end{table*}

Table~\ref{tab:gnfw_results} summarizes the results of the GNFW parameter inference
from the harmonic-space kSZ measurements.
For each galaxy sample, we fit the measured $C_\ell^{g e}$ using a four-parameter GNFW model,
allowing the overall electron density normalization $\rho_0$, inner slope $\alpha$, outer slope $\beta$,
and two-halo amplitude to vary $A_{k2h}$, while fixing the core scale and transition parameter
to $x_c = 0.7$ and $\gamma=-0.5$.
The fits are performed using \texttt{dynesty}, and we quote posterior means and symmetric
$68\%$ confidence intervals.

Across all samples, the harmonic-space covariance is sufficiently diagonal that the $\chi^2$ values closely follow expectations for $\nu = 30$ degrees of freedom. The resulting SNR values show only a weak dependence on the stellar-mass threshold, consistent with the competing effects of increasing halo mass and decreasing number density. The ELG sample similarly yields a robust detection. 
Although the individual GNFW parameters exhibit substantial degeneracies,
the overall goodness-of-fit and detection significance are stable across samples.
These best-fit GNFW models are subsequently used as templates for comparisons with
real-space measurements. We foresee their applicability in accounting for the effect of baryonic feedback on galaxy–galaxy lensing analyses.

\subsection{Null tests}
\label{sec:cl_ksz_null}

To validate the robustness of our measurement of the galaxy-electron cross-power
spectrum, $C_\ell^{\,g\tau}$, we perform two null tests that are designed to destroy
any true correlation between the galaxy catalog and the reconstructed
electron-velocity field while preserving the survey geometry.

\subsubsection{Random angular offsets}

In the first test, we add, to every galaxy in the catalog, a random angular
offset drawn uniformly from the interval $[0,20]$ arcmin in both right ascension
and declination.  This procedure preserves the overall sky mask but erases any
true correlation with the velocity field.  We then recompute $C_\ell^{\,g\tau}$ for
each of our five samples: ELGs with $\log_{10}(M_\star/M_\odot) > 9.5$ and BGS galaxies with
$\log_{10}(M_\star/M_\odot) > 10.0, 10.5, 11.0,$ and $11.25$.

\subsubsection{Velocity-shuffling test}

In the second test, we randomly permute the velocities of the galaxy catalog
while keeping the positions fixed.  This process destroys the true
velocity-density relation while leaving the angular selection function
unchanged.  For each shuffled catalog, we again measure $C_\ell^{\,g\tau}$.

\begin{figure*}[t]
    \centering
    \includegraphics[width=0.48\linewidth]{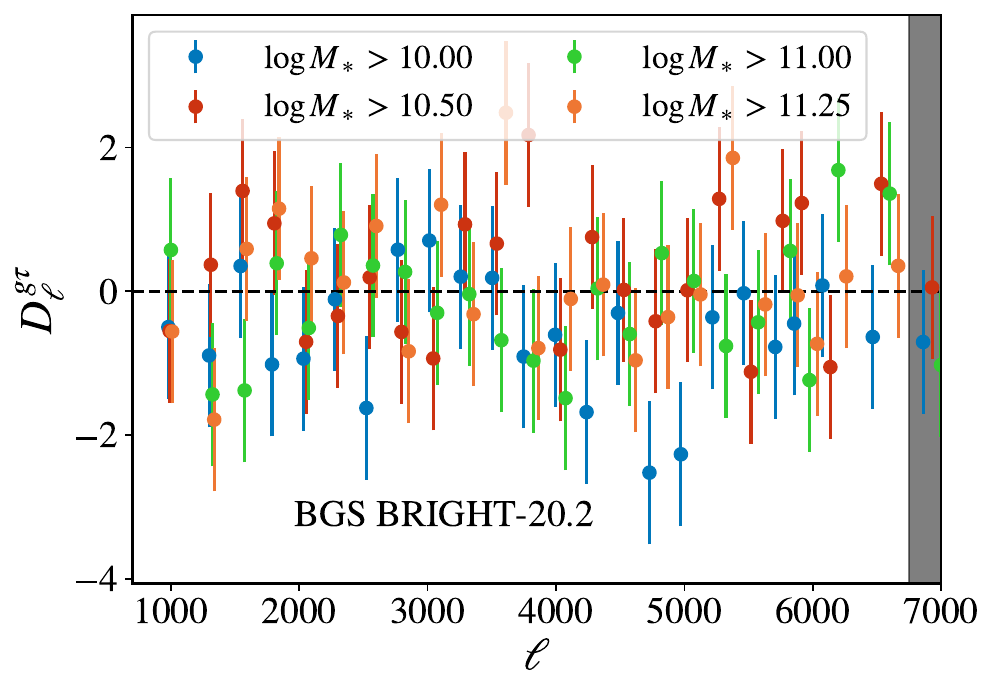}
    \includegraphics[width=0.47\linewidth]{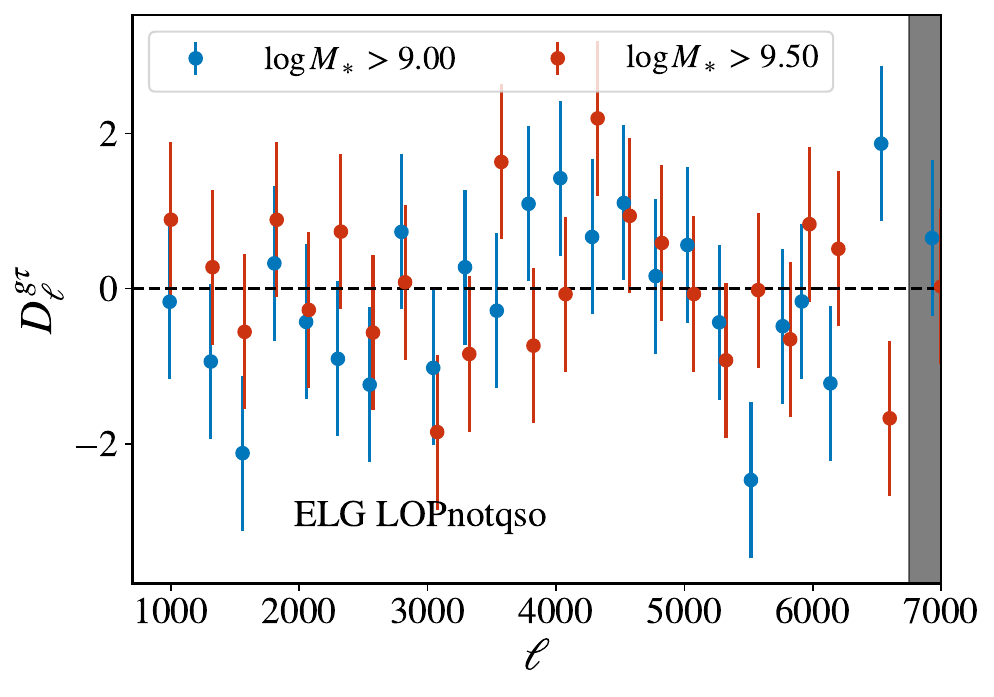}
    \includegraphics[width=0.48\linewidth]{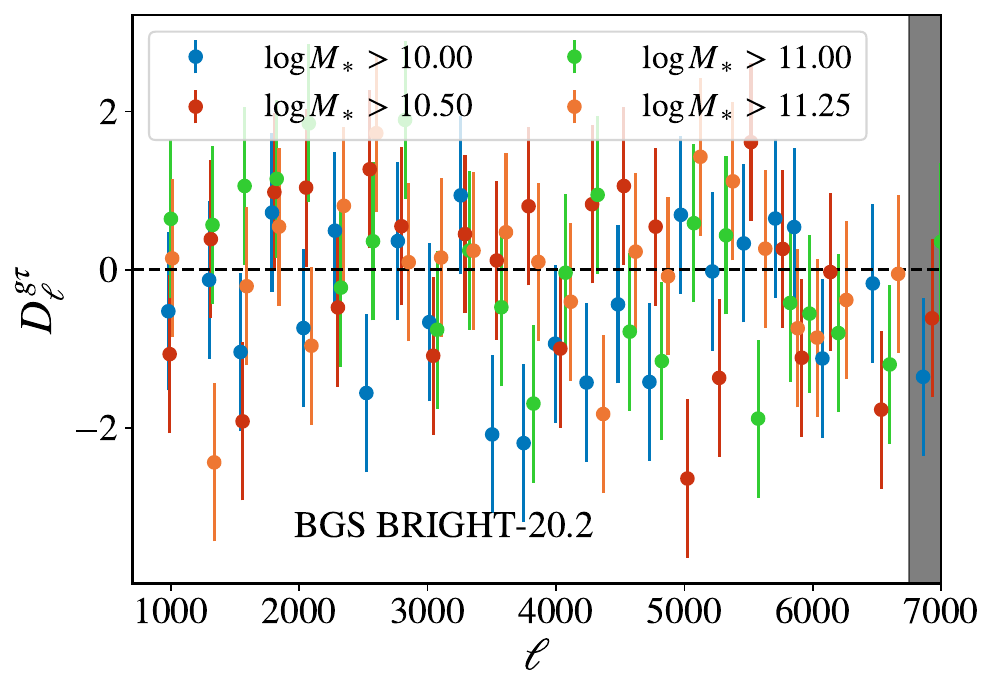}
    \includegraphics[width=0.47\linewidth]{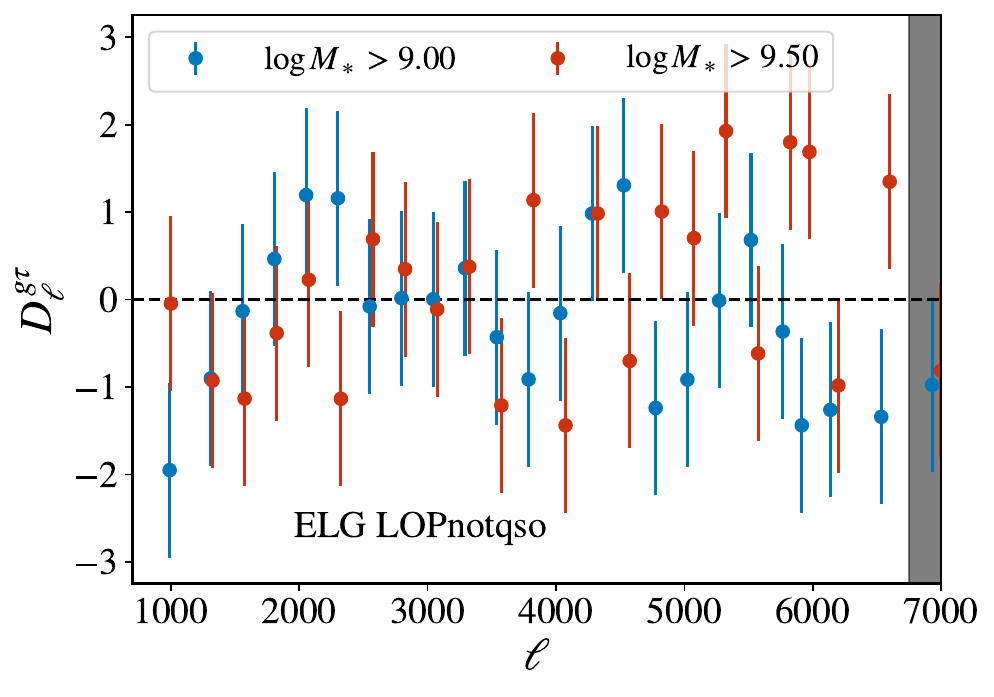}
    \caption{
    Null tests of the $D_\ell^{\,g\tau}$ measurement.    
    \textit{Left:} BGS samples with $M_\ast > 10.0, 10.5, 11.0,$ and $11.25$.
    \textit{Right:} ELGs with $M_\ast > 9.0$ and $9.5$.
    For each sample, we show the result of two null tests:
    (i) \textit{top:} applying a random angular offset of $[0,20]$ arcmin to each galaxy, and
    (ii) \textit{bottom:} shuffling the velocities among galaxies.  
    Both procedures destroy any true correlation between galaxies and the
    reconstructed electron-velocity field while preserving the survey geometry.
    The resulting spectra are consistent with zero, with PTEs reported in the
    text. Both tests remain consistent with a null signal.
    }
\label{fig:cl_ksz_null}
\end{figure*}

Because a single shuffling introduces substantial realization noise, a more
robust implementation would involve repeating this procedure a large number of
times (in our case 1000) and averaging the resulting null spectra to
suppress noise. These are shown in Fig.~\ref{fig:cl_ksz_null}.

\begin{table}
\centering
\begin{tabular}{l|ccc|ccc}
\hline
 & \multicolumn{3}{c}{Random positions} & \multicolumn{3}{c}{Shuffled velocities} \\
Sample 
& $\chi^2_{\rm null}$ & dof & PTE 
& $\chi^2_{\rm null}$ & dof & PTE \\
\hline

ELG, $\log M_\star > 9.0$ 
& 35.11 & 34 & 0.42 
& 34.36 & 34 & 0.45 \\

ELG, $\log M_\star > 9.5$ 
& 39.54 & 34 & 0.24 
& 36.82 & 34 & 0.34 \\

BGS, $\log M_\star > 10.0$ 
& 31.05 & 34 & 0.61 
& 40.09 & 34 & 0.22 \\

BGS, $\log M_\star > 10.5$ 
& 26.92 & 34 & 0.80 
& 37.62 & 34 & 0.31 \\

BGS, $\log M_\star > 11.0$ 
& 37.42 & 34 & 0.32 
& 33.73 & 34 & 0.48 \\

BGS, $\log M_\star > 11.25$ 
& 29.77 & 34 & 0.68 
& 30.67 & 34 & 0.63 \\
\hline
\end{tabular}
\caption{
Results of null tests in harmonic space.
For each galaxy sample we report the $\chi^2_{\rm null}$, number of degrees of
freedom, and probability-to-exceed (PTE) for two null tests: stacking on random
sky positions and shuffling reconstructed velocities.
The random-position null test yields consistently acceptable PTE values across
all samples.
The velocity-shuffling null test shows significant deviations from the null
hypothesis for the lowest-mass BGS samples, indicating sensitivity to the inner-halo
dynamics and satellite populations.
}
\label{tab:cl_ksz_null}
\end{table}

The results are summarized in Table~\ref{tab:cl_ksz_null}.
Both the random-position as well as the velocity-shuffling null tests yield uniformly acceptable PTE values across all
samples, indicating that the covariance estimation and noise modeling are well
under control and that no spurious large-scale correlations are present. The real-space null tests are shown in App.~\ref{app:ksz_stack_null}.

\section{Conclusions}
\label{sec:discussion}

In this work, we have presented the first spectroscopic stacked detections of the kinematic Sunyaev--Zel’dovich (kSZ) signal for both emission line galaxies (ELGs) and bright galaxy sample (BGS) galaxies from DESI, using CMB temperature and lensing data from ACT DR6. By combining velocity reconstruction, stacked temperature measurements, and harmonic-space analyses, we achieve a high signal-to-noise of 7 and 9 for the ELGs and BGS, respectively, 
and provide a unified characterization of the gas distribution around star-forming and low-redshift galaxies across a wide range of physical scales.

This analysis represents the first kSZ-based constraints on the circumgalactic gas content of spectroscopically selected blue, star-forming galaxies, ELGs. The combination of real-space stacking, harmonic-space modeling, stellar-mass binning using DESI value-added catalogs, and joint interpretation with CMB lensing provides a comprehensive framework for studying baryonic feedback across galaxy populations. The GNFW fits presented here enable direct forward modeling in future galaxy-galaxy lensing and kSZ analyses involving DESI ELG and BGS samples.

In configuration space, we detect the stacked kSZ signal for both ELGs and BGS galaxies with high significance (Fig.~\ref{fig:ksz_stack}), and demonstrate the robustness of the measurement through a suite of null tests based on velocity shuffling and random offsets (Fig.~\ref{fig:ksz_stack_null}). We further validate the velocity reconstruction by examining the line-of-sight velocity distribution as a function of redshift, finding no evidence for redshift-dependent biases or edge effects, aside from the expected increase in noise at the sample boundaries due to reduced number density (Fig.~\ref{fig:velz}). Splitting the samples by stellar mass, we observe systematic variations in the stacked signal amplitude that reflect differences in the typical host halo mass (Fig.~\ref{fig:ksz_stack_mass}).

In parallel, we measure the galaxy-CMB lensing cross-correlation $\kappa(\theta)$ and use it to infer halo occupation parameters via an emulator trained on \textsc{AbacusSummit} simulations (Fig.~\ref{fig:kappa}). Fixing these HOD parameters, we model the kSZ signal using a generalized NFW gas profile and perform parametric fits in harmonic space. The resulting $C_\ell^{\pi\Theta}$ measurements are well described by a nearly diagonal covariance matrix (Fig.~\ref{fig:cl_ksz}), consistent with Gaussian expectations, and exhibit no significant residual correlations beyond those predicted by the Knox approximation. Null tests in harmonic space further confirm the absence of spurious signal (Fig.~\ref{fig:cl_ksz_null}).

An important feature of the present analysis is the joint use of CMB
lensing-derived matter profiles alongside the kSZ measurements (see also Ref.~\citep{Hadzhiyska:2025mvt}). This enables
an inference of a gas-fraction proxy that is less sensitive to the mean halo
mass of the galaxy sample, since both observables are convolved with the same
HOD and are both (approximately) proportional to mass. While residual dependence on the velocity reconstruction coefficient $r$, stellar fraction, and satellite fraction
remains, this sensitivity is milder than the direct dependence on halo mass
that limited earlier comparisons to simulations.

Combining the kSZ and CMB lensing measurements, we construct an approximate cumulative tracer of the gas fraction as a function of scale (Fig.~\ref{fig:fgas}). Across both tracers, we find low enclosed gas fractions near the characteristic virial radius, with values well below the cosmic baryon fraction, indicating substantial redistribution of gas to larger radii. BGS consistently exhibit low gas fractions ($\tilde f_{\rm gas} (\Omega_m/\Omega_b) \approx 0.3$), consistent with strong AGN feedback processes. A mild dependence on stellar mass is also observed, with higher-mass selections showing systematically higher gas fractions near $R_{\rm vir}$, in qualitative agreement with expectations from gravitational potential depth.

Our results are broadly consistent with previous kSZ measurement works, indicating strong
feedback in group-sized halos \citep{2024arXiv240707152H,2025arXiv250319870R,Hadzhiyska:2024ecq,Hadzhiyska:2025mvt}. In particular, as also seen in the companion DR2
LRG analysis \citep{Qu2026}, the measured profiles lie significantly below the standard
Battaglia 16 gas density profile, suggesting strong feedback. The low gas
fractions inferred for the BGS samples ($\sim 0.3$ near the virial radius)
point in the same direction. Interestingly, we find that ELGs exhibit higher
gas fractions than BGS despite residing in lower-mass halos, which is consistent with the hypothesis of a reduced feedback
efficiency near the peak of the stellar-to-halo mass relation
($M_{\rm halo} \sim 10^{12}\,h^{-1}M_\odot$), where halos are too massive to be
dominated by supernova feedback, yet not massive enough for strong AGN-driven
gas expulsion. However, the uncertainties on the ELG-hosting halo masses make it hard to draw definite conclusions from this finding. If confirmed with
higher signal-to-noise and well-constrained HOD, we could establish an important link between galaxy
evolution, feedback processes, and the spatial distribution of circumgalactic gas.

The measurement of gas density around galaxies spanning a vast range of masses and redshifts, in this work and the companion paper \citep{Qu2026}, together with measurements of the thermal SZ effect (tSZ) of the same samples \cite{Liu:2025zqo}, brings us an important step forward to understanding the complex processes of baryonic feedback and galaxy evolution.


\begin{acknowledgments}

We thank Nick Battaglia, Elisabeth Krause, Henry Liu, Daisuke Nagai, Uros Seljak, Kritti Sharma, and Martin White for enlightening discussions during the preparation of this manuscript.

B.H. is supported by the Institute of Astronomy, University of Cambridge, and the Kavli Institute of Cosmology Cambridge. 
S.F. is supported by Lawrence Berkeley National Laboratory and the Director, Office of Science, Office of High Energy Physics of the U.S. Department of Energy under Contract No.\ DE-AC02-05CH11231.

This material is based upon work supported by the U.S. Department of Energy (DOE), Office of Science, Office of High-Energy Physics, under Contract No. DE–AC02–05CH11231, and by the National Energy Research Scientific Computing Center, a DOE Office of Science User Facility under the same contract. Additional support for DESI was provided by the U.S. National Science Foundation (NSF), Division of Astronomical Sciences under Contract No. AST-0950945 to the NSF’s National Optical-Infrared Astronomy Research Laboratory; the Science and Technology Facilities Council of the United Kingdom; the Gordon and Betty Moore Foundation; the Heising-Simons Foundation; the French Alternative Energies and Atomic Energy Commission (CEA); the National Council of Humanities, Science and Technology of Mexico (CONAHCYT); the Ministry of Science, Innovation and Universities of Spain (MICIU/AEI/10.13039/501100011033), and by the DESI Member Institutions: \url{https://www.desi.lbl.gov/collaborating-institutions}. Any opinions, findings, and conclusions or recommendations expressed in this material are those of the author(s) and do not necessarily reflect the views of the U. S. National Science Foundation, the U. S. Department of Energy, or any of the listed funding agencies.

The authors are honored to be permitted to conduct scientific research on I'oligam Du'ag (Kitt Peak), a mountain with particular significance to the Tohono O’odham Nation.

We acknowledge the use of public ACT data products made available through the National Energy Research Scientific Computing Center (NERSC), a U.S. Department of Energy Office of Science User Facility operated under Contract No. DE-AC02-05CH11231.

\end{acknowledgments}

\section*{Data Availability}
The data products associated with this work are 
publicly available on Zenodo at \url{https://doi.org/10.5281/zenodo.19160138}.

%




\appendix

\section{Foreground tests}
\label{app:hilc_vs_single}

\begin{figure}[t]
    \centering
    \includegraphics[width=0.98\linewidth]{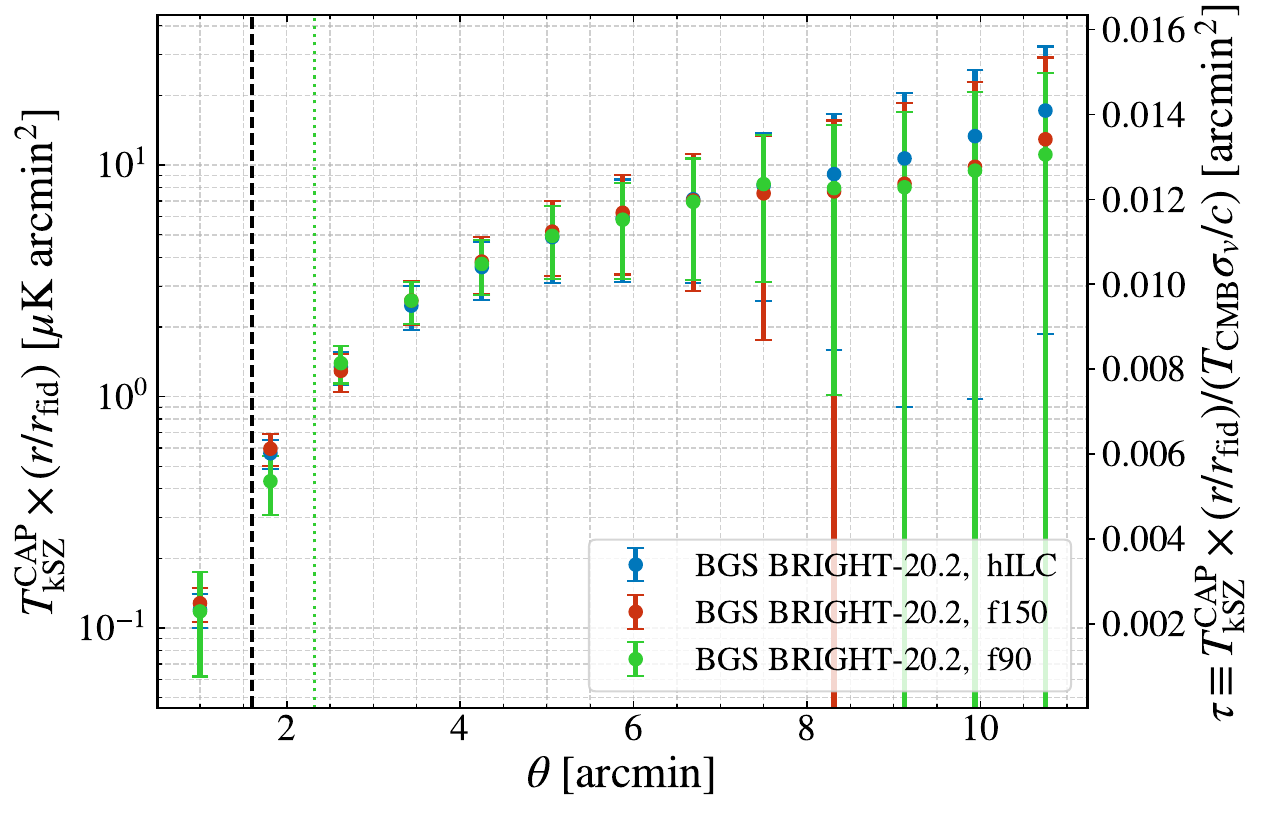}
    \caption{Real-space stacking measurement of the kSZ-galaxy cross-correlation using the full BGS sample, comparing the hILC map with the single-frequency ACT DR6 \texttt{f090} (90 GHz) and \texttt{f150} (150 GHz) maps. The three measurements are highly consistent across radial bins. The hILC uncertainties are smallest, reflecting the optimal combination of frequency channels in the ILC construction. The slight downward fluctuation in the second radial bin of the 90 GHz measurement ($\sim1\sigma$) is not present in the hILC result and is consistent with residual foreground contamination in the single-frequency map.}
    \label{fig:foreground}
\end{figure}

In this appendix, we test the robustness of our fiducial kSZ measurement to possible foreground contamination and map-making systematics. Our baseline analysis uses the harmonic ILC (hILC) map, which optimally combines frequency channels to minimize foreground power while preserving the CMB signal. However, the harmonic-space construction and scale-dependent weighting could in principle redistribute residual foregrounds or introduce subtle biases. To verify that this procedure does not affect our results, we repeat the measurement using the single-frequency \texttt{f090} and \texttt{f150} maps. Because the kSZ effect is frequency-independent (in contrast to the tSZ signal), the recovered cross-correlation amplitude should be consistent across frequencies if foreground contamination is under control. The comparison is shown in Fig.~\ref{fig:foreground}.

We find that the kSZ cross-correlation signal is consistent between the hILC map and the single-frequency \texttt{f090} and \texttt{f150} measurements. The close agreement among these independent reconstructions demonstrates that our kSZ estimator is robust to foreground contamination and insensitive to the specific details of the ILC construction. While the 90 GHz map shows a modest ($\sim1\sigma$) downward fluctuation in the second radial bin, this feature is not observed in the 150 GHz or hILC measurements and is therefore consistent with statistical variation or mild residual foreground contamination. The hILC result, which we adopt as fiducial, exhibits the smallest uncertainties as expected from the optimal frequency combination performed by the ILC.

\section{Reconstructed line-of-sight velocity}
\label{app:velz}

\begin{figure}[t]
    \centering
    \includegraphics[width=0.95\linewidth]{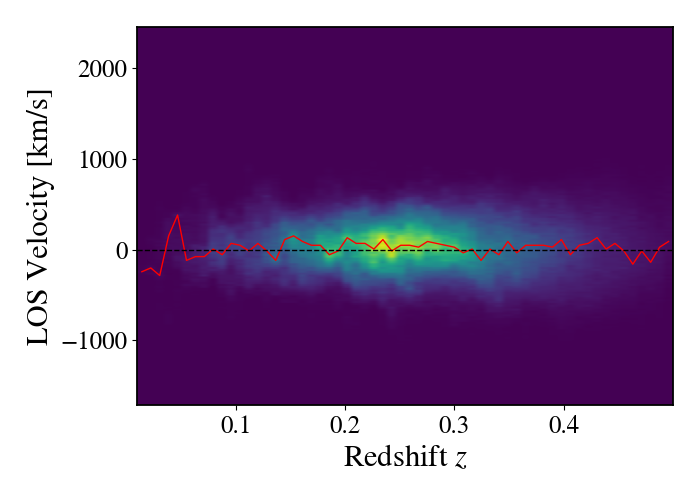}
    \includegraphics[width=0.95\linewidth]{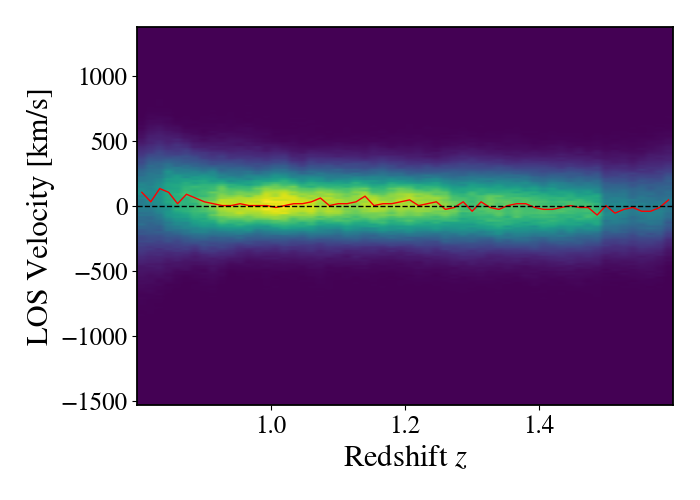}
    \caption{
    Reconstructed LOS velocity as a function of redshift for BGS (top) and ELG (bottom).
    Each panel shows a two-dimensional histogram of the \textsc{pyrecon} LOS velocity field versus galaxy redshift, using the reconstruction settings adopted in the kSZ analysis (smaller smoothing radius for ELGs; default \texttt{BGS\_BRIGHT-20.2} settings for BGS; \texttt{MultiGrid} mode; \texttt{RecSym} convention). The red curve indicates the mean reconstructed velocity in narrow redshift bins. In both samples, the mean remains consistent with zero at all redshifts, indicating no detectable redshift-dependent biases or edge effects. Increased scatter near the redshift boundaries reflects the lower number density of galaxies in those regions.}
    \label{fig:velz}
\end{figure}

To assess potential redshift-dependent systematics in the reconstructed velocities, we examine the line-of-sight (LOS) velocity field produced by \textsc{pyrecon} using the reconstruction settings adopted in our stacked kSZ analysis. For the ELG sample, we employ a slightly smaller smoothing scale, while for the BGS we retain the default parameters used for the \texttt{BGS\_BRIGHT-20.2} selection. In both cases, we use the \texttt{MultiGrid} mode and the \texttt{RecSym} convention.

Fig.~\ref{fig:velz} shows two-dimensional histograms of reconstructed LOS velocity versus redshift for ELGs and BGS galaxies. The overplotted mean velocity in narrow redshift bins provides a sensitive diagnostic of residual biases: in the absence of systematics, the mean reconstructed LOS velocity is expected to be consistent with zero at all redshifts. We find no evidence for redshift-dependent biases or edge effects in either tracer. As expected, the velocity field appears somewhat noisier near the redshift boundaries of each sample due to the reduced galaxy number density, but the mean remains consistent with zero across the full redshift range.

These tests demonstrate that the adopted reconstruction settings do not introduce detectable redshift-dependent biases that could contaminate our stacked kSZ measurements.

\section{Correlation structure of the harmonic-space covariance}
\label{app:cl_ksz_cov}

\begin{figure}
\centering
\includegraphics[width=0.95\linewidth]{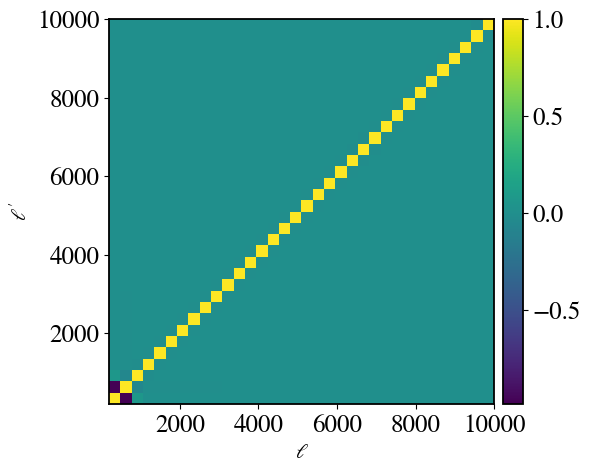}
\caption{
Correlation matrix of the harmonic-space kSZ power spectrum $C_\ell^{\pi\Theta}$ over the range $0 \le \ell \le 10^4$.
The covariance is highly diagonal, closely matching the expectations from the Knox approximation.
Noticeable correlations appear only on very large angular scales ($\ell \lesssim 100$), which contribute negligibly to the total signal-to-noise of the measurement.
}
\label{fig:cl_ksz_cov}
\end{figure}

In Fig.~\ref{fig:cl_ksz_cov} we show the correlation matrix of the harmonic-space kSZ power spectrum,
$C_\ell^{\pi\Theta}$, over the multipole range $0 \le \ell \le 10^4$. The correlation matrix is calculated from the full covariance obtained with \texttt{NaMaster} (see Section~\ref{sec:covariance_harmonic}).

The covariance is found to be extremely diagonal across nearly the entire multipole range relevant for the measurement. This behavior closely follows the expectations from the Knox approximation for Gaussian fields (see Eq.~\ref{eq:knox}), indicating that mode coupling induced by masking, filtering, or reconstruction effects is negligible for the scales dominating the signal-to-noise.

Small departures from diagonality appear only on the largest angular scales, at $\ell \lesssim 100$. However, these modes contribute negligibly to the total signal-to-noise of the kSZ measurement, and their impact on parameter inference is therefore minimal. This has also been reported in the kSZ harmonic space analysis of \cite{Harscouet:2025pwl, FrankLRG}, providing additional validation of this covariance structure.

\begin{figure}[t]
    \centering
    \includegraphics[width=0.95\linewidth]{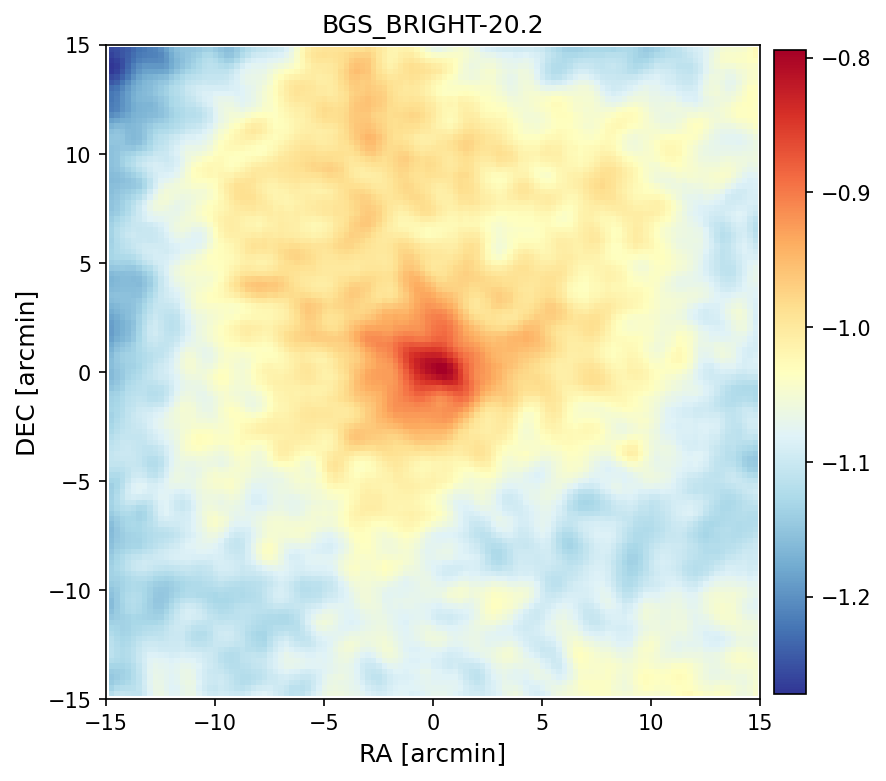}
    \includegraphics[width=0.95\linewidth]{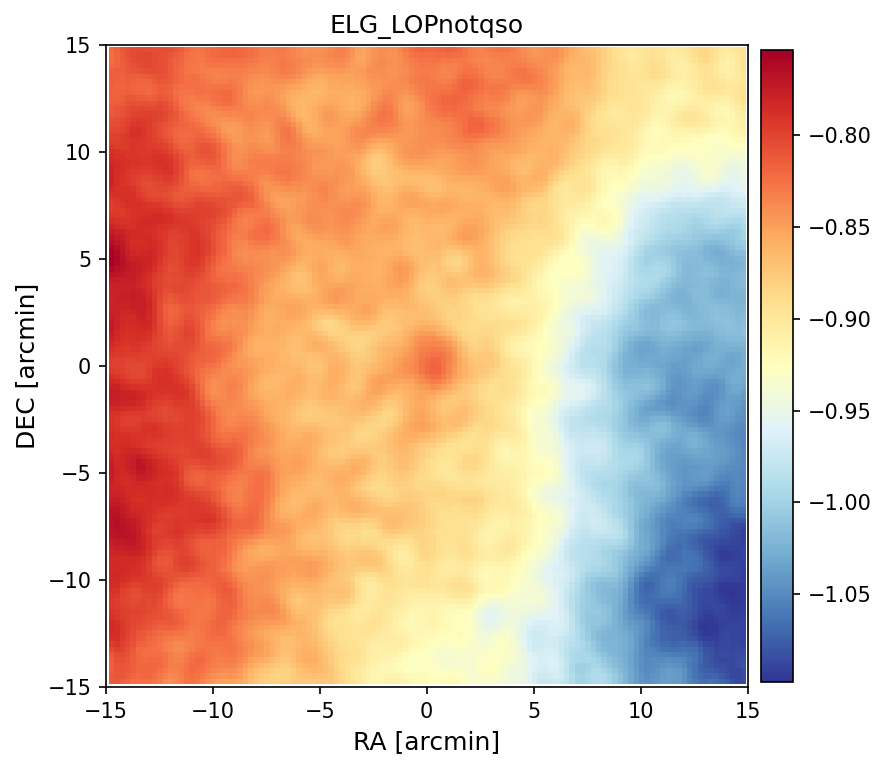}
    \caption{
    Velocity-weighted stacked CMB temperature cutouts (kSZ maps) for the BGS
    (\textit{top}) and ELG (\textit{bottom}) samples.
    Each pixel shows the mean CMB temperature increment weighted by the
    line-of-sight peculiar velocity of the galaxies, isolating the kSZ signal
    and providing a direct, pre-filtering view of the optical-depth field.
    The central feature reflects the kSZ imprint from the gas associated with
    each sample, while primary CMB fluctuations dominate on large angular
    scales, manifesting as the coherent large-scale modes visible in both
    panels.
    The contrast between the two panels reflects the angular-size difference
    expected from the BGS and ELG redshift distributions: the more nearby BGS
    halos are clearly more extended on the sky than the higher-redshift ELG
    halos, consistent with their respective effective virial radii
    (see Fig.~\ref{fig:ksz_stack}).
    }
    \label{fig:resMap}
\end{figure}

Fig.~\ref{fig:resMap} shows the velocity-weighted stacked CMB cutouts for the
BGS (top) and ELG (bottom) samples, providing a direct image-space view of the
kSZ signal prior to any aperture-photometry transformation.
The central feature traces the mean optical depth of the gas associated with
each galaxy population, while the surrounding large-scale fluctuations are
dominated by the primary CMB, whose power exceeds the kSZ signal on scales
beyond a few arcminutes.
The ACT DR6 beam visibly smooths the signal on $\mathcal{O}(1)$-arcminute scales, and the
two panels illustrate a clear angular-size difference: the more nearby BGS
halos appear noticeably extended, whereas the higher-redshift ELG halos are
largely unresolved at this resolution, consistent with their much smaller
effective virial radii.

\section{Posterior degeneracies of the GNFW parameters}
\label{app:corner}

\begin{figure}
\centering
\includegraphics[width=\linewidth]{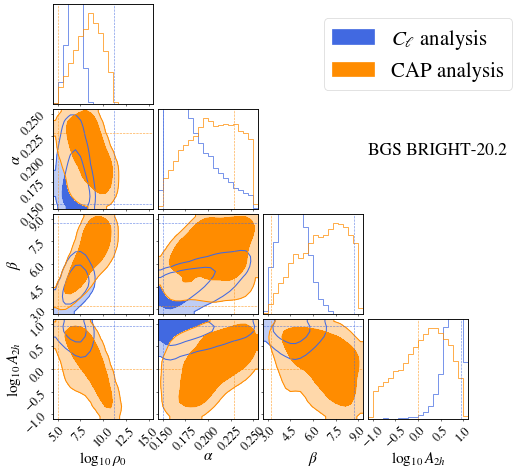}
\includegraphics[width=\linewidth]{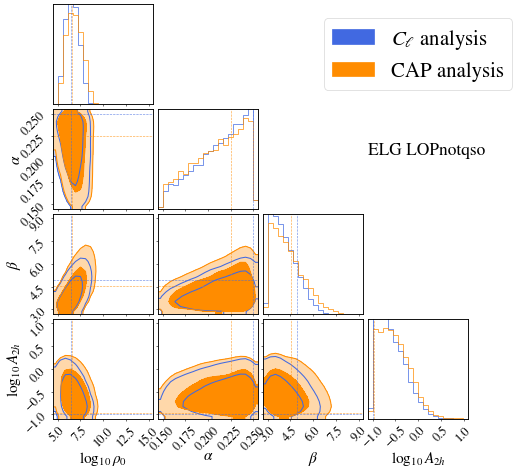}
\caption{
Corner plots showing the marginalized posterior distributions of the four free
GNFW parameters for representative samples.
The top panel corresponds to BGS ($\log M_\star > 10.0$) and the bottom panel to
ELG ($\log M_\star > 9.0$).
In each panel, we show both the harmonic-space (blue) and real-space posteriors (orange).
The parameters are the normalization $\rho_0$, the transition parameter $\alpha$,
the outer slope $\beta$, and the two-halo amplitude $A_{\rm k2h}$.
Contours denote the 68\% and 95\% credible regions.
}
\label{fig:corner_gnfw}
\end{figure}

In this Appendix, we illustrate the posterior structure and parameter degeneracies
of the GNFW model by showing representative corner plots from the \texttt{dynesty}
analysis in both harmonic and real space.
The top panel of Fig.~\ref{fig:corner_gnfw} shows the BGS sample with
$\log M_\star > 10.0$, while the bottom panel shows the ELG sample with
$\log M_\star > 9.0$.
Similar behavior is observed for the other mass bins.

In general, we expect the information contained in the two approaches to be equivalent, and this is broadly confirmed by Tables~\ref{tab:gnfw_results} and~\ref{tab:CAP_gnfw}, which show that while there remain $\mathcal{O}(1\sigma)$ differences in the recovered parameters, the significance of the detection is consistent between real and harmonic space, as one might expect. Some differences in the SNR and in the detailed posteriors remain, which can be attributed to several factors: (1) the relatively low significance of the detection combined with the large number of free parameters can lead to volume effects in the posterior; (2) degeneracies between the GNFW parameters can cause the chains to explore different parts of parameter space in real versus harmonic space; and (3) the binning of the data, which is necessarily different in the two approaches, makes each measurement sensitive to slightly different features of the underlying gas profiles. These considerations imply that minor differences between the posteriors are expected and do not indicate a failure of either method.

The four free parameters in our analysis are the normalization $\rho_0$,
the transition parameter $\alpha$, the outer slope $\beta$, and the two-halo
amplitude $A_{\rm k2h}$.
The most prominent feature of the posterior is a degeneracy between
$\rho_0$ and $\beta$.
Physically, $\rho_0$ sets the overall amplitude of the gas density profile,
while $\beta$ controls how rapidly the profile falls off at large radii.
From Eq.~(\ref{eq:gnfw_rho}), the outer regions scale approximately as
$\rho_{\rm gas} \propto \rho_0\, r^{-\beta}$ (up to fixed factors),
so an increase in the normalization can be partially compensated by a steeper
outer slope.
Over the limited radial range and signal-to-noise probed by the current data,
these two parameters therefore trade off against each other, producing an
extended degeneracy direction in the posterior.

With the present SNR, the GNFW parameters are only weakly constrained
individually.
The normalization $\rho_0$ and outer slope $\beta$ are the best determined,
as they most directly control the overall amplitude and effective shape of the
stacked signal.
By contrast, the transition parameter $\alpha$ and the two-halo amplitude
$A_{\rm k2h}$ remain largely unconstrained, with broad marginalized posteriors.
This reflects the limited sensitivity of the current measurements to the detailed
transition between the inner and outer profile regions and to the large-scale
two-halo contribution.
As the SNR improves in future analyses, these degeneracies will shrink and the
constraints on all GNFW parameters will tighten accordingly.

As shown in Fig.~\ref{fig:corner_gnfw}, the posterior contours obtained from the real- and harmonic-space fits are nearly identical for the ELG samples, indicating that the two estimators constrain the GNFW parameter space in a fully consistent manner. For the BGS samples, however, we observe a visible offset between the real- and harmonic-space contours. This difference arises from a strong degeneracy between $\log_{10}\rho_0$ and $\beta$, combined with the limited constraining power of the current data. Because the likelihood surface is relatively shallow along this degeneracy direction, small differences in how the information is projected in real versus harmonic space can shift the posterior mean. With increased signal-to-noise, the contours would contract and the two estimators would converge. Importantly, once this degeneracy is taken into account, the real- and harmonic-space constraints for BGS remain statistically consistent with each other.

\section{Real-space GNFW fits}
\label{app:real_snr}

\begin{figure}
\centering
\includegraphics[width=\linewidth]{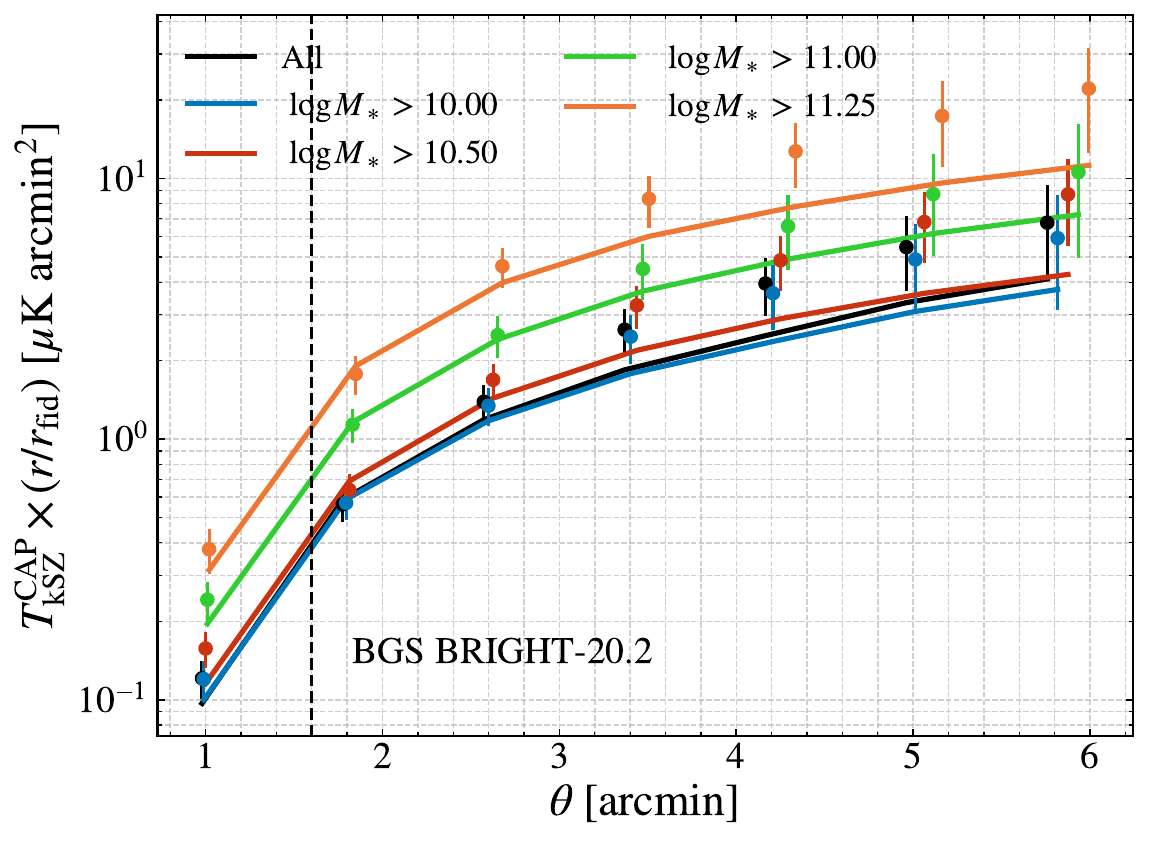}
\includegraphics[width=\linewidth]{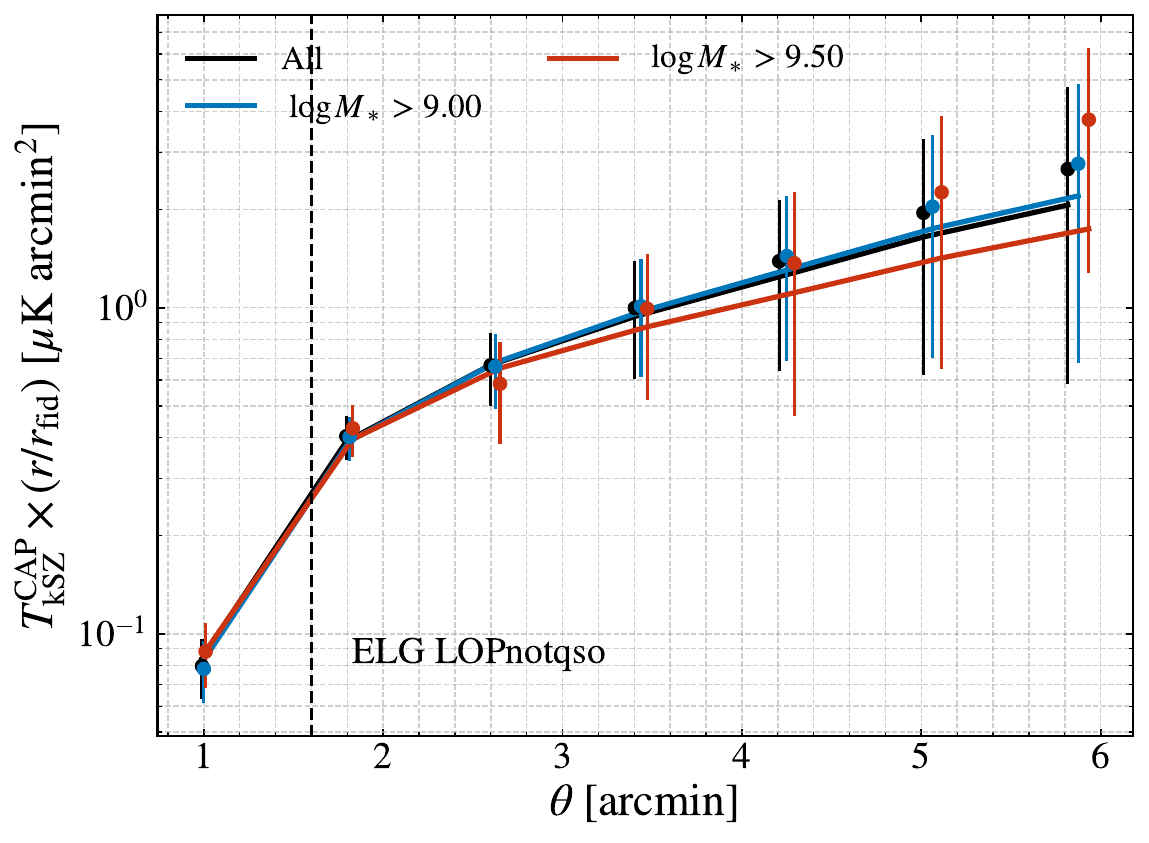}
\caption{Direct real-space GNFW fits to the stacked CAP profiles for representative samples.
The top panel shows BGS, and the bottom panel shows ELG.
Solid curves correspond to the best-fit real-space GNFW model obtained from the full covariance fit, while points denote the measured stacked kSZ profiles.
The fits provide an excellent description of the data across all radial bins.
Note that the data points are strongly correlated, so the visual scatter does not reflect independent measurements.
The corresponding signal-to-noise ratios for all samples are reported in Table~\ref{tab:CAP_gnfw}. Overall, the fits match the data very well, with the ELG sample at $\log M_\ast > 9.5$ being the most discrepant (but still within error bars).
}
\label{fig:CAP_gnfw}
\end{figure}

In this appendix, we perform a direct real-space analysis of the stacked kSZ signal by fitting the GNFW density profile introduced in Sec.~\ref{sec:gnfw_fit} to the measured CAP profiles. The model is evaluated directly in configuration space and compared to the stacked kSZ measurements for each galaxy sample using the full real-space covariance matrix. We carry out this procedure for the full ELG and BGS samples, as well as for the stellar-mass subsamples (two bins for ELG and four bins for BGS), resulting in a total of eight independent fits.

The best-fit parameters and corresponding signal-to-noise ratios are summarized in Table~\ref{tab:CAP_gnfw}, while the resulting model curves are shown in Fig.~\ref{fig:CAP_gnfw}. The agreement between the best-fit GNFW profiles and the data is excellent across all samples: the model captures both the overall amplitude and the radial dependence of the stacked signal. Although the fits formally involve 7 data points, the reduced $\chi^2$ values are significantly below unity. This reflects the fact that the real-space covariance matrix is highly correlated between radial bins, so that the effective number of independent degrees of freedom is substantially smaller than the number of data points. As in harmonic space, we observe non-trivial degeneracies between the GNFW parameters, which are illustrated explicitly in Appendix~\ref{app:corner}.

We note that the fit for the $\log M_\star > 10.0$ sample appears noticeably worse than for the other samples. We attribute this primarily to the second radial bin, which is systematically lower than in the corresponding bins of the other samples.
A similar behaviour is observed when comparing measurements of the full BGS sample using the hILC map (our fiducial CMB map) with those obtained from the single-frequency maps at 90\,GHz and 150\,GHz (see Fig.~\ref{fig:foreground}) In particular, the second radial bin shows a discrepancy between the 90\,GHz and 150\,GHz measurements. This suggests that residual foreground contamination in the hILC map may be affecting the fiducial measurement.

In principle, the information content in real and harmonic space is identical. Indeed, comparing Table~\ref{tab:CAP_gnfw} with the harmonic-space results in Table~\ref{tab:gnfw_results}, we find that the recovered SNR values are consistent between the two approaches. However, in the presence of parameter degeneracies, as is the case for our GNFW model, the posterior geometry can differ slightly between configuration and harmonic space. This leads to small shifts in the inferred parameter means, which can be seen by comparing Tables~\ref{tab:CAP_gnfw} and \ref{tab:gnfw_results}, as well as in the posterior contours presented in Appendix~\ref{app:corner}. Overall, the two analyses provide mutually consistent descriptions of the data.

\begin{table*}
\centering
\begin{tabular}{lccccccc}
\hline\hline
Sample &
$\log_{10}\rho_0$ &
$\alpha$ &
$\beta$ &
$\log_{10}A_{\rm k2h}$ &
$\chi^2_{\rm null}$ &
$\chi^2_{\rm bf}$ &
SNR \\
\hline

ELG (all) &
$6.23 \pm 0.58$ &
$0.214 \pm 0.026$ &
$3.96 \pm 0.72$ &
$-0.74 \pm 0.20$ &
56.71 &
0.64 &
7.49 \\[3pt]

ELG ($\log M_\star>9.0$) &
$6.24 \pm 0.60$ &
$0.214 \pm 0.026$ &
$3.98 \pm 0.72$ &
$-0.73 \pm 0.21$ &
53.23 &
0.73 &
7.25 \\[3pt]

ELG ($\log M_\star>9.5$) &
$6.32 \pm 0.62$ &
$0.213 \pm 0.027$ &
$4.05 \pm 0.80$ &
$-0.77 \pm 0.19$ &
46.56 &
6.91 &
6.30 \\[3pt]

BGS (all) &
$5.81 \pm 0.58$ &
$0.210 \pm 0.026$ &
$3.79 \pm 0.61$ &
$-0.29 \pm 0.41$ &
78.38 &
8.95 &
8.33 \\[3pt]

BGS ($\log M_\star>10.0$) &
$5.87 \pm 0.59$ &
$0.210 \pm 0.026$ &
$3.88 \pm 0.66$ &
$-0.29 \pm 0.40$ &
76.61 &
6.78 &
8.36 \\[3pt]

BGS ($\log M_\star>10.5$) &
$5.78 \pm 0.55$ &
$0.212 \pm 0.025$ &
$3.82 \pm 0.61$ &
$-0.39 \pm 0.37$ &
80.77 &
13.07 &
8.23 \\[3pt]

BGS ($\log M_\star>11.0$) &
$6.24 \pm 0.67$ &
$0.212 \pm 0.026$ &
$4.07 \pm 0.75$ &
$-0.55 \pm 0.32$ &
68.28 &
3.59 &
8.04 \\[3pt]

BGS ($\log M_\star>11.25$) &
$5.93 \pm 0.61$ &
$0.214 \pm 0.025$ &
$3.75 \pm 0.59$ &
$-0.40 \pm 0.40$ &
58.17 &
6.71 &
7.17 \\

\hline\hline
\end{tabular}
\caption{
Best-fit GNFW parameters obtained from direct real-space fits to the CAP profiles for the ELG and BGS samples. We report the posterior mean and $68\%$ confidence interval for the four free GNFW parameters, along with $\chi^2$ values for the null hypothesis and best-fit model, and the corresponding signal-to-noise ratio (SNR). All fits formally have 7 degrees of freedom (number of data points). The reduced $\chi^2$ values are significantly below unity because the real-space covariance matrix is highly correlated between radial bins; therefore the effective number of independent degrees of freedom is substantially smaller than the number of data points.
}
\label{tab:CAP_gnfw}
\end{table*}

\section{Real-space null tests}
\label{app:ksz_stack_null}

\begin{figure}[t]
    \centering
    \includegraphics[width=\linewidth]{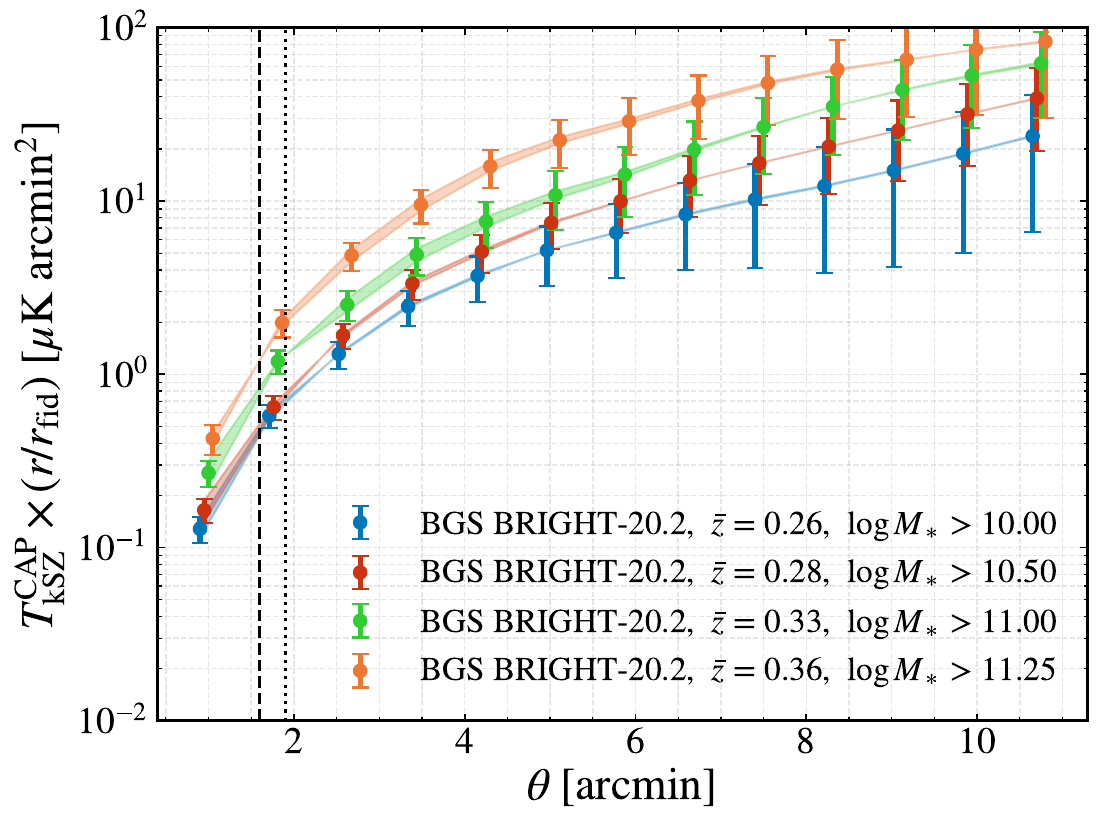}
    \includegraphics[width=\linewidth]{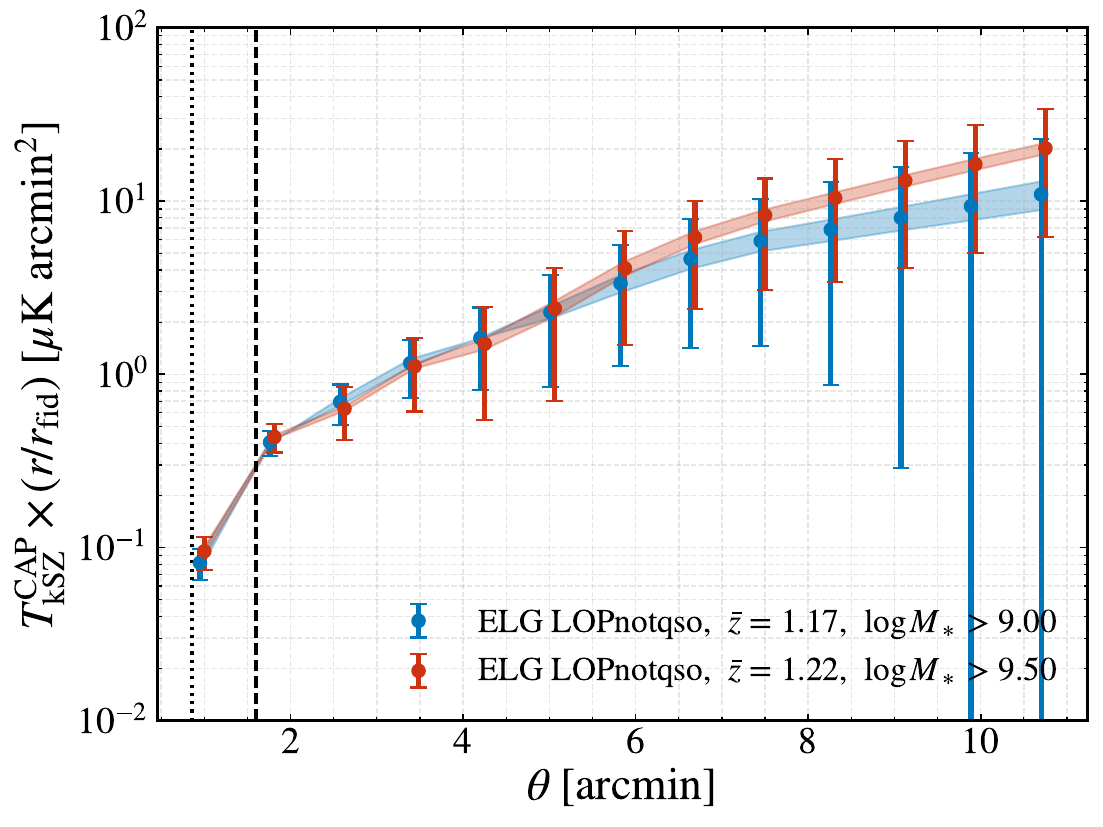}
    \caption{
    Real-space velocity shuffling null test for the stacked kSZ profiles of the BGS (top) and ELG (bottom) samples.
    Each panel shows the measured stacked kSZ profile (colored points with error bars) together with
    the mean null signal obtained from $10{,}000$ velocity-shuffling realizations (colored band).
    The band width corresponds to the added and subtracted mean from the velocity-shuffling realizations. This representation is chosen so that one can see by eye how large the bias is relative to the error bars of the measurement.
    For all samples the band lies well within the error bars,
    indicating no evidence for spurious correlations induced by the reconstruction or stacking
    procedure.}
    \label{fig:ksz_stack_null}
\end{figure}

In the main text (Fig.~\ref{fig:cl_ksz_null}) we presented two null tests of the
$C_\ell^{\,g\tau}$ measurement based on (i) applying random angular offsets to the
galaxy positions and (ii) shuffling the reconstructed line-of-sight velocities.
Here we repeat these tests in \emph{real space} using the stacked kSZ profiles.
The procedure is identical in spirit to the harmonic-space tests, but with two
key differences: (1) the signal is now the aperture-photometry kSZ profile
$r/r_{\rm fid}$ measured around each galaxy sample, and (2) for the velocity
shuffling test we generate $10{,}000$ independent shufflings of the LOS velocities.
A larger number of realizations is required in real space because the null
distribution is inherently noisier point-by-point than in the harmonic
representation.

For each of the ELG and BGS mass-selected subsamples, we compute the stacked
kSZ profile using the true velocities (our ``data'' measurement), and then
compute $10{,}000$ null profiles obtained by randomly permuting the velocities
among the galaxies. We show in Fig.~\ref{fig:ksz_stack_null} the mean of these
null realizations as a colored band around the measured profile. The closer that this averaged curve is to being zero (i.e., the smaller the band around the measured profiles), the smaller the potential bias from foregrounds and outliers is.

Across all ELG and BGS mass bins, the measured profiles lie well within
the error bars. Overall, these real-space null tests reinforce the conclusions from the
harmonic-space analysis: the reconstructed velocities and stacked pipeline do not introduce detectable systematic deviations.


\bibliography{refs,Misha_DESI_supporting_papers2025-05-11}{}
\bibliographystyle{prsty}



\end{document}